\documentclass[fleqn,usenatbib,useAMS]{mnras}

\usepackage{newtxtext,newtxmath}

\usepackage[T1]{fontenc}
\usepackage{ae,aecompl}

\DeclareRobustCommand{\VAN}[3]{#2}
\let\VANthebibliography\thebibliography
\def\thebibliography{\DeclareRobustCommand{\VAN}[3]{##3}\VANthebibliography}

\usepackage{graphicx}	
\usepackage{amsmath}	
\usepackage{gensymb}
\usepackage[export]{adjustbox}

\title[Ejected stars from NGC 2264]{Constraining the initial conditions of NGC 2264 using ejected stars found in \textit{Gaia} DR2}

\author[C. Schoettler, R. J. Parker and J. de Bruijne]{Christina Schoettler,$^{1}$\thanks{E-mail: cschoettler1@sheffield.ac.uk}
Richard J. Parker$^{1}$\thanks{Royal Society Dorothy Hodgkin Fellow}
and Jos de Bruijne$^{2}$
\\
$^{1}$Department of Physics and Astronomy, The University of Sheffield, Hicks Building, Hounsfield Road, Sheffield S3 7RH, UK\\
$^{2}$Science Support Office, Directorate of Science, European Space Research and Technology Centre (ESA/ESTEC)\\
Keplerlaan 1, NL-2201 AZ Noordwijk, The Netherlands}

\date{Accepted 2021 November 23. Received 2021 November 12; in original form 2021 June 18}

\pubyear{2021}

\begin{document}
\label{firstpage}
\pagerange{\pageref{firstpage}--\pageref{lastpage}}
\maketitle

\begin{abstract}
Fast, ejected stars have been found around several young star-forming regions, such as the Orion Nebula Cluster (ONC). These ejected stars can be used to constrain the initial density, spatial and kinematic substructure when compared to predictions from $N$-body simulations. We search for runaway and slower walkaway stars using \textit{Gaia} DR2 within 100 pc of NGC 2264, which contains subclustered regions around higher-mass OB-stars (S\,Mon, IRS\,1 and IRS\,2). We find five runaways and nine walkaways that trace back to S\,Mon and six runaways and five walkaways that trace back to IRS\,1/2 based on their 3D-kinematics. We compare these numbers to a range of $N$-body simulations with different initial conditions. The number of runaways/walkaways is consistent with initial conditions with a high initial stellar density ($\sim$10 000 M$_{\sun}$ pc$^{-3}$), a high initial amount of spatial substructure and either a subvirial or virialised ratio for all subclusters. We also confirm the trajectories of our ejected stars using the data from \textit{Gaia} EDR3, which reduces the number of runaways from IRS\,1/2 from six to four but leaves the number of runaways from S\,Mon unchanged. The reduction in runaways is due to smaller uncertainties in the proper motion and changes in the parallax/distance estimate for these stars in \textit{Gaia} EDR3. We find further runaway/walkaway candidates based on proper motion alone in \textit{Gaia} DR2, which could increase these numbers once radial velocities are available. We also expect further changes in the candidate list with upcoming \textit{Gaia} data releases.

\end{abstract}





\section{Introduction}

Stars moving at peculiar velocities above those in their immediate surroundings were discovered over 60 years ago and termed runaway (RW) stars \citep{RN325, RN255, RN67}. Their trajectories can often be traced back to close-by star-forming regions where many are thought to have been born before being ejected \citep[e.g.][]{RN312,RN309}. One of the causes of these ejections are dynamical interactions within the regions \citep[dynamical ejection scenario (DES),][]{RN189}, caused by close encounters between a binary and at least a single star. 

The dynamical interactions are an outcome of the dynamical evolution of these star-forming regions. During this evolution the regions can undergo a rapid change in their spatial and kinematic structure in a time period as short as a few Myr \citep[e.g.][]{RN4, RN258}, which can erase this initial structure \citep[e.g.][]{RN14,RN4,RN248}, making it difficult to infer the initial conditions of these regions.

Further effects of the dynamical evolution are a reduction in stellar density \citep[e.g.][]{RN277,RN8}, even though these regions still have higher densities than those in the Galactic field \citep{RN25, RN59}. We also expect the destruction of primordial binaries/multiples \citep[e.g.][]{RN277,RN8}, dynamical mass segregation \citep[e.g.][]{2007ApJ...655L..45M,RN15,2009MNRAS.396.1864M,2009MNRAS.400..657M,RN5} and detrimental effects on protoplanetary discs and young planetary systems, i.e. disruption or destruction of the discs and planetary systems during a denser phase \citep[e.g.][]{RN271,RN272,RN269,2016ApJ...828...48V,2019MNRAS.485.4893N}. 

Ejected stars are now often categorised according to their peculiar velocity with RWs typically stars with a velocity exceeding $\sim$30-40\,km\,s$^{-1}$ \citep[e.g. ][]{RN255, RN67, RN276, RN50, RN190, RN137,2011A&A...530L..14B,RN293} reaching up to $\sim$400--500\,km\,s$^{-1}$ \citep[e.g.][]{RN241,2020MNRAS.495.3104S}. The second group are the slower walkaway (WW) stars moving at least at $\sim$5\,km\,s$^{-1}$ \citep{RN137, RN136}. 

Both RW and WW ejections are mainly caused by the DES or the binary supernova scenario \citep[BSS,][]{RN67}. In the BSS the starting point is a close binary with a massive primary star, which undergoes a core-collapse supernova. This can cause a disruption of the binary and the ejection of the less massive companion star at velocities up to the previous orbital velocity. Unlike the DES, which can occur right from early ages of a young star-forming region, the BSS happens only after a few Myr, when the most massive stars start to reach the end of their main-sequence lives. 

Historically, RWs were often brighter, higher mass (OB) stars, as these stars were easier to observe compared to stars with lower mass, however, this has changed over the past few years \citep[e.g.][]{RN263,RN252,2019JKAS...52...57Y,2019ApJ...884....6M, 2020MNRAS.495.3104S,2020AN....341..908B,2020AJ....159..272P}. The advent of the \textit{Gaia} telescope and in particular the second data release (DR2) in April 2018 \citep{RN308, RN238} has allowed the discovery of a much larger number of RWs, WWs, hyper-runaway and even faster hypervelocity stars than ever before \citep[e.g.][]{RN316,RN320,RN315,RN313,RN319,RN323,2018ApJ...866...39B,2020MNRAS.493.1512R,2021ApJS..252....3L}. The improvement in accurate astrometry continues in the most recent \textit{Gaia} Early Data Release 3 (EDR3) to provide ever more accurate data for an ever expanding number of stars \citep{2020arXiv201201533G}.


In \citet{2020MNRAS.495.3104S}, we used \textit{Gaia} DR2 observations of stars in the vicinity (within 100 pc) of the ONC to search for RWs and WWs. We found nine RWs and 24 WWs in the observational data, which we compared to $N$-body simulations that were tailored to a set of initial conditions based on other constraints from the literature. We showed that the ejected stars from the ONC were consistent with those predicted from simulations with an initially moderate amount of spatial substructure and an initially subvirial ratio up to an age of 2.4 Myr, which is consistent with the age estimate for the ONC from other sources.


In this paper, we use \textit{Gaia} DR2 and EDR3 observations to search for RW and WW stars around NGC 2264 and we compare these numbers to $N$-body simulations with different initial conditions to constrain the initial conditions of NGC 2264. NGC 2264 is another region within 1 kpc to Earth showing on-going star formation, which should allow us to probe the ejected population down to sub-solar masses. The region is less centrally concentrated with obvious subclustering and fewer studies have investigated the initial conditions of this star-forming region. A detailed study of the kinematics will allow us to more broadly test our approach using RW stars to constrain initial conditions.  

This paper is organised as follows. In Section 2, we describe our target and the \textit{Gaia} DR2 data selection process. Section 3 describes the \textit{Gaia} DR2 data analysis process. In Section 4, we present the results of our search. This is followed by a brief description of our $N$-body simulation set-up and predictions for RWs/WWs from different sets of initial conditions in Section 5. In Section 6, we provide a discussion of our results and we conclude this paper in Section 7. 

\section{Target and data selection}

\subsection{Search target}
We are searching for RW and WW stars from the young star-forming region NGC 2264, which is located in the Monoceros OB1 association cloud complex. Recent estimates put this region at distances ranging from approx. 720 pc (719 pc: \citeauthor{2019A&A...630A.119M} \citeyear{2019A&A...630A.119M}; 723 pc: \citeauthor{2018A&A...618A..93C} \citeyear{2019A&A...630A.119M}) up to 750 pc \citep[][]{RN264}. All of these estimates used \textit{Gaia} DR2 and are subject to considerable error margins ($\pm$16 pc to $\pm$60 pc). Another recent estimate by \citet{2018ApJ...855...12Z} used Near-Infrared (NIR) extinction from 2MASS photometry to derive a much larger distance estimate of 1.2 kpc. Older estimates also encompassed a large range from 667 pc \citep{2002A&A...389..871D, 2012yCat....102022D} and 760 pc \citep{2008hsf1.book..966D} up to 913 pc \citep{2009AJ....138..963B} with similarly high margins of error.

NGC 2264 is not a centrally concentrated cluster, but spatially elongated along a NW-SE orientation with different subclusters spread along $\sim$8 pc \citep{2015AJ....149..119T, 2020A&A...636A..80B}. 
\citet{2017A&A...601A.101Z} suggested a diameter of $\sim$39 arcmin, which translates to 8-14 pc depending on the distance used to convert from arcmin to parsec. The northern region is dominated by S\,Monocerotis (S\,Mon), an O7 spectroscopic binary \citep[$\sim$40-50 M$_{\sun}$;][]{2018ApJS..235....6T, 2019A&A...630A.119M}. The southern region consists of two subclusters. These subclusters have differing designations in literature but are most often referred to as NGC 2264-C or IRS\,1 and NGC 2264-D, IRS\,2 or the Spokes cluster \citep{2006ApJ...636L..45T}.  IRS\,1 is thought to centre around a B2-type star with a mass estimate of $\sim$10 M$_{\sun}$ \citep[IRAS~06384+0932;][]{1972ApJ...172L..55A,1998ApJ...492L.177T,2006A&A...445..979P}. IRS\,2 is thought to centre around a young Class 1 type source, which is possibly a B-type binary with a primary star of $\sim$8 M$_{\sun}$  \citep[IRAS~06382+0939;][]{1988ApJ...335..150C,1989ApJ...345..906M, 2006ApJ...636L..45T}. 

\citet{RN25} suggested that NGC 2264 appears to be a cluster containing at least two different areas of increased surface density in a hierarchical structure. \citet{2019MNRAS.490.2521H} calculated a surface density for the whole region of NGC 2264 of $\sim$7 stars pc$^{-2}$ and calculated a radius for the region of $\sim$3.8 pc, which is roughly in line with the diameter values from other authors. Slightly higher surface density values were recorded by \citet{2014ApJ...794..124R}, who showed that most of the young stellar objects (YSOs) in NGC 2264 were found in regions with densities above $\sim$10 stars pc$^{-2}$ with a peak between $\sim$10--25 stars pc$^{-2}$. \citet{2006A&A...445..979P} suggested a stellar density for the southern region around IRS\,1/2 of $\sim$80 stars pc$^{-2}$ based on a figure from \citet{1993ApJ...408..471L}. \citet{2013ApJ...772...81M} calculated a value for the stellar density of $\sim$30 stars pc$^{-2}$ for the southern region.

There is very little foreground extinction and the dark cloud located directly in the background reduces the contamination from non-member stars \citep{1954ApJ...119..483H,2018A&A...609A..10V}. While this is helpful when identifying member stars still located in NGC 2264, it makes it more difficult to find stars ejected in radial direction moving away behind the cluster.

The age of the cluster differs across the different regions of NGC 2264. The average age of the cluster is suggested to be 3--5 Myr with an age spread of 4--5 Myr, with the region around S\,Mon containing older stars whereas the two subclusters in the south are thought to be younger as a result of sequential star formation \citep{2008hsf1.book..966D,2008MNRAS.386..261M,2009AJ....138.1116S, 2009MNRAS.399..432N, 2018A&A...609A..10V}. A recent study by \citet{2019MNRAS.490.2521H} used a sample of stars spread across the whole region of NGC 2264 and calculated an average age of $\sim$2 Myr. \citet{2014ApJ...787..108G} also calculated ages for stars in NGC 2264 with the mean value suggesting an age of $\sim$2.6 Myr.
\citet{2007A&A...464..983P} calculated very young ages for the southern subclusters IRS\,1 and IRS\,2 of $\sim$0.1 Myr, whereas \citet{2013ApJ...772...81M} calculated a median age of $\sim$1 Myr for this region.

All of these age estimates suggest that star formation activity started first in the northern region where S\,Mon is now located over 5 Myr ago and in the southern region more recently $\sim$1.5--2 Myr ago. Furthermore, in the Cone nebula located around IRS\,1 there are also a number of embedded sources further suggesting a young age \citep{2018A&A...609A..10V}.

\begin{table*}
	\centering
	\caption{NGC 2264 centre parameters used in the analysis from \citet{RN264}}
	\label{tab:Centre_para}
	\begin{tabular}{lccc}
		\hline
		 & S\,Mon & IRS\,1 & IRS\,2 \\
		\hline	
		Right ascension [RA] (ICRS) $\alpha_{0}$ & 6h 40m 50s  & 6h 41m 07s & 6h 41m 01s  \\ 
		Declination [Dec] (ICRS) $\delta_{0}$ & 09$\degree\,$51$^\prime\,$03$^{\prime\prime}$ &  09$\degree\,$28$^\prime\,$09$^{\prime\prime}$  & 
		09$\degree\,$35$^\prime\,$56$^{\prime\prime}$  \\ 
		Proper motion RA ${\mu}_{{\alpha }^{\star},0}$ (mas\,yr$^{-1}$) & -1.62\,$\pm$\,0.08 & -2.05\,$\pm$\,0.18  & -2.29\,$\pm$\,0.14 \\
		Proper motion Dec ${\mu}_{{\delta}_{0}}$ (mas\,yr$^{-1}$) & -3.71\,$\pm$\,0.07  & -3.90\,$\pm$\,0.09  & -3.61\,$\pm$\,0.08 \\
		RV (\,km\,s$^{-1}$) & 15.8,$\pm$\,2.9  & 15.8,$\pm$\,2.9  & 15.8,$\pm$\,2.9 \\
		\hline
		Adopted distance (pc) & 738\,$\pm$\,23 & 736\,$\pm$\,23 & 748\,$\pm$\,24\\
        \hline
	\end{tabular}
\end{table*}

In this paper, we separately analyse the northern (S\,Mon) and the two southern (IRS\,1 and IRS\,2) subclusters of NGC 2264 due to the difference in age and search for RW and WW stars that have been ejected from all three. We use an upper age of 5 Myr for the search around S\,Mon and 2 Myr for the search around IRS\,1 and IRS\,2, thereby covering most of the available age estimates. We use the distances to these subclusters from \citet{RN264} for our analysis, as shown in Table~\ref{tab:Centre_para}. These distances are at the upper end of the distance estimates for NGC 2264 from \textit{Gaia} DR2 measurements and have considerable uncertainties.

\citet{2012A&A...540A..83T} used IR-luminosity functions to estimate the size of the stellar population in NGC 2264 within their search fields that contain all three of the regions of interest in this paper (S\,Mon, IRS\,1 and IRS\,2). Their results suggested that the whole cluster contains 1436\,$\pm$242 members. They also provided an estimate of the stellar mass of the region using a simple assumption that each star has a mass of 0.5 M$_{\sun}$, which they adopted from \citet{2007AJ....134..411M}. This resulted in their stellar mass estimate of 718 $\pm$121 M$_{\sun}$ \citep{2012A&A...540A..83T}. 

This mass estimate means that NGC 2264 is considerably less massive than the ONC. This lower mass estimate reduces the required escape velocity from the cluster compared to the value for the ONC \citet{RN322} and therefore also affects our choice for the lower velocity limit for walkaway stars \citep{2020MNRAS.495.3104S}. Given the much lower mass, we choose the original velocity of 5\,km\,s$^{-1}$ suggested in \citet{RN137} as our lower velocity limit for WW stars and consider RWs to have velocities exceeding 30 \,km\,s$^{-1}$.

There has not yet been any comprehensive search for RW/WW stars from NGC 2264, however, \citet{2020A&A...643A.138M} found two potential RW 2D-candidates that appear to have been ejected from the northern region of the cluster: \textit{Gaia} DR2 3326734332924414976 and \textit{Gaia} DR2 3326951215889632128. We will seek to confirm these stars in our analysis.

In \citet{2021MNRAS.501L..12S}, we searched for circumstellar discs around RW/WW stars from the ONC, as well as around future and past visitors to determine the resilience of protoplanetary discs to multiple encounters within dense environments. For NGC 2264, we repeat this search in the Simbad/VizieR databases \citep{2000A&AS..143....9W, 2000A&AS..143...23O} for studies that have covered this cluster and search for circumstellar discs to possibly increase our sample size of ejected star-disc systems that have or will encounter a second dense star-forming region.

\subsection{Data selection and filtering}

We use the centre parameters for position and velocity as defined in \citet{RN264} with 100 pc as the outer boundary for our search region. Choosing an outer boundary for our search region, both in the observations and the simulations, allows us to compare similar volumes of space. 100 pc translates to $\sim$8$\degree$ around each of the centres of the three subclusters evaluated here. Instead of parallax, we use the distances from the catalogue by \citet{RN305} and reduce the sample by selecting sources that are within $\pm$100 pc of each of the centre distances, as shown in Table~\ref{tab:Centre_para}.

\citet{2020MNRAS.496.4701J} suggested central velocity parameters for NGC 2264, which are close to the values of \citet{RN264} in proper motion ($\sim$-1.95 mas\,yr$^{-1}$ and $\sim$-3.76 mas\,yr$^{-1}$, when converted from km s$^{-1}$ using the distance to S\,Mon in Table~\ref{tab:Centre_para}). However, the values differ more in the radial velocity (RV), where the higher \citet{2020MNRAS.496.4701J} value of $\sim$20.3 km s$^{-1}$ is closer to the higher radial velocities quoted in other studies \citep{2005A&A...438.1163K,2006ApJ...648.1090F}. We have run our analysis with this higher RV as well as the one in \citet{RN264} and find only minor differences in the results. We use the RV from \citet{RN264} throughout this analysis to be consistent with the choice of the other velocity parameters for the centre regions.

As in \citet{2020MNRAS.495.3104S}, we filter our data for astrometric and photometric quality. We use the re-normalised unit weight error (RUWE) \citep[technical note GAIA-C3-TN-LU-LL-124-01,][]{LL:LL-124} for one of the two quality cuts. When plotting our data set as described in the technical note and as shown in \citet{2020MNRAS.495.3104S}, we find that an upper RUWE limit of 1.3 to be the best fitting value.

For the photometric quality cut, we keep those sources within the excess noise filter (flux excess factor $E$) range suggested in \citet{RN307} and \citet{2018A&A...616A..10G}:
\begin{equation}\label{eq:Phot_excess}
    1.0 + 0.015(G_{\text{BP}}-G_{\text{RP}})^2 < E < 1.3 + 0.06(G_{\text{BP}}-G_{\text{RP}})^2
\end{equation}

\noindent With this filter, sources that show spurious photometry (mainly faint sources located in dense, crowded areas) are removed \citep{RN307,2018A&A...616A..10G}. 

\section{\textit{Gaia} DR2 Data analysis}

For our data analysis, we adjust the approach shown in \citet{2020MNRAS.495.3104S}, which followed \citet{RN264} and \citet{2019MNRAS.487.2977G}. We incorporate the changes described in \citet{2020RNAAS...4..116V}, which take into account the effects of individual stars' RV on their proper motion when converting ICRS coordinates and velocities into a Cartesian coordinate system.


For stars without RV, we use the RV of the cluster centre for this conversion. We then use the transformation matrix described in \citet{2020RNAAS...4..116V} to convert both the positions and velocities to be cluster-aligned and then shift the origin of the coordinate system to the cluster centre.

We first remove the Sun's peculiar motion relative to the Local Standard of Rest (LSR) \citep[using velocity values from][]{2010MNRAS.403.1829S} from the velocity parameters of the three NGC 2264 subcluster centres (see Table~\ref{tab:Centre_para}) and from the velocities of all stars in our data set. We then convert these velocities into cluster-centred velocities and apply a rest frame that is centred on the subclusters by subtracting the central velocity parameters. We define the xy-plane as a projected representation of the positions on the sky and the radial direction is represented by the z-direction.

\subsection{Search procedure}\label{Search_proc}

In our search for RW (velocity > 30\,km\,s$^{-1}$) and WW (velocity: 5--30\,km\,s$^{-1}$) candidates, we first trace back their positions in the xy-plane for up to 2 and 5 Myr depending on the subcluster using the converted proper motions. We define a cluster boundary based on existing member lists and once the backwards path of a star crosses this boundary, a star becomes a 2D-candidate. We use the time at which this path intersects the boundary as the minimum time since ejection, i.e. the flight time.

As we are interested in ejections from the subclusters, we define a radius of 2 pc in the xy-direction as our subcluster radius. When considering the whole star-forming region, this will amount to it extending across $\sim$2 pc in the x-direction (right ascension), while extending to over $\sim$8 pc in the y-direction (declination), due to the NW-SE elongation of the whole cluster. This 2 pc radius corresponds to an angular size of $\sim$0.16\degree\,around the subcluster centre positions, which are located at the origin in the rest-frame for each converted data set. 

\begin{figure}
    \centering
    \includegraphics[width=1.0\linewidth]{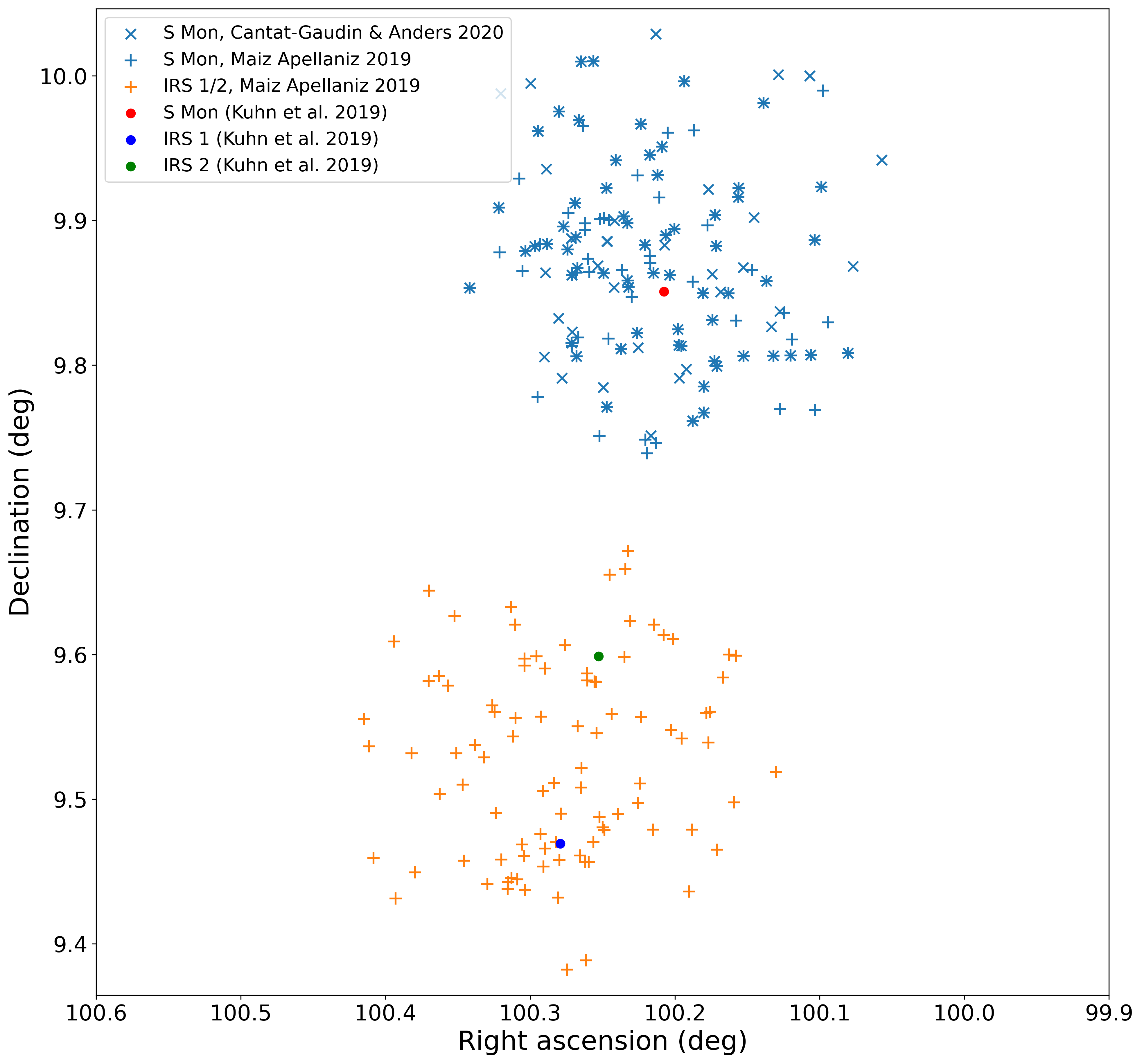}
    \caption{\textit{Gaia} DR2 right ascension and declination of NGC 2264 members identified in \citet[][members with 100 percent probability, only around S\,Mon, blue ``x'']{2020A&A...633A..99C} and \citet[][North cluster: blue ``+'', South cluster: orange ``+'']{2019A&A...630A.119M}. The centres of S\,Mon, IRS\,1 and IRS\,2 are plotted based on their positions in \citet{RN264}.}
    \label{fig:2264_ra_dec}
\end{figure}

Fig.~\ref{fig:2264_ra_dec} illustrates this choice, where we plot cluster members identified in \citet[][members with 100 per cent probability, only around S\,Mon, no members identified around IRS\,1/2]{2020A&A...633A..99C} and in \citet[][North and South cluster]{2019A&A...630A.119M} using their \textit{Gaia} DR2 measurements for the right ascension and declination.
These regions are each well enclosed by a circle with a radius of $\sim$0.16\degree. 

\begin{figure}
    \centering
    \includegraphics[width=1.0\linewidth]{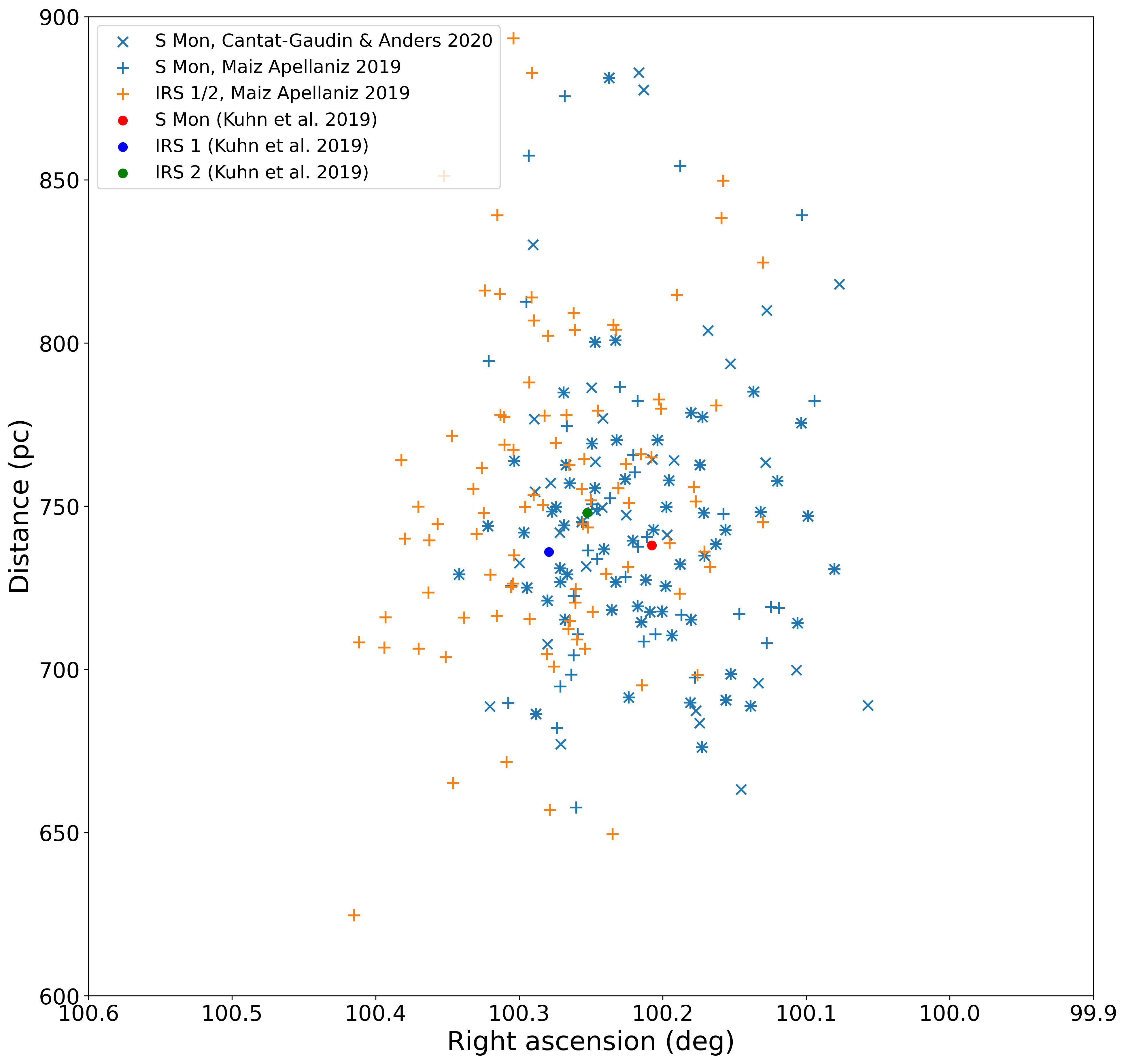}
    \caption{\textit{Gaia} DR2 right ascension and distance (inverted parallax) of NGC 2264 members identified in \citet[][members with 100 percent probability, only around S\,Mon, blue ``x'']{2020A&A...633A..99C} and \citet[][North cluster: blue ``+'', South cluster: orange ``+'']{2019A&A...630A.119M}. The centres of S\,Mon, IRS\,1 and IRS\,2 are plotted based on positions in \citet{RN264}.}
    \label{fig:2264_distance}
\end{figure}

To constrain the size of NGC 2264 in the radial direction (distance), one of the issues is the uncertainties in the distance of the subcluster centres \citep[][]{RN264}, another is the lack of information about their radial extent. The available membership lists could be helpful in constraining the radial extent of NGC 2264. In Fig.~\ref{fig:2264_distance}, we again plot the members from \citet{2019A&A...630A.119M} and \citet{2020A&A...633A..99C} using their \textit{Gaia} DR2 right ascension and distance (inverted parallax). There appears to be a higher stellar surface density between 700--800 pc. This could suggest a radial extent of NGC 2264 of several tens of pc with a centre distance of 740--750 pc but could also be driven by the parallax errors.

As Fig.~\ref{fig:2264_distance} does not allow us to put better constraints on the size in radial direction, we use the subcluster centre distance estimates in \citet[][736--748 pc]{RN264}, which show upper uncertainties of 23--24 pc in the subcluster centres. We add these uncertainties to the assumed 2 pc on-the-sky cluster radius turning our chosen search radius in the radial direction to 25--26 pc depending on subcluster. However, we do not suggest that this value represents the actual radial extent of NGC 2264. The circular cluster boundary used in the xy-plane turns to an elliptical cluster boundary in the xz- and yz-plane with a semi-minor axis of 2 pc (in the x and y axes) and a semi-major axis of 25--26 pc (in the z axis) depending on the subcluster.

Only $\sim$2 per cent of the stars in our NGC 2264 data set have RVs in \textit{Gaia} DR2, which makes the search for ejected stars using 3D kinematics difficult. For our 2D-candidates without RV, we have searched through the Simbad/VizieR databases \citep{2000A&AS..143....9W, 2000A&AS..143...23O} and find several additional RVs in \citet{2016A&A...586A..52J}, \citet{2019AJ....157..196K} and individual measurements in other sources. We also cross-match positions with Lamost DR5 data \citep{2019yCat.5164....0L} via VizieR (using a target match radius of 2 arcsec) to complete our data set with RV measurements from secondary literature sources. 

We then trace back any 2D-candidates with RV also in the xz- and yz-directions and if they positively trace back in these two additional planes, the 2D-candidate becomes a 3D-candidate. For the trace back, we consider errors in velocity and distance, due to their larger values. However, we do not consider the errors from \textit{Gaia} DR2 for the on-the-sky positions as these are considerably smaller. In practice, this approach leads to any star becoming a candidate if it traces back using a combination of highest, average, lowest velocities and distances (e.g. a star will become a candidate if it traces back using its proper motion plus related uncertainty, its radial velocity minus related uncertainty and its average distance).

We do not consider the influence of the Galactic potential during the trace-back and assume constant proper motion (linear trajectory) once the star has been ejected. As mentioned in \citet{2020ApJ...900...14F}, this can introduce a systematic error, which increases with distance of the ejected star to the cluster. However, we consider the effect of these systematics negligible in comparison to the larger errors in distance and RV. \citet{2018MNRAS.476..381W} compared the results of linear trace-back trajectories to those using an epicycle approximation for the Scorpius-Centaurus OB association. Their results indicated that there are no significant differences in the velocity dispersion distributions between these two approaches and that effects from the Galactic potential only start to dominate at older ages than those covered in our analysis of NGC 2264. Furthermore, \citet{2021MNRAS.508.4952D} recently also suggested that linear trajectories provided robust results for their analysis of RWs from the Carina Arm with flight times below $\sim$4 Myr.

For all our 2D-candidates, we search for mass estimates, ages and spectral types from literature sources. We find mass estimates for most of our candidates in the StarHorse database \citep[][]{2019A&A...628A..94A}, where we choose the 50th percentile and in the TESS Input Catalogue  \citep[v8.0,][]{2019AJ....158..138S}. We also find additional masses for individual stars from other literature sources. For any stars with more than one mass estimate, we quote the whole range of the values in the data tables.

\subsection{Constructing the CAMD}\label{CMD_constr}

Not all stars that we can trace back to the subclusters of NGC 2264 have originated in those regions. In order to identify ejected stars from our search targets, we use a colour--absolute magnitude diagram (CAMD) and PARSEC isochrones \citep[version 1.2S,][]{RN225} to differentiate older stars from those young enough to have originated in the cluster. The construction of the CAMD follows the procedure described in \citet{2020MNRAS.495.3104S}.

Our data set covers a region with a radius of 100 pc around the respective subcluster centres, so we convert the apparent $G$-band magnitude of each star $G$ to its absolute magnitude $M_{\rm{G}}$ using its distance $r$ in pc \citep{RN305} in equation 2 in \citet{RN303}:
\begin{equation}\label{Distance_Modulus}
    M_{\rm{G}} = G +5 - 5 \log_{10} r - A_{\rm{G}}\,.
\end{equation}

\noindent This equation also includes a correction for the extinction in the $G$-band, $A_{\rm{G}}$; we denote extinction-corrected absolute magnitudes as $M_{\rm{G,0}}$. We also correct $G_{\rm{BP}}-G_{\rm{RP}}$ for reddening using $E(G_{\rm{BP}}-G_{\rm{RP}})$ and denote this as $(G_{\rm{BP}}-G_{\rm{RP}})_0$.

\subsubsection{Extinction and reddening correction}

The \textit{Gaia} DR2 catalogue includes values for extinction and reddening, but only for a small subset of all sources in the data set. For the stars with missing values, we estimate these correction values following the approach described in \citet{2020MNRAS.495.3104S}, which is based on \citet{2018A&A...620A.172Z}. The missing values for a star are estimated from averages of \textit{Gaia} DR2 extinction and reddening values from surrounding stars within a distance of 10 pc. 

\subsubsection{Age estimate using PARSEC isochrones}

We use an upper age limit of 5 Myr for S\,Mon and 2 Myr for IRS\,1 and IRS\,2 and only stars that are younger than these ages are considered as ejected stars from these subclusters. We use PARSEC isochrones \citep[version 1.2S,][]{RN225}  in order to allow us to separate the stars in our data set into two age brackets. Younger stars are either fully located above the isochrones or when located below the isochrone have error bars crossing the age boundary. We download data\footnote{http://stev.oapd.inaf.it/cgi-bin/cmd} to produce an isochrone using a linear age of 2 and 5 Myr and select an Initial Mass Function (IMF) option similar to the one we use in our simulations, i.e. a combination of \citet{2001ApJ...554.1274C} and \citet{RN204}. 

We do not adjust for extinction in the isochrone, as we consider this in the stellar data and use the \citet{2018A&A...619A.180M} passbands. Other passband options available for \textit{Gaia} DR2 are \citet{2018A&A...617A.138W} and \citet{2018A&A...616A...4E}. Regardless of which of these three passbands options we apply to our data, the results do not change. The isochrones chosen for our analysis \citep[passbands from][]{2018A&A...619A.180M} result in a better fit to the higher-mass end of the CAMD, where the stars have already reached the main-sequence. 

Not much is known about the metallicity of NGC 2264. \citet{2020A&A...634A..34B} gave an abundance ratio [Fe/H] = 0.11, but used only one star to determine this value. The average value given in \citet{2017A&A...601A..70S} was [Fe/H] = -0.06. \citet{2000ApJ...533..944K} suggested [Fe/H] = -0.15, but calculated this value from only three stars. \citet{2014A&A...561A..93H} re-evaluated the observational data of \citet{2000ApJ...533..944K} and applied restrictions on certain parameters to get an abundance [Fe/H] = -0.13. This value was also found by \citet{2016A&A...585A.150N}. We use the abundance ratio [Fe/H] = -0.13 for our isochrones, which results in a metallicity Z = 0.011. Using a different value such as [Fe/H] = -0.06 \citep{2017A&A...601A..70S} has no effect on the results as the position of the isochrone on the CAMDs barely changes.

\begin{figure*}
    \centering
    \begin{minipage}[t]{1.0\columnwidth}
        \centering
        \vspace{0pt}
    	\includegraphics[width=1.0\columnwidth]{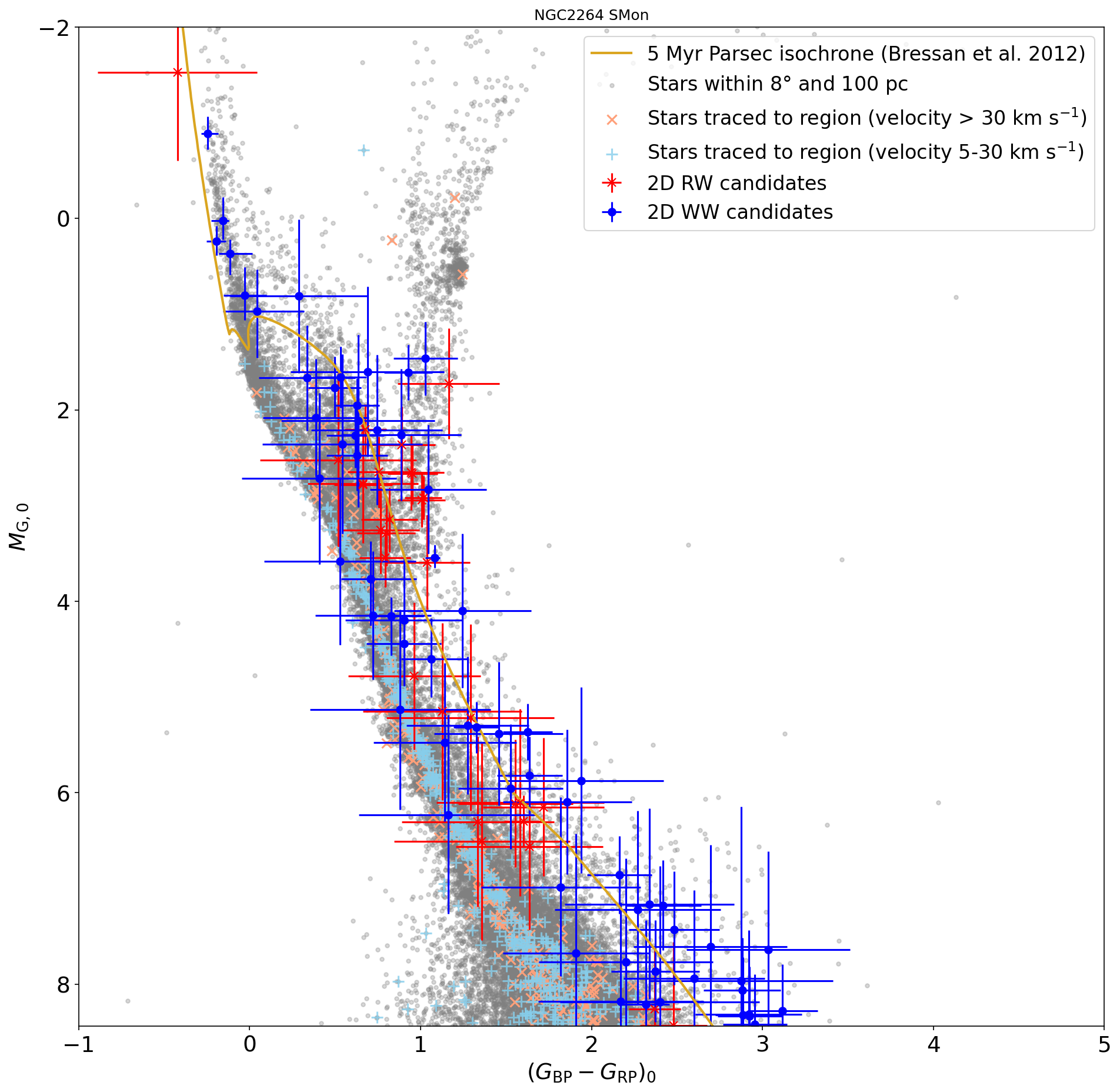}
    \end{minipage}
    \hfill
    \begin{minipage}[t]{1.0\columnwidth}
        \centering
        \vspace{0pt}
        \includegraphics[width=1.0\columnwidth]{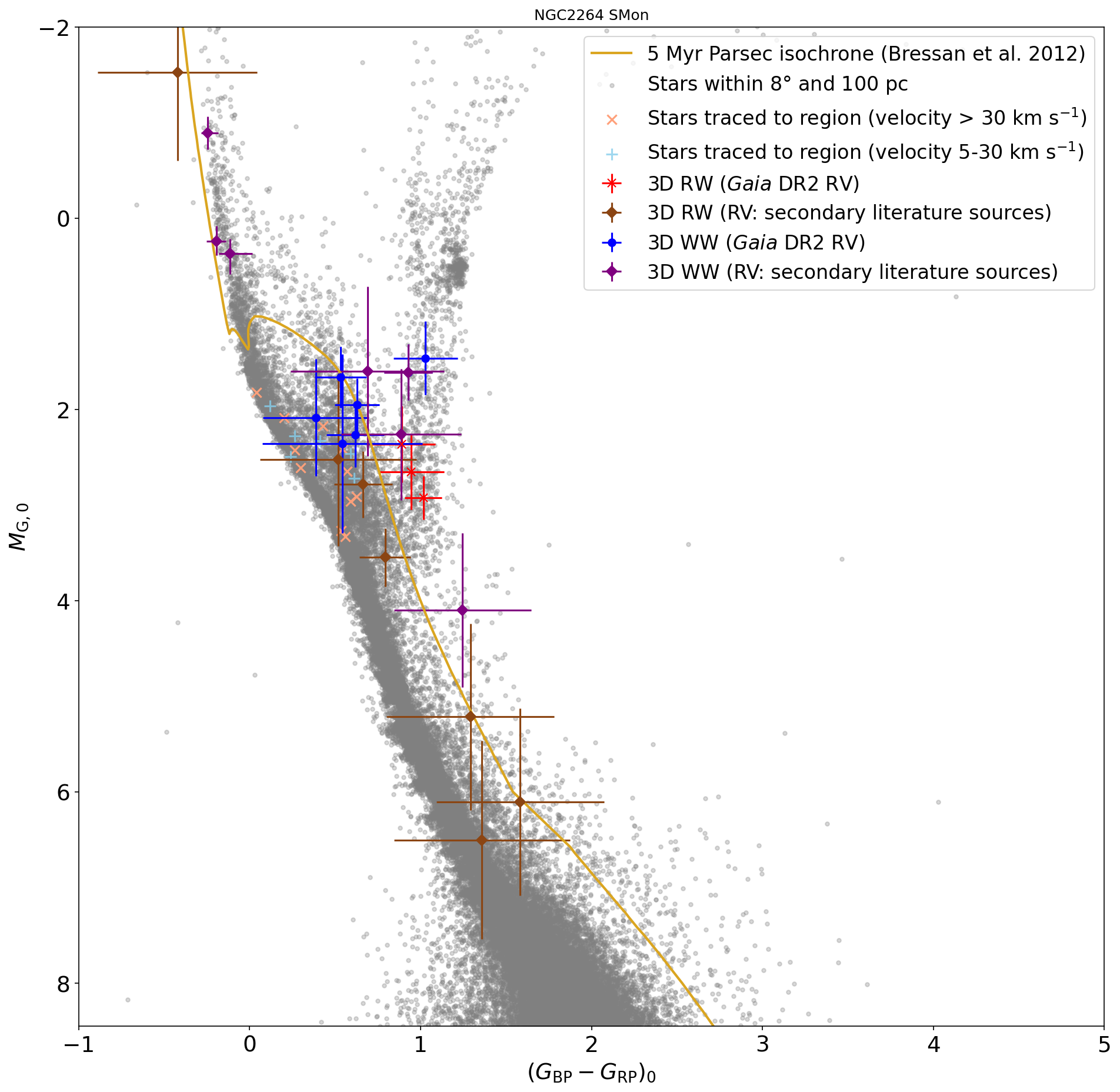}
    \end{minipage}
    \caption{CAMDs showing all 2D (left) and 3D (right) RW (> 30\,km\,s$^{-1}$) and WW (5--30\,km\,s$^{-1}$) candidates that can be traced back to the cluster around S\,Mon (RW: red ``x'', WW: blue square). The diagrams are magnitude-limited to show the range of -2\,mag < $M_{\rm{G,0}}$ < 8.5\,mag, which corresponds to a $G$-magnitude $\approx$ 18\,mag at the fainter end. At this faint apparent magnitude value, the typical uncertainties in the astrometry increase quickly. The CAMDs include a large number of stars that are traced back to the cluster around S\,Mon but that are fully located (considering errors) along the MS underneath the 5 Myr isochrone (RW-velocity: orange ``x'', WW-velocity: lightblue ``+''). These stars are likely too old to have been born in S\,Mon. The photometric errors for many of our identified candidates are rather large and predominantly driven by the errors in the extinction and reddening. Some candidates sit below the 5 Myr isochrone but due to the large errors, they might be younger than their position suggests. In the right plot, we show the 3D-candidates with RV from \textit{Gaia} DR2 (RW: red ``x'', WW: blue square) and from secondary literature sources (RW: brown ``x'', WW: purple dot).}
    \label{fig:RW_WW_SMon}
\end{figure*}

\subsection{Error calculation}

\subsubsection{Astrometric errors}\label{Astrometric error}

The errors in the velocity of our RW and WW candidates are calculated using an approach based on  \citet{RN264}. We change how the astrometric errors are calculated compared to \citet{2020MNRAS.495.3104S} due to the inclusion of RVs in the transformation from ICRS to the Cartesian coordinate system.

The basis is a covariance matrix as used in  \textit{Gaia} DR2 for the astrometric solution \citep[Equation B.3,][]{RN307}. This original covariance matrix considers only the proper motion errors and their correlation. We extend this matrix with the radial velocity errors, but do not add any correlations as the radial velocities and proper motions are not correlated. 

We then convert them into errors in our Cartesian coordinate system. The internal \textit{Gaia} DR2 uncertainties are multiplied with a correction factor of 1.1, as suggested in \citet{RN307}. We also use this correction factor on the radial velocity errors. We do not correct for systematic errors in the proper motions but include the errors in the motion of the centre of the cluster. Finally, all velocities are converted into km\,s$^{-1}$ using distances $r$ and $\kappa$\,=\,4.74 (conversion factor from mas\,yr$^{-1}$ to km\,s$^{-1}$).

\subsubsection{Photometric errors}\label{Photometric error}

For the calculation of the photometric errors in G-magnitude, $G_{\rm{BP}}$ and $G_{\rm{RP}}$, we follow the same approach as used in \citet{2020MNRAS.495.3104S}. 

\textit{Gaia} DR2 does not directly provide errors for the above quantities, as the error distributions are only symmetric in flux space \citep{2018gDR2.reptE..14H}. 
We use equation 5.26 in the Gaia DR documentation \citep{2018gDR2.reptE...5B}, which allows us to calculate errors in $G$:
\begin{equation}
    \sigma_G = \sqrt{\left(1.0857 \cfrac{\sigma_{\,\bar{I}}}{\bar{I}} \right)^2 + \left(\sigma_{G_{\rm{0}}} \right)^2}
\end{equation}
\noindent Here, $\sigma_{\,\bar{I}}$ represents the error on the G-band mean flux, $\bar{I}$ represents the weighted mean flux and $\sigma_{G_{\rm{0}}}$ is the passband error in the zero point $G_{\rm{0}}$.

We use the passband errors in the zero points in the VEGAMAG system \citep{2018A&A...616A...4E} and calculate the errors for $G$-magnitude, $G_{\rm{BP}}$ and $G_{\rm{RP}}$. We then use the distances from \citet{RN305} to convert the apparent $G$-magnitude errors into absolute $M_{\rm{G}}$-errors and also consider the extinction and distance errors in this final error value. We also calculate the $G_{\rm{BP}} - G_{\rm{RP}}$ errors, which include the reddening errors. 

\begin{table*}
    \renewcommand\arraystretch{1.2}
	\centering
	\caption{S\,Mon 3D-RW and WW stars sorted by decreasing 3D-velocity. Column 2+3: velocity in S\,Mon rest frame [rf]; Column 3: RV sources - $^{a}$\textit{Gaia} DR2, $^{b}$\citet{2016A&A...586A..52J}, $^{c}$\citet{2019yCat.5164....0L}, $^{d}$\citet{2019AJ....157..196K}, $^{e}$\citet{1992A&AS...95..541F}, 
	$^{f}$\citet{1995A&AS..114..269D};  Column 4: minimum flight time since ejection (crossing of search boundary); Column 5: age from PARSEC isochrones \citep{RN225}; Column 6--7: from literature sources -
	$^{1}$\citet{2019yCat.5164....0L}, $^{2}$\citet{2015A&A...581A..66V}, $^{3}$\citet{2017A&A...599A..23V}, $^{4}$\citet{2014A&A...570A..82V}, $^{5}$\citet{2019AJ....157..196K}, $^{6}$\citet{2004A&A...417..557L},
	$^{7}$\citet{1972A&AS....7...35K},
	$^{8}$\citet{1993yCat.3135....0C}, 
	$^{9}$\citet{1985cbvm.book.....V}, 
	$^{10}$\citet{2001A&A...373..625P}, 
	$^{11}$\citet{2002AJ....123.1528R}, 
	$^{12}$\citet{2019AJ....158..138S},
	$^{13}$\citet{2019A&A...628A..94A}, 
	$^{14}$\citet{2018A&A...609A..10V}.}
	\label{tab:RWC_2D_SMon}
	\begin{tabular}{lcccccc} 
		\hline
		\textit{Gaia} DR2 source-id & 2D-velocity rf & Radial velocity rf &  Flight time & Iso. age &  Mass & Spectral type  \\
		 & (km\,s$^{-1}$) & (km\,s$^{-1}$) &  (Myr) & (Myr) &  (M$_{\sun}$) & \\
		\hline
		3D-RW stars \\
		\hline
    3132380637509393920 & 54.2  $\pm$0.7   & -34.2 $\pm$6.3$^{a}$      & 0.9**   &  1.1\,$^{+1.0}_{-0.6}$     & 1.1$^{13}$ &G9$^{1}$ \\
    3326893624672681216 & 17.5  $\pm$1.1   & 59.5  $\pm$2.9$^{b}$      & 0.2   &   31.0\,$^{+19.0}_{-26.0}$    &  0.7$^{12,13}$ &-\\
    3326630811329448576 & 8.4   $\pm$0.6   & 54.9  $\pm$15.3$^{d}$      & 0.8**  &  >0.5   & 2.3--2.9$^{12,13}$& B2$^{8}$ \\
    3132474680112352128 & 42.7  $\pm$0.6   & -18.1 $\pm$6.0$^{c}$      & 1.3   & 10.0\,$^{+5.0}_{-6.0}$  & 1.0$^{12,13}$   & G5/6$^{1}$ \\
    3331678394335985152 & 41.1  $\pm$0.7   & 21.3  $\pm$4.0$^{a}$      & 2.0   &  1.8\,$^{+3.2}_{-1.3}$     &1.1$^{13}$ &  -\\
    3134176728413264896 & 41.9  $\pm$0.6   & 8.0   $\pm$3.1$^{a}$      & 0.6***   &   1.8\,$^{+3.2}_{-1.3}$ &1.1$^{13}$    & - \\
    3326654725707134464 & 8.6   $\pm$0.5   & 34.1  $\pm$7.4$^{c}$     & 1.1***  &   12.0\,$^{+7.0}_{-11.0}$    & 1.1--1.2$^{12,13}$ & F9$^{1}$ \\
    3326632082639739264 & 27.7  $\pm$0.7   & 21.6  $\pm$2.9$^{b}$      & 0.3**   &   5.0\,$^{+35.0}_{-4.3}$     & 0.9--1.1$^{12,13}$ &-\\
    3326908781612828544 & 30.5  $\pm$0.8   & 17.4  $\pm$2.9$^{b}$      & 0.1   &    5.0\,$^{+35.0}_{-4.0}$   & 0.9--1.0$^{12,13}$ & -\\
    3131997187129420672 & 16.5  $\pm$1.1  & 28.8  $\pm$12.2$^{c}$     & 3.8   & 10.0\,$^{+3.0}_{-6.0}$     &   1.2--1.3$^{12,13}$ & F8$^{1}$ \\
		\hline

        3D-WW stars \\
		\hline
    3134179335455713408 & 11.5  $\pm$0.7   & 26.5  $\pm$6.6$^{a}$      & 2.1&    7.0\,$^{+3.0}_{-4.0}$     &  1.2--2.1$^{12,13}$  &-\\
    3327008867233046528 & 6.7   $\pm$0.8   & -24.7 $\pm$3.8$^{e}$      & 1.1 & >0.7 & 3.3$^{13}$ & B5$^{9}$\\    
    3331597816450524288 & 22.4  $\pm$0.6   & -9.0  $\pm$3.2$^{a}$      & 3.5& 5.0\,$\pm$2.0     & 1.2--1.5$^{12,13}$  & - \\
    3132933764876638848 & 19.1  $\pm$0.5   & -13.6 $\pm$3.1$^{a}$      & 2.6*&  0.7\,$^{+1.3}_{-0.5}$       & 1.1$^{13}$  & G5$^{1}$ \\
    3326713442204844160 & 8.1   $\pm$0.5   & -16.1 $\pm$9.4$^{b}$      & 0.1 & >2.5 & 2.0$^{12,13}$ & A2/3$^{7}$\\
    3351602404024775168 & 16.8  $\pm$0.6   & 6.3   $\pm$3.0$^{a}$      & 2.1*&  0.3\,$^{+0.7}_{-0.1}$       & 1.1$^{13}$  & G5/K1$^{1}$ \\
    3326938567209095936 & 8.7   $\pm$0.6   & 15.5  $\pm$3.0$^{b}$      & 0.2 & 3.0\,$^{+4.0}_{-2.7}$ & 2.1$^{12,13}$ & B8$^{9}$\\ 
    3327203588170236672 & 13.3  $\pm$0.6   & -7.2  $\pm$5.8$^{a}$      & 2.4&  10.0\,$^{+5.0}_{-6.0}$     &  1.5--1.9$^{12,13}$  &-\\
    3351770835461871360 & 14.6  $\pm$0.7   & -0.2  $\pm$5.6$^{a}$      & 3.1&  5.0\,$^{+3.0}_{-2.0}$     &  1.5--1.6$^{12,13}$  &-\\
    3326693857153492736 & 4.2   $\pm$0.5   & 11.7  $\pm$2.9$^{d}$    & 0.6** & 1.3\,$^{+13.7}_{-1.0}$& 0.8--1.3$^{4,13,14}$ & -\\ 
    3326576656084295296 & 10.7  $\pm$0.6   & 3.4   $\pm$7.2$^{c}$      & 1.0 &   1.8\,$^{+8.0}_{-1.5}$    & 1.0$^{13}$  & K1$^{1}$ \\
    3326739933562218496 & 1.5   $\pm$0.5   & 8.8   $\pm$9.3$^{a}$      & in cluster &10.0\,$^{+8.0}_{-9.0}$ & 0.9--2.1$^{4,12,13}$ & K0$^{4,11}$\\ 
    3326740693772293248 & 6.0   $\pm$0.8   & 0.2   $\pm$2.9$^{f}$      & in cluster & >2.0 & 2.3$^{13}$ & A1$^{10}$ \\
    \hline
    \multicolumn{7}{l}{\parbox[t]{14cm}{*age estimate is smaller than the flight time; **more likely from IRS\,1, ***more likely from IRS\,2}}
\end{tabular}
\end{table*}  

Stars with extinction $A_{\rm{G}}$ and reddening $E(G_{\rm{BP}}-G_{\rm{RP}})$ values in \textit{Gaia} DR2 data also have upper and lower percentile values provided. We use these to calculate upper and lower errors. For all other stars with correction values using averages, we calculate extinction and reddening errors using the standard deviation.

\section{Results from \textit{Gaia} DR2}

We analyse the \textit{Gaia} DR2 trace-backs separately for S\,Mon and IRS\,1/2, due to the difference in age estimates for these regions, requiring different PARSEC isochrones and time limits for the flight times to be applied. While the regions encompassed by the search radius around IRS\,1 and IRS\,2 overlap in our analysis, we draw the search radius separately around each of the subcluster centres and trace back stars to each of them. However, when we compare the results from IRS\,1/2 to the simulations later, we consider them as one subcluster having evolved together.

\subsection{RW/WW stars from S Mon}\label{SMon_RW_WW}

Fig.~\ref{fig:RW_WW_SMon} shows the CAMDs for the RW/WW candidates that can be traced back to S\,Mon in the past 5 Myr. On the left, we see all 2D-candidates with absolute magnitudes of -2\,mag < $M_{\rm{G,0}}$ < 8.5\,mag. We find further candidates below the 8.5\,mag limit, however, the uncertainties for these candidates are too large for them to be included in our results. On the right in this figure, we see all 3D-candidates within the same magnitude range. In addition to 3D-candidates traced back using \textit{Gaia} DR2 RV, we also trace back several 3D-candidates using RVs from secondary literature sources.

We can trace back 30 2D-RW and 65 2D-WW candidates to S\,Mon. Not all of these 2D-candidates turn into 3D-candidates. Several of them are missing RVs, so their full status is unclear until this velocity measurement is available. Of the 30 2D-RW candidates, 21 candidates have RV measurements. Of these, ten candidates also trace back in 3D, three with RVs from \textit{Gaia} DR2, seven with RVs from secondary literature sources. 

Table~\ref{tab:RWC_2D_SMon} gives information about the candidates, that trace in all three dimensions, while Tables~\ref{tab:RWC_2D_SMon_app} and \ref{tab:WWC_2D_SMon_app} in Appendix~\ref{2D-app} provide the information for the 2D-candidates (and non-3D trace-backs), such as their 2D-velocity and RV in our chosen reference frame, flight time since ejection and an age estimate based on PARSEC isochrones. We have also searched for additional information in the literature about our sources and have found mass estimates and spectral types for several sources, however only a few independent age estimates. We have not included these independent age estimates in the tables but comment where they are not within our age estimates from the PARSEC isochrones.

\begin{figure*}
    \centering
    \begin{minipage}[t]{1.0\columnwidth}
        \centering
        \vspace{0pt}
    	\includegraphics[width=1.0\columnwidth]{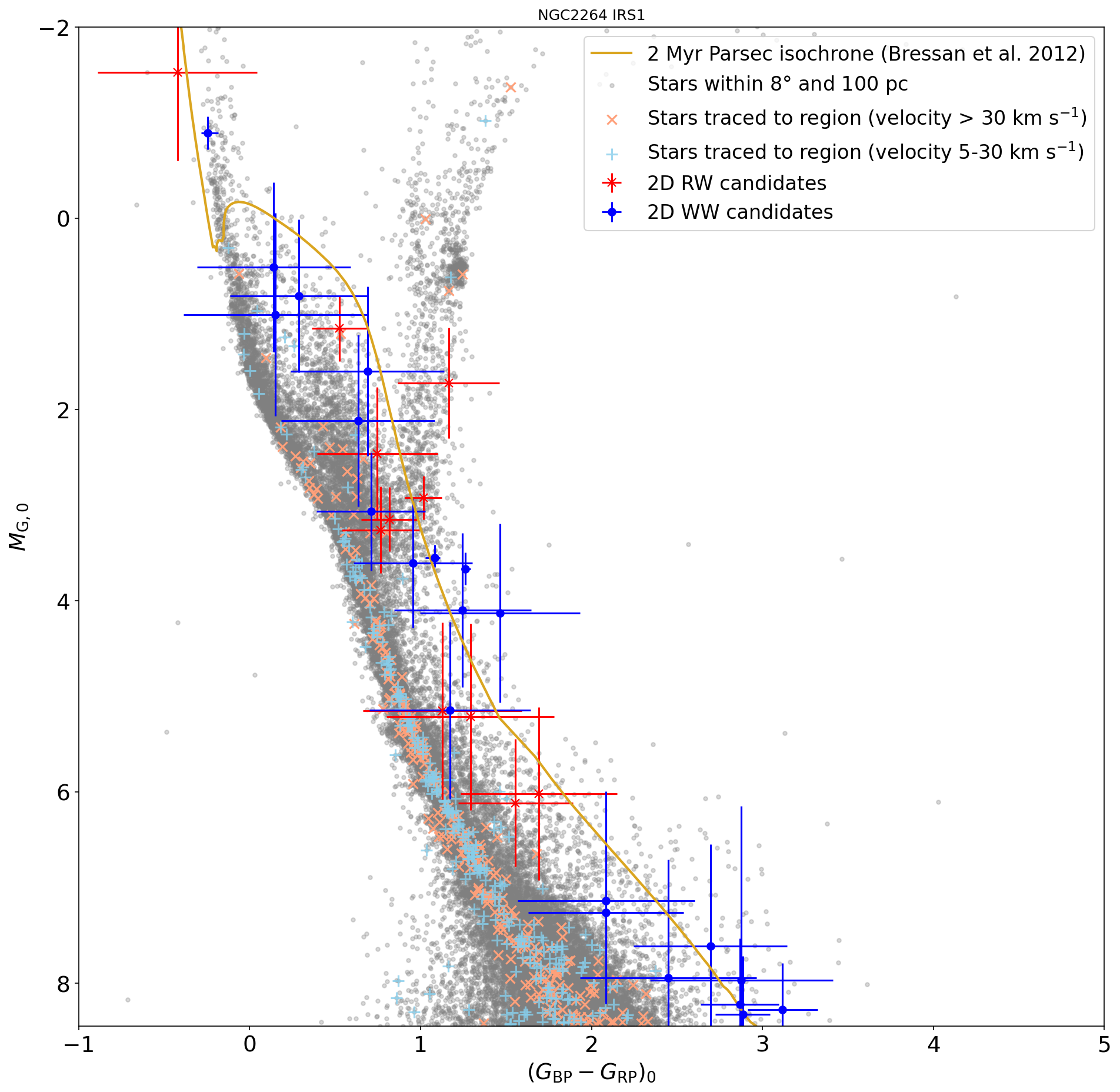}
    \end{minipage}
    \hfill
    \begin{minipage}[t]{1.0\columnwidth}
        \centering
        \vspace{0pt}
        \includegraphics[width=1.0\columnwidth]{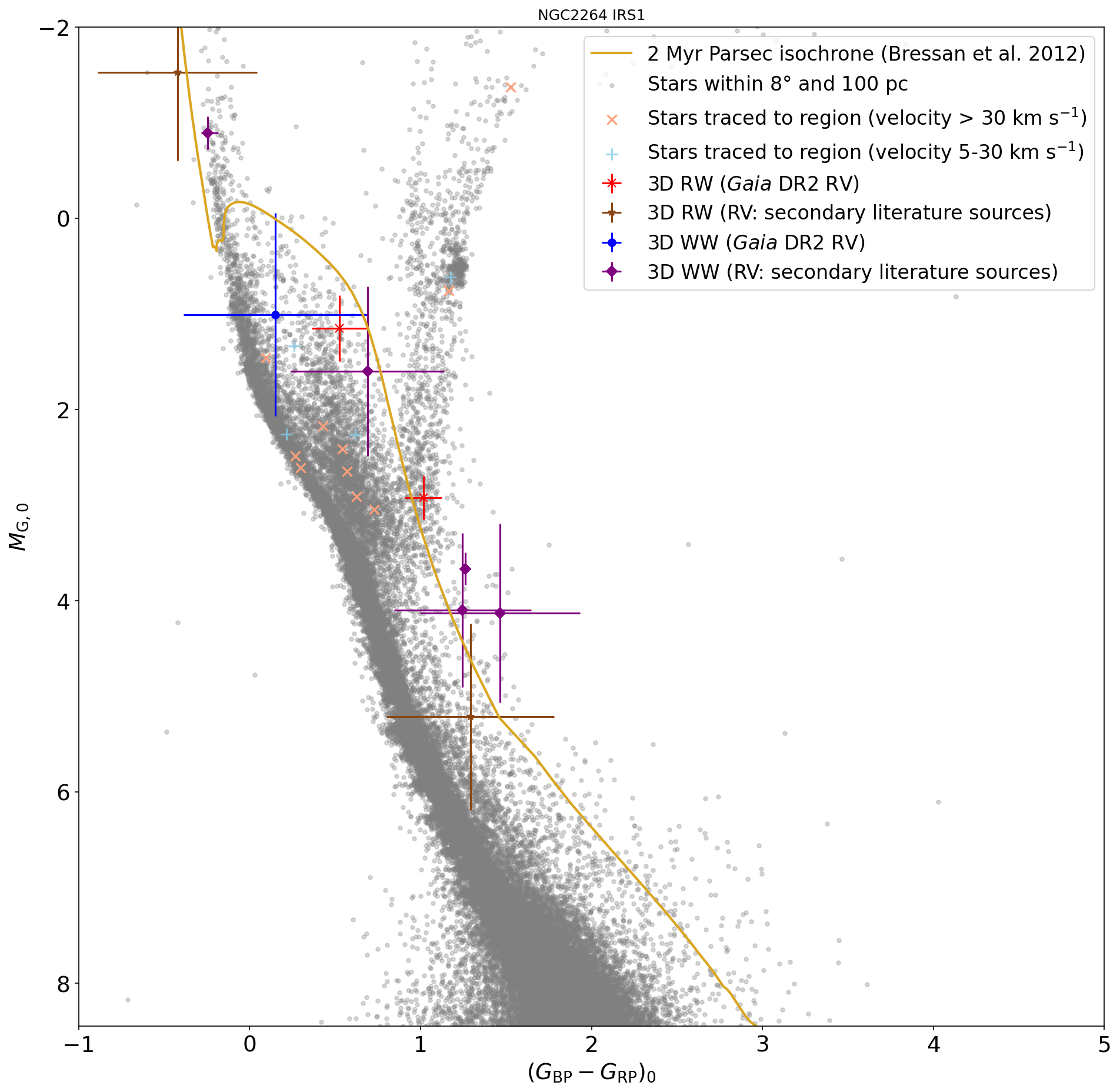}
    \end{minipage}
    \begin{minipage}[t]{1.0\columnwidth}
        \centering
        \vspace{0pt}
    	\includegraphics[width=1.0\columnwidth]{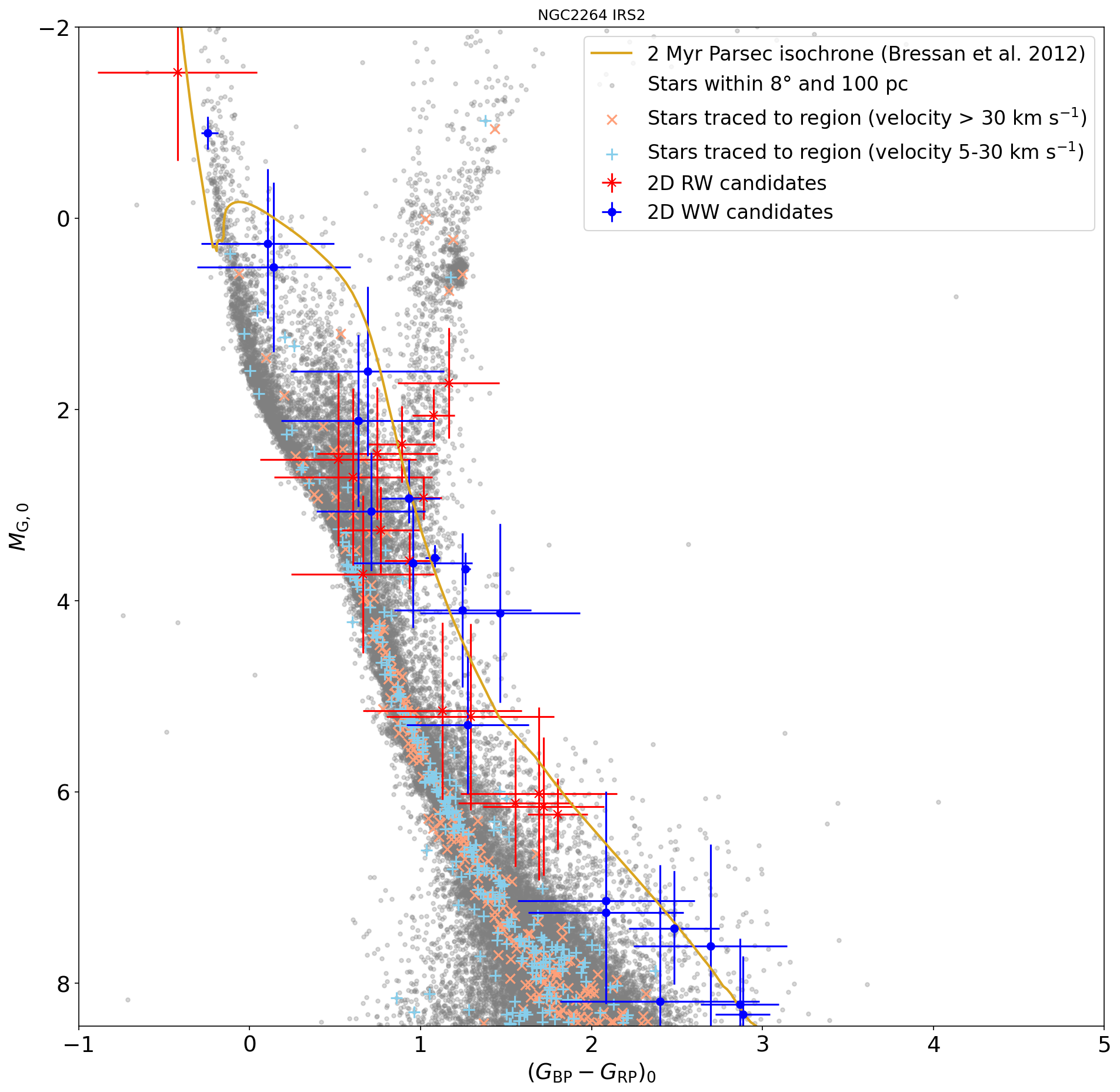}
    \end{minipage}
    \hfill
    \begin{minipage}[t]{1.0\columnwidth}
        \centering
        \vspace{0pt}
        \includegraphics[width=1.0\columnwidth]{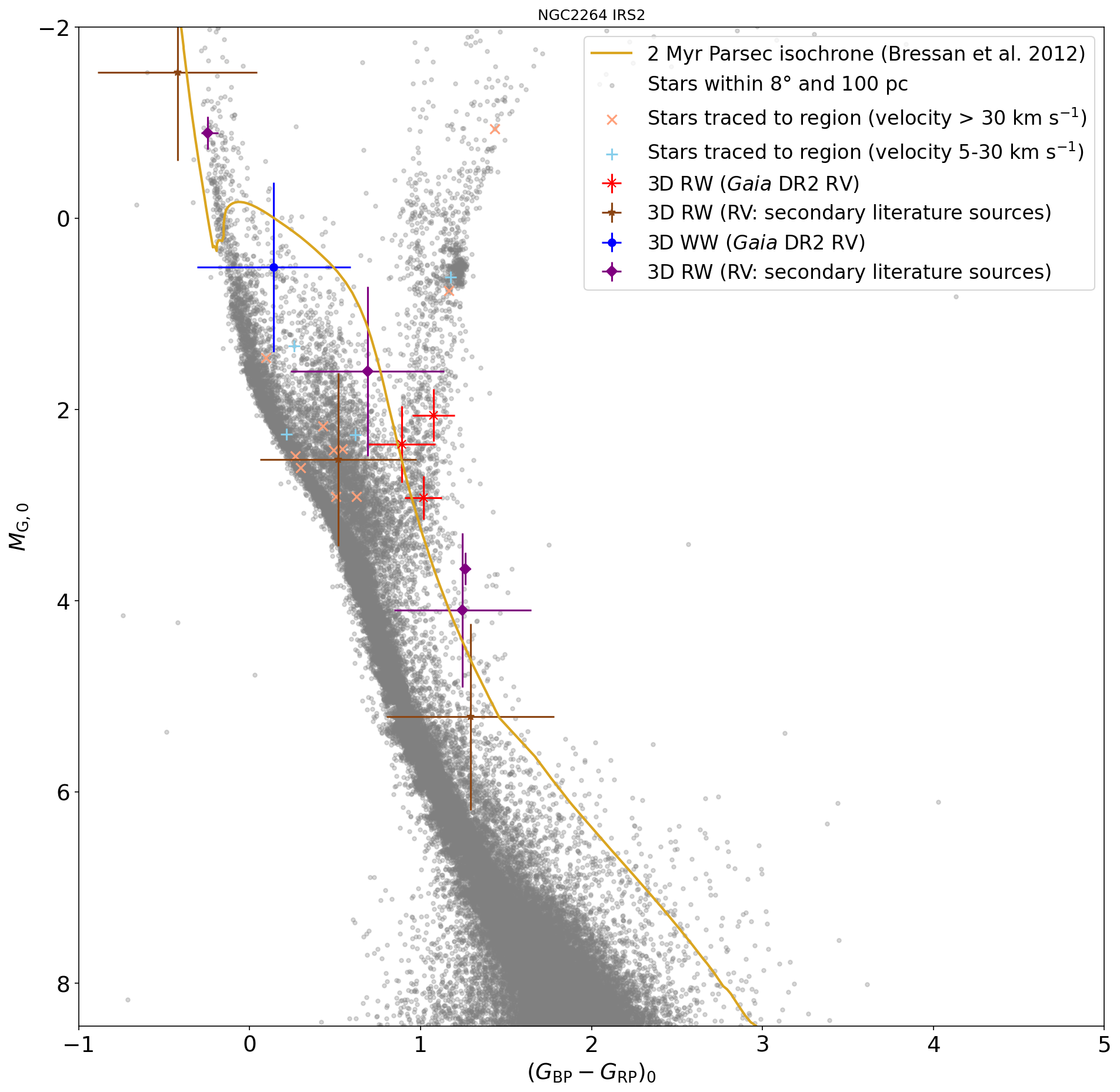}
    \end{minipage}
    \caption{CAMDs showing all 2D (left) and 3D (right) RW (> 30\,km\,s$^{-1}$) and WW (5--30\,km\,s$^{-1}$) candidates that can be traced back to the cluster around IRS\,1 (top) and IRS\,2 (bottom) (RW: red ``x'', WW: blue square). The diagrams are magnitude-limited to show the range of -2\,mag < $M_{\rm{G,0}}$ < 8.5\,mag, which corresponds to a $G$-magnitude $\approx$ 18\,mag at the fainter end. At this faint apparent magnitude value, the typical uncertainties in the astrometry increase quickly. The CAMDs include a large number of stars that are traced back to the clusters around IRS\,1 or IRS\,2 but that are fully located (considering errors) along the MS underneath the 2 Myr isochrone (RW-velocity: orange ``x'', WW-velocity: lightblue ``+''). These stars are likely too old to have been born in IRS\,1/2. The photometric errors for many of our identified candidates are rather large and predominantly driven by the errors in the extinction and reddening. Some candidates sit below the 2 Myr isochrone but due to the large errors, they might be younger than their position suggests. In the right plot, we show the 3D-candidates with RV from \textit{Gaia} DR2 (RW: red ``x'', WW: blue square) and from secondary literature sources (RW: brown ``x'', WW: purple dot).}
    
    \label{fig:RW_WW_IRS_2}
.\end{figure*}

The brightest 3D-RW candidate is HD 262042 (\textit{Gaia} DR2 3326630811329448576), which is only a WW when considering its proper motion and turns into a RW due to its large RV \citep{2019AJ....157..196K}. It has already reached the MS, which makes an age estimate with isochrones more difficult. The isochrone analysis provides a minimum age of $\sim$0.5 Myr, however \citet{2012AstL...38..694G} suggested an age of 159 Myr. It has an estimate for its mass of $\sim$2.3--2.9 M$_{\sun}$ \citep{2019AJ....158..138S,2019A&A...628A..94A} and a B2 spectral type \citep{1993yCat.3135....0C}. The mass estimate is low for the given spectral type, however the spectral type identification is quite old. This star also traces back to IRS\,1 and IRS\,2 and we discuss how we handle this issue in section~\ref{Discuss_double_trace_back}.

The faintest 3D-RW candidate is CSIMon-012143 (\textit{Gaia} DR2 3326893624672681216) with an absolute magnitude of $\sim$6.5 mag and a mass estimate of $\sim$0.7 M$_{\sun}$ \citep{2019AJ....158..138S,2019A&A...628A..94A}. It is located underneath the 5 Myr isochrone but its error bars make it possibly young enough to have originated from S\,Mon. It is the second fastest 3D-RW from this region and the fastest one that only traces back to S\,Mon with a space velocity of $\sim$62 km\,s$^{-1}$ in our reference frame.

Our fastest 3D-RW candidate is \textit{Gaia} DR2 3132380637509393920 with a space velocity of $\sim$64 km\,s$^{-1}$. It has an absolute magnitude of $\sim$2.9 mag, a mass estimate of $\sim$1.1 M$_{\sun}$ \citep{2019A&A...628A..94A} and a G9 spectral type \citep{2019yCat.5164....0L}, however this star also traces back to IRS\,1/2.


Of the 65 2D-WW candidates, only 18 candidates have RV measurements. 13 of these candidates trace back to S\,Mon in 3D, six of which with an RV measurement from secondary literature sources. We have three 3D-WW candidates that have already reached the main sequence (MS) since S\,Mon formed. 

The brightest of these three MS-trace-backs is HD 47662 (\textit{Gaia} DR2 3327008867233046528), which has an absolute magnitude of $\sim$-0.9 mag, a mass estimate of $\sim$3.3 M$_{\sun}$ \citep{2019A&A...628A..94A} and a spectral type of B5 \citep{1985cbvm.book.....V}. Our isochronal age analysis provides a minimum age estimate of $\sim$0.7 Myr, however \citet{RN263} suggested an age of $\sim$50 Myr. The next brightest is BD+10~1222 (\textit{Gaia} DR2 3326740693772293248) with an absolute magnitude of $\sim$0.2 mag, a mass estimate of $\sim$2.3 M$_{\sun}$ \citep{2019A&A...628A..94A} and an A1 spectral type \citep{2001A&A...373..625P}. The minimum age estimate from our isochrones is $\sim$2.0 Myr and \citet{2007AJ....134..999D} suggested a very similar age of $\sim$2.1 Myr. Finally, we have HD 261737 (\textit{Gaia} DR2 3326713442204844160) with an absolute magnitude of $\sim$0.4 mag and an age estimate of > $\sim$2.5 Myr, a mass estimate of 2.0 M$_{\sun}$ \citep{2019AJ....158..138S,2019A&A...628A..94A} and a A2/3 spectral type \citep{1972A&AS....7...35K}. For this star, we have not found an independent age estimate.

Of these three bright 3D-WW candidates, only the brightest (\textit{Gaia} DR2 3327008867233046528) also traces back to IRS\,1 and 2, the other two are trace-backs to S\,Mon only.



The faintest 3D-WW candidate is Cl* NGC 2264 VAS 204 (\textit{Gaia} DR2 3326693857153492736) with an absolute magnitude of $\sim$4.1 mag. It has a mass estimate of 0.8--1.3 M$_{\sun}$ \citep{2018A&A...609A..10V, 2019A&A...628A..94A}, but no spectral type information. This is one of the few stars for which we have been able to find an independent age estimate. Our age estimate from PARSEC isochrones is $\sim$1.3 Myr, the age estimate from \citet{2018A&A...609A..10V} is $\sim$1.7 Myr, which is within the uncertainties of our result.

We trace back a total of ten 3D-RW candidates to S\,Mon, however, we find five of these candidates also trace back to either IRS\,1 or IRS\,2, which could reduce the total number. At WW-velocities, we trace back 13 3D-candidates, one of which also traces back to IRS\,1. We find two MS stars, one with RW-velocity the other with WW-velocity, which based on independent age estimates are too old to have originated in S\,Mon. We exclude these two stars from further analysis.

\subsection{RW/WW stars from IRS 1 and IRS 2}

\begin{table*}
    \renewcommand\arraystretch{1.2}
	\centering
	\caption{IRS\,1/2 3D-RW and WW stars sorted by decreasing 3D-velocity. Column 2+3: velocity in respective IRS rest frame [rf]; Column 3: RV sources - $^{a}$\textit{Gaia} DR2, $^{b}$\citet{2016A&A...586A..52J}, $^{c}$\citet{2019yCat.5164....0L}, $^{d}$\citet{2019AJ....157..196K}, $^{e}$\citet{1992A&AS...95..541F}, 
	$^{f}$\citet{1995A&AS..114..269D}; Column 4: minimum flight time since ejection (crossing of search boundary); Column 5: age from PARSEC isochrones \citep{RN225}; Column 6: Subcluster identification; Column 7--8: from literature sources - $^{1}$\citet{2019yCat.5164....0L}, $^{2}$\citet{2015A&A...581A..66V}, $^{3}$\citet{2017A&A...599A..23V}, $^{4}$\citet{2014A&A...570A..82V}, $^{5}$\citet{2019AJ....157..196K}, $^{6}$\citet{2004A&A...417..557L}, $^{7}$\citet{1972A&AS....7...35K},
	$^{8}$\citet{1993yCat.3135....0C}, $^{9}$\citet{1985cbvm.book.....V}, 
	$^{10}$\citet{2001A&A...373..625P}, $^{11}$\citet{2002AJ....123.1528R}, 
	$^{12}$\citet{2019AJ....158..138S}, $^{13}$\citet{2019A&A...628A..94A}, 
	$^{14}$\citet{2018A&A...609A..10V}.}
	\label{tab:RWC_2D_IRS}
	\begin{tabular}{lcccccccc} 
		\hline
		\textit{Gaia} DR2 source-id & 2D-velocity rf & Radial velocity rf &  Flight time & Iso. age & Subcluster & Mass &  Spectral type  \\
		 & (km\,s$^{-1}$) & (km\,s$^{-1}$) & (Myr) & (Myr) &  & (M$_{\sun}$) & \\
		\hline
		3D-RW stars \\
		\hline
        3132688088448708736  & 53.8  $\pm$1.2    & 69.6   $\pm$4.1$^{a}$     & 0.9   &   0.4\,$^{+0.5}_{-0.3}$     & IRS\,2  &       1.1$^{13}$&        K1$^{1}$ \\
        3132380637509393920  & 55.0   $\pm$0.8    & -33.8  $\pm$6.3$^{a}$      & 0.9**   &  1.1\,$^{+1.0}_{-0.6}$      & IRS\,2  &       1.1$^{13}$&          G9$^{1}$ \\
        3132380637509393920  & 53.9  $\pm$0.9    & -33.7  $\pm$6.3$^{a}$      & 0.9   &    1.1\,$^{+1.0}_{-0.6}$   & IRS\,1  &      1.1$^{13}$&          G9$^{1}$ \\
        3326630811329448576 & 9.2   $\pm$0.7   & 55.1  $\pm$15.3$^{d}$      & 0.4**  &  >0.5    & IRS\,2& 2.3--2.9$^{12,13}$& B2$^{8}$  \\
        3326630811329448576 & 7.8   $\pm$0.9   & 55.1  $\pm$15.3$^{d}$      & 0.2  & >0.5   & IRS\,1& 2.3--2.9$^{12,13}$& B2$^{8}$  \\
        3134176728413264896  & 43.4  $\pm$0.8    & 8.3    $\pm$3.1$^{a}$      & 0.5   &    1.8\,$^{+3.2}_{-1.3}$   & IRS\,2  &     1.1$^{13}$&   -        \\
        3326632082639739264  & 29.1  $\pm$0.8    & 21.8   $\pm$2.9$^{b}$     & 0.1**    &   5.0\,$^{+35.0}_{-4.3}$     & IRS\,2  &       0.9--1.1$^{12,13}$ &  - \\
        3326654725707134464  & 10.9  $\pm$0.7    & 34.2   $\pm$7.4$^{c}$      & 0.8   &  12.0\,$^{+7.0}_{-11.0}$      & IRS\,2  &     1.1--1.2$^{12,13}$ & F9$^{1}$ \\
        3326632082639739264  & 27.7  $\pm$1.0    & 21.9   $\pm$2.9$^{b}$        & 0.1   &  5.0\,$^{+35.0}_{-4.3}$      & IRS\,1  &       0.9--1.1$^{12,13}$ &   -       \\
        3134140405870541568  & 17.4  $\pm$1.1   & -27.9  $\pm$20.0$^{a}$       & 1.7   & 3.5\,$\pm$1.5       & IRS\,1  &      1.2--1.9$^{12,13}$&      F6/7$^{1}$   \\ 
    \hline

        3D-WW stars \\
		\hline
        3327008867233046528 & 7.0   $\pm$1.1   & -24.6 $\pm$3.8$^{e}$      & 1.7*& >0.7     & IRS\,1  & 3.3$^{13}$ & B5$^{9}$\\		
        3327008867233046528 & 6.0   $\pm$0.9   & -24.6 $\pm$3.8$^{e}$    & 1.7*& >0.7     & IRS\,2 & 3.3$^{13}$ & B5$^{9}$\\
        3326938567209095936 & 9.7   $\pm$0.9   & 15.6  $\pm$3.0$^{b}$    & 0.7*&3.0\,$^{+4.0}_{-2.7}$     & IRS\,1  & 2.1$^{12,13}$ & B8$^{9}$\\
        3326938567209095936 & 9.0   $\pm$0.8   & 15.6  $\pm$3.0$^{b}$      & 0.6* &3.0\,$^{+4.0}_{-2.7}$     & IRS\,2 &2.1$^{12,13}$ & B8$^{9}$  \\
        3326685512032888320 & 9.5   $\pm$0.8   & 13.3  $\pm$10.4$^{a}$      & 1.7   &    3.0\,$^{+3.0}_{-1.5}$   & IRS\,2  &   1.2--2.1$^{4,12,13}$   & F5/G0$^{4,6}$ \\ 
        3326693857153492736 & 6.0   $\pm$0.7   & 11.8  $\pm$2.9$^{d}$    & 0.3**& 1.3\,$^{+13.7}_{-1.0}$     & IRS\,2 & 0.8--1.3$^{4,13,14}$ & K6$^{4}$\\     
        3326704238089925120 & 10.5  $\pm$0.8   & -7.9  $\pm$2.9$^{d}$      & 0.1   &  0.5\,$^{+3.0}_{-0.4}$     & IRS\,1  &  0.6--1.0$^{4,13}$   &  M0$^{2,4}$ \\
        3326693857153492736 & 4.7   $\pm$0.8   & 11.9  $\pm$2.9$^{d}$      & 0.2& 1.3\,$^{+13.7}_{-1.0}$     & IRS\,1 &0.8--1.3$^{4,13,14}$ & K6$^{4}$\\ 
        3326695991753074304 & 6.1   $\pm$1.2   & 5.0   $\pm$2.9$^{b}$     & in cluster & 0.6  $\pm$0.1    & IRS\,1 & 0.3$^{4}$ & M3$^{4}$ \\ 
        3326695991753074304 & 4.9   $\pm$1.0   & 5.0   $\pm$2.9$^{b}$     & in cluster & 0.6  $\pm$0.1     & IRS\,2 & 0.3$^{4}$&  M3$^{4}$\\
        3326682552799138176 & 0.8 $\pm$0.9 & -6.7 $\pm$11.5$^{a}$      & in cluster   &   5.0\,$^{+5.0}_{-3.1}$    & IRS\,1  &   - &-\\
        \hline
        \multicolumn{8}{l}{\parbox[t]{15cm}{*more likely from S\,Mon,**more likely from IRS\,1, ***more likely from IRS\,2}}
	\end{tabular}
\end{table*}

Fig.~\ref{fig:RW_WW_IRS_2} shows the CAMDs with all 2D- and 3D-RW and WW candidates for the subclusters IRS\,1 (first row) and IRS\,2 (second row). On the left, we identify all 2D-candidates that we can trace back to either region, whereas on the right are the 3D-trace-backs, where RVs are considered as well. The magnitude range is the same as for S\,Mon. We also find further candidates with absolute magnitudes fainter than 8.5\,mag, however do not identify these here due to the very large uncertainties. 

We find eleven 2D-RW candidates that trace back to IRS\,1 and 18 2D-RW candidates that trace back to IRS\,2. Several of these candidates trace back to both regions and we will comment on this in section~\ref{Discuss_double_trace_back}. In 3D, we can trace back four RW-candidates to IRS\,1, two of these with secondary RVs. To IRS\,2, we can trace back six 3D RW-candidates, three of these with secondary RVs. Table~\ref{tab:RWC_2D_IRS} identifies all 3D-RW and WW stars traced back to these subclusters and we also identify the subcluster to which these are traced back. The information for the 2D-candidates is located in Appendix~\ref{2D-app} in Tables~\ref{tab:RWC_2D_IRS_app} and \ref{tab:WWC_2D_IRS_app}.

As mentioned in section~\ref{SMon_RW_WW}, the brightest 3D-RW candidate HD 262042 (\textit{Gaia} DR2 3326630811329448576) traces back to all three regions and is a MS star. The next brightest 3D-RW candidate is TYC 747-2093-1 (\textit{Gaia} DR2 3134140405870541568) and only traces back to IRS\,1 with a flight time of $\sim$1.7 Myr. It has an absolute magnitude of $\sim$1.2 mag and an isochronal age estimate of $\sim$3.5 Myr, which is just at our upper age boundary of 2 Myr when considering the errors. Its spectral type is F6/7 based on information in \citet{2019yCat.5164....0L} and its mass estimate is $\sim$1.2--1.9 M$_{\sun}$ \citep{2019AJ....158..138S,2019A&A...628A..94A}. 
The fastest 3D-RW candidate from IRS\,1 is the same as from S\,Mon, \textit{Gaia} DR2 3132380637509393920 with a space velocity of $\sim$64 km\,s$^{-1}$.

The second brightest and fastest 3D-RW candidate only tracing back to IRS\,2 is \textit{Gaia} DR2 3132688088448708736. Its space velocity is $\sim$88 km\,s$^{-1}$ and it has an absolute magnitude of $\sim$2.1 mag. Its estimated isochronal age is $\sim$0.4 Myr, its mass estimate is $\sim$1.1 M$_{\sun}$ (source) and the spectral type is K1 \citep{2019yCat.5164....0L}.

The faintest 3D-RW candidate CSIMon-005775 (\textit{Gaia} DR2 3326632082639739264) traces back to all three regions analysed here, however its flight time to S\,Mon is $\sim$0.2 Myr longer than to IRS\,1/2. Its absolute magnitude is  $\sim$5.2 mag and its age estimate is $\sim$5.0 Myr with a large error putting it into the age range of IRS\,1/2. 

At WW-velocities, we trace back 21 2D-candidates to IRS\,1 and 20 2D-candidates to IRS\,2. Of these 2D-candidates, six trace back to IRS\,1 in 3D, five of them with secondary RV measurements. Five trace back to IRS\,2 in 3D, four of them with secondary RVs. 

The brightest 3D-WW candidate is the same for IRS\,1 and IRS\,2 as for S\,Mon, HD 47662 (\textit{Gaia} DR2 3327008867233046528). This star already has reached the MS and it is possibly too old to have originated from NGC 2264 based on its age estimate of 50 Myr from \citet{RN263}. The second brightest 3D-WW candidate tracing only back to IRS\,1 is NGC 2264 118 (\textit{Gaia} DR2 3326682552799138176). Its absolute magnitude is $\sim$1.0 mag and its age estimate is $\sim$5.0 Myr and it is only young enough to possibly have originated from IRS\,1 given its age errors. It is also one of the slowest 3D-WW stars, with a proper motion far below the WW-velocity and also still located within the cluster, so could possibly still be bound to it when considering its large RV errors.

Only one 3D-WW star traces solely back to IRS\,2, NGC 2264 189 (\textit{Gaia} DR2 3326685512032888320). It has an absolute magnitude of $\sim$0.5 mag and an estimated age of $\sim$3.0 Myr from the PARSEC isochrones with errors large enough to put it in the age range of IRS\,2. We found information on the spectral type in two literature sources with it being either a F5 \citep{2004A&A...417..557L} or a G0 spectral type \citep{2014A&A...570A..82V}. In \citet{2014A&A...570A..82V}, we also find a mass estimate of $\sim$2.1 M$_{\sun}$ and an age estimate $\sim$4.7 Myr \citep{2014A&A...570A..82V}, which is consistent with our age estimate when considering the errors. However this independent age estimate puts this star outside the age range for IRS\,2. 

The faintest 3D-WW candidate V*~V500 Mon (\textit{Gaia} DR2 3326704238089925120) traces back in 2D to both IRS\,1 and 2, however in 3D it only traces back to IRS\,1. It has an absolute magnitude of $\sim$4.1 mag and an age estimate from PARSEC isochrones of $\sim$0.5 Myr. This age is consistent with the estimate from \citet{2014A&A...570A..82V} of $\sim$0.7 Myr.

We trace back a total of ten 3D-RW candidates to IRS\,1 and IRS\,2, however find three of these candidates trace back to both regions. At WW-velocities, we trace back eleven 3D-candidates, four of which also trace back to S\,Mon and two common trace-back between IRS\,1 and 2. We find two MS stars, one at RW-velocities the other at WW-velocities, which based on independent age estimates are too old to have originated in IRS\,1 or 2. We exclude these two stars from further analysis.

\subsection{Confirming previously identified ejected stars}

\citet{2020A&A...643A.138M} found two ejected stars from NGC 2264 and suggested they are more likely to originate from the northern region around S\,Mon. These two stars are \textit{Gaia} DR2 3326734332924414976 and \textit{Gaia} DR2 3326951215889632128. 

We confirm that \textit{Gaia} DR2 3326951215889632128 also traces back to S\,Mon in our analysis with a 2D WW-velocity of $\sim$14 km\,s$^{-1}$, however this star does not have a RV-measurement, so cannot be confirmed in 3D. It is located very close to S\,Mon and is within the age range of this subcluster based on the isochrone analysis, so could either originate in S Mon or possibly be a future visitor depending on the direction of its RV. 

\textit{Gaia} DR2 3326734332924414976 also traces back to S\,Mon in 2D with WW-velocities in our analysis, but it is again missing the RV. When plotted on the CAMD, it appears too old to have originated in S\,Mon as it is located right on the MS with very small photometry errors taken directly from \textit{Gaia} DR2. The age estimate from its position on the MS puts it at an age of at least $\sim$15 Myr but a different method to determine its age might change our age estimate.

\subsection{3D-candidates with a protoplanetary disc}

In \citet{2021MNRAS.501L..12S}, we searched for evidence of the presence of protoplanetary discs around RWs and WWs (and past/future visitors) from the ONC identified in \citet{2020MNRAS.495.3104S}. We showed that in addition to some of the protoplanetary discs surviving the ejection of their host stars from their birth clusters, these star-disc systems can encounter a second dense star-forming environment and possibly emerge with an intact disc from this encounter. 

We repeat this analysis for NGC 2264 and search through the Simbad/VizieR databases \citep{2000A&AS..143....9W, 2000A&AS..143...23O}. In Appendix Table~\ref{tab:Past_visitors_list}, we show the older past visitors to the regions within the past 2 Myr (IRS\,1 and IRS\,2) and 5 Myr (S\,Mon) and in Tables~\ref{tab:Visitors_list_fut_RW} and \ref{tab:Visitors_list_fut_WW} we show all future visitors to all subclusters of NGC 2264 up to a cluster age of 10 Myr. We have searched through the databases for any evidence of protoplanetary discs around the identified RW/WWs. We were successful in this endeavour for several of our ejected 3D-trace-back stars from the region. Also, one of the future visitors to the region shows clear evidence of a disc.

\textit{Gaia} DR2 3326704238089925120 is our faintest 3D-WW trace-back to IRS\,1. It appears in \citet{2002AJ....123.1528R} with an IR-excess value $\Delta$(I$_{C}$-K) = 0.32 mag. The authors used the IR-excess limit stated in \citet{1998AJ....116.1816H}, where a value of $\Delta$(I$_{C}$-K) > 0.30 mag is a clear indicator of a protoplanetary disc. This star also appears in \citet{2007ApJ...671..605C} with a value of [3.6 \micro m] - [8 \micro m] > 1 mag. A mid-IR colour index above 1 mag is suggested by \citet{2006ApJ...646..297R} as an indicator of a disc.

\textit{Gaia} DR2 3326695991753074304 traces back in 3D to both IRS\,1 and 2 as a WW star and according to \citet{2018A&A...609A..10V}, who used IR- and UV-excess to identify disc-bearing, accreting objects, it is a disc-bearing star showing evidence of on-going accretion. It has also been previously identified as an accreting CTTS by \citet{2014A&A...570A..82V}.

\textit{Gaia} DR2 3326693857153492736 is a 3D-WW tracing back to all three regions. It appeared in \citet{2018A&A...609A..10V}, but it was considered a non-accreting source there. However according to \citet{2014A&A...570A..82V} and \citet{2017A&A...599A..23V} it was suggested to be a CTTS, which could be disc-bearing and accreting.

\textit{Gaia} DR2 3326687878559500288 could be a future visitor to IRS\,1 and 2 approaching these subclusters at a WW-velocity of $\sim$10 km\,s$^{-1}$. It is suggested to be an accreting CTTS in \citet[][based on H$\alpha$]{2006ApJ...648.1090F}, \citet{2015A&A...581A..66V} and \citet{2017A&A...599A..23V}. Its estimated age from the PARSEC isochrones is 7.0\,$\pm$6.5 Myr. Its maximum flight time since ejection from its birth cluster is therefore $\sim$13.5 Myr, if we consider an ejection right after formation. Given its velocity and this maximum flight time, it could have travelled as far as $\sim$130 pc during this time. Its velocity and distance measurements show large margins of error so we do not attempt to locate its possible origin. It has a K1.5 spectra type \citep{2018A&A...620A..87M} and due to its close proximity to NGC 2264, it is typically associated with the cluster, even though it was not born there based on its kinematics.

\textit{Gaia} DR2 3326689807000188032 is a visitor, which is currently passing through IRS\,1 and IRS\,2. It appears in \citet{2014A&A...570A..82V} and \citet{2017A&A...599A..23V} identified as a WTTS. As such it is no longer accreting, but with an age of $\sim$2.1--10 Myr, it would still be young enough to possibly have a debris disc.

\section{RW and WW stars from \textit{N}-body simulations}

We now use $N$-body simulations to predict the number and velocities of the RW and WW stars that we can expect to find in the observational data.

\citet{2006ApJ...648.1090F} suggested that the current subclustering observable in NGC 2264 is likely a remnant of spatial substructure that was initially present in this cluster. They also found that the kinematics of the region correlate with the spatial substructure. \citet{2015AJ....149..119T} refined the results of \citet{2006ApJ...648.1090F} by using further RV measurements. They found further velocity substructure, i.e. groups with different velocities and suggested that NGC 2264 is possibly a collection of star-forming clumps instead of a dense bound cluster. \cite{2018MNRAS.476.3160C} studied the spatial and kinematic structure in the larger Monoceros region of which NGC 2264 is a part. They confirmed the previous results on the presence of kinematic substructure.

Based on these literature sources, we suspect that NGC 2264 started with some level of initial spatial and kinematic substructure as traces of both appear to be still present at current times. This substructure cannot be created by dynamical evolution, only erased \citep[e.g.][]{RN14,RN4,RN248}. As we have no specific information on the exact level of substructure, we run a larger set of different initial conditions based on combinations of different initial density, spatial and kinematic substructure.

\subsection{Simulation set-up}

In $N$-body simulations, spatial substructure can be created using fractal distributions \citep[][]{RN14}. In this approach, the degree of substructure can be defined by a single parameter, which is the fractal dimension $D$. In our $N$-body simulations, we use several different initial fractal dimensions, which are shown in Table~\ref{tab:Init_cond}. This table provides an overview of all the initial condition combinations used in our $N$-body simulations. The Simulation IDs are used to refer to our different sets of initial conditions and are a number combination of the fractal dimension, initial virial ratio and initial radius. We also provide initial median stellar densities for each of the initial condition combinations in Table~\ref{tab:Init_cond}.

A fractal dimension of $D$ = 1.6 represents a high amount of initial spatial substructure, whereas $D$ = 3.0 is representative of a completely smooth distribution. For lower initial density simulations with an initial radius of 5 pc, we use a value of $D$ = 2.0 to represent a moderate amount of substructure and do not use a fractal dimension of $D$ = 3.0. The lower initial stellar density combined with a smooth initial spatial substructure represented by $D$ = 3.0 results in a very slow dynamical evolution of the simulated region and would only produce RWs at older cluster ages, if any are produced at all.

Fractals can also be used to set up the initial kinematic substructure. Stars that are located close to each other have correlated velocities, whereas stars at a larger distance can have very different velocities \citep{RN27,RN14}. In our simulations, the velocities are scaled in such a way that the regions are either initially subvirial with a virial ratio $\alpha_{\text{vir}}$ = 0.3 or initially virialised with a virial ratio $\alpha_{\text{vir}}$ = 0.5. The virial ratio is $\alpha_{\text{vir}}$ = $ T/|\varOmega|$, with $T$ representing the total kinetic energy and $|\varOmega|$ as the total potential energy of all stars. 

More details about the construction of the fractals in the simulations can be found in \citet{RN14} and also \citet{RN5, RN1}.

\begin{table}
	\centering
	\caption{Overview of all initial condition combinations used in our $N$-body simulations. Column 1 shows the Simulation ID, which is a number combination of the fractal dimension, initial virial ratio and initial radius; Column 2 shows the fractal dimension $D$; Column 3 shows the initial virial ratio $\alpha_{\text{vir}}$; Column 4 shows the initial radius of the simulated regions $r_F$; Column 5 shows the initial median stellar density $\tilde{\rho}$; Column 6 shows the initial (observable) median stellar surface density $\tilde{\sum}$.}
	\label{tab:Init_cond}
	\begin{tabular}{lccccc} 
		\hline
		Sim. ID & $D$ & $\alpha_{\text{vir}}$ & $r_F$  & $\tilde{\rho}$ & $\tilde{\sum}$  \\
		 & & & (pc) & (M$_{\sun}$ pc$^{-3}$) & (stars pc$^{-2}$) \\
		\hline
		16-03-1 & 1.6 & 0.3 & 1 & 10 000 & 3 000\\
		30-03-1 & 3.0 & 0.3 & 1 &150 & 400 \\
		16-05-1 & 1.6 & 0.5 & 1 & 10 000  & 3 000\\
		30-05-1 & 3.0 & 0.5 & 1 & 150 & 400 \\
		16-03-5 & 1.6 & 0.3 & 5 &70 & 100 \\
		20-03-5 & 2.0 & 0.3 & 5 & 10 & 40\\
		16-05-5 & 1.6 & 0.5 & 5 & 70 & 100\\
		20-05-5 & 2.0 & 0.5 & 5 & 10 & 40\\
		\hline
	\end{tabular}
\end{table}

The number of systems in our simulations is much smaller than those in \citet{2020MNRAS.495.3104S} and is comparable to the number used in \citet{RN309}. The number of members in NGC 2264 is suggested to be $\sim$1400, distributed across the northern and southern regions \citep{2012A&A...540A..83T}. For simplicity, we assume that half of the stars are located in the northern region around S\,Mon and the other half in the southern regions around IRS\,1 and IRS\,2, which are subcluster with overlapping boundaries. For our simulations, we use a number of 725 systems per simulation with masses for these systems sampled randomly from a \citet{RN203} IMF. We use a range of stellar masses between 0.1 M$_{\sun}$ and 50 M$_{\sun}$. This upper mass limit is consistent with recent estimates for the primary star in S\,Mon \citep[$\sim$40-50 M$_{\sun}$;][]{2018ApJS..235....6T, 2019A&A...630A.119M}.


The Maschberger IMF is a combination of the power-law slope of Salpeter (\citeyear{RN204}) for stars more massive than 1 M$_{\sun}$ combined with a Chabrier (\citeyear{RN200}) lognormal IMF approximation for lower-mass stars:
\begin{equation}
    p(m) \propto \cfrac{\left(\cfrac{m}{\mu}\right)^{-\alpha}}{\left(1+\left(\cfrac{m}{\mu}\right)^{1-\alpha}\right)^\beta}.
    \label{eq:MaschbergerIMF}
\end{equation}
\noindent In this probability density function $\alpha$ = 2.3 (power-law exponent for higher-mass stars), $\beta$ = 1.4 (describing the IMF slope for lower-mass stars) and $\mu$ = 0.2 (average stellar mass). 

In our simulations, we also include primordial binaries. We define the binary fraction $f_{\text{bin}}$ depending on the mass of the primary star $m_{\text{p}}$. This fraction is defined as:
\begin{equation}
   f_{\text{bin}} = \frac{B}{S+B}.
\end{equation}

In this fraction, S and B represent the number of single or binary systems that are present in the simulations, respectively. There are no primordial higher-order multiple systems (triples or quadruples). Table~\ref{tab:Bin_frac} shows $f_{\text{bin}}$ depending on the primary star's mass, Table~\ref{tab:Bin_sep} provides the corresponding binary separations.

These binary fractions and separations are similar to those in the Galactic field \citep{2006A&A...458..461K,2007AJ....134.2272R,RN8, RN257}. \citet{2016ApJ...821....8K} suggested that the binary fractions in NGC 2664 are similar to those in the ONC and that both are consistent with how binaries are distributed in the field.

\begin{table}
	\centering
	\caption{Binary fractions in the $N$-body simulations. Column 1 shows the mass range based on the mass of the primary star; Column 2 shows the binary fraction $f_{\text{bin}}$.}
	\label{tab:Bin_frac}
	\begin{tabular}{lcl} 
		\hline
		$m_{\text{p}}$ [M$_{\sun}$] & $f_{\text{bin}}$ & Source \\
		\hline
		0.10 $\leq m_{\text{p}}$  < 0.45 & 0.34 & \citet{2012ApJ...758L...2J}\\
		0.45 $\leq m_{\text{p}}$  < 0.84 & 0.45 & \citet{1992ASPC...32...73M}\\
		0.84 $\leq m_{\text{p}}$  < 1.20 & 0.46 & \citet{2012ApJ...745...24R}\\
		1.20 $\leq m_{\text{p}} \leq$ 3.00 & 0.48 & \citet{2012MNRAS.422.2765D,2014MNRAS.437.1216D}\\
		$m_{\text{p}}$  > 3.00 & 1.00 & \parbox[t]{4cm}{\citet{1998AJ....115..821M}\\\citet{2007AA...474...77K}} \\		
		\hline
	\end{tabular}
\end{table}

The secondary star in our primordial binaries is assigned a mass $m_{\text{s}}$ based on a flat mass ratio distribution. This type of distribution is observed in the field and many star-forming regions \citep[e.g.][]{2011ApJ...738...60R, 2013A&A...553A.124R, 2013MNRAS.432.2378P}. 
The binary mass ratio $q$ is: 
\begin{equation}
    q = \cfrac{m_{\text{s}}}{ m_{\text{p}}}\,.
\end{equation}

The secondary stars in our simulations are allowed to have a mass lower than the primary stars, which are limited to masses $\geq$ 0.1 M$_{\sun}$. The secondary stars can have a mass of $m_{\text{p}}$ > $m_{\text{s}}\geq$ 0.01 M$_{\sun}$, so brown dwarfs (BDs) are possible as secondaries in our primordial binaries, however not as primary or single stars. 

The binary fractions used in the simulations result in an average total number of stars in our simulations of $\sim$1000 stars, average cluster masses of $\sim$730 M$_{\sun}$ and an escape velocity of $\sim$2--3\,km\,s$^{-1}$, which confirms our choice of lower walkaway limit of 5\,km\,s$^{-1}$. This cluster mass is similar to that suggested by \citet{2012A&A...540A..83T}, However, we have considerably fewer stars contributing to this mass estimate, highlighting that the choice of an average mass of 0.5 M$_{\sun}$ in \citet{2012A&A...540A..83T} does not replicate results achieved with applying an IMF. 


The initial separation in our binaries (the semi-major axis) is shown in Table~\ref{tab:Bin_sep} and differs for different primary masses. It is based on a log-normal distribution and the mean values for the binary separation $\bar{a}$ in astronomical units (au) and the variance are shown in the table and follow observations of field binaries \citep[e.g.][]{2013ARA&A..51..269D,2014MNRAS.442.3722P}.
Binaries with orbital periods $\lessapprox$0.1 au have initially circular orbits in line with what we find in observations. For binaries with larger orbital periods than 0.1 au, we draw initial eccentricity values randomly from a flat distribution  \citep[e.g.][]{2013ARA&A..51..269D,2014MNRAS.442.3722P,2019MNRAS.485L..48W}. We refer to \citet{2014MNRAS.442.3722P} for further information on the set-up of the binary systems.

\begin{table}
	\centering
	\caption{Mean binary separations in the $N$-body simulations. Column 1 shows the mass range based on the mass of the primary star; Column 2 shows the mean binary separation $\bar{a}$; Column 3 represents the variance $\sigma_{\textrm{log}\,\bar{a}}$ of the log-normal fit to the binary separation distributions}
	\label{tab:Bin_sep}
	\begin{tabular}{lccl} 
		\hline
		$m_{\text{p}}$ [M$_{\sun}$] & $\bar{a}$ [au] & $\sigma_{\textrm{log}\,\bar{a}}$ & Source \\
		\hline
		0.10 $\leq m_{\text{p}}$ < 0.45 & 16 & 0.80 & \citet{2012ApJ...758L...2J} \\ 
		0.45 $\leq m_{\text{p}}$ < 1.20 & 50 & 1.68 & \citet{2012ApJ...745...24R}\\
		1.20 $\leq m_{\text{p}} \leq$ 3.00 & 389 & 0.79 & \citet{2014MNRAS.437.1216D}\\
		$m_{\text{p}}$ > 3.00 & \parbox[t]{1.01cm}{\centering \"Opik~law\\(0-50)}& -- & \parbox[t]{2.8cm}{\citet{1924PTarO..25f...1O},\\\citet{2013AA...550A.107S}} \\
		\hline
	\end{tabular}
\end{table}

\subsubsection{Alternative set-up of binary systems}\label{sec:alt_set-up}

For the simulations described above, we set up our binary systems based on observed binary fractions and separations in the Galactic field. Binaries in NGC\,2264 have been suggested to exhibit similar characteristics \citep{2016ApJ...821....8K}. \citet{2014MNRAS.442.3722P} and \citet{RN8} also showed that the binary population in most star-forming regions was unlikely to have been altered much from the birth population, so a field-like population was likely to be similar to the birth population. However, \citet{RN261} suggested that the initial (primordial) binary fraction could be as high as 100 per cent, and that binaries can form based on a universal initial period distribution \citep{RN261,2011A&A...529A..92K}. We investigate this binary system set-up with 20 additional simulations.

The above simulations are also set-up with a flat mass ratio distribution for all stellar masses based on observations. This choice appears justified even for the high-mass stars. However, observations in the Small Magellanic Cloud suggested that high-mass stars are predominately found in equal-mass binary systems \citep {2006ApJ...639L..67P}. This would suggest a mass ratio of 1 for high-mass stars, instead of a flat mass ratio distribution. Shortly thereafter, \citet{2006A&A...457..629L} investigated these results and found that they could be due to selection bias.

Past research also indicated that the mass ratio distribution showed peaks, but could be considered generally flat \citep[see][and references therein]{2013ARA&A..51..269D}. \citet{2012Sci...337..444S} found that the intrinsic distribution of mass ratios for high-mass SBs was essentially flat. And \citet{2012A&A...538A..74P} showed that among visual binaries a flat distribution matched observations of high-mass stars in the Carina region. Nevertheless, we also investigate an alternative mass ratio for high-mass stars (>8\,M$_{\sun}$) and define these binaries to have an equal mass ratio in a second set of 20 additional simulations.

To determine, if the initial conditions used for the set-up of our binary systems may have an effect on the number of RWs, we run the two alternative set-ups of the binary systems described above. We do this for one of our original initial condition sets and choose ID 16-05-1 ($D$ = 1.6, $\alpha_{\text{vir}}$ = 0.5, radius 1pc).

For the first alternative set-up with Simulation ID 16-05-1-100-bin, we define that all stars in the simulations are initially in binaries, i.e. $f_{\text{bin}}$=1.00 \citep[as suggested in ][]{RN261}. We reduce the number of systems in the simulations from 725 to 610 as this ensures similar cluster masses and therefore escape velocities. We adjust the separation distribution to one that is also based on \citet{RN261}. We use the approximation for the period distribution used in \citet{2014MNRAS.442.3722P}, expressed in terms of the semi-major axes \textit{a}:
\begin{equation}
    f (\text{log}_{10}a) = \eta \cfrac{\text{log}_{10}a - \text{log}_{10}a_{min}}{\sigma + (\text{log}_{10}a - \text{log}_{10}a_{min})^2} .
\end{equation}

In the above equation $\text{log}_{10}a$ is the logarithm of the semi-major axis expressed in au. $\text{log}_{10}a_{min}$ = -2, so $a_{min}$ = 0.01 au. The constants are $\sigma$ = 77 and $\eta$ = 5.25. These are derived from the constants adopted for the period generating function by \citet{RN261} and \citet{2011A&A...529A..92K} when they fitted their simulations to observed period distributions in pre-MS stars.

For the second alternative set-up with Simulation ID 16-05-1-eq-mass, we keep the original binary fractions and separations as in 16-05-1 and the flat mass ratio for binaries with primary masses <8\,M$_{\sun}$. However, we change the mass ratio to unity for binaries with a primary mass >8\,M$_{\sun}$.

For all simulations, we use software from the \texttt{Starlab} environment: the $N$-body integrator \texttt{kira} and the stellar and binary evolution package \texttt{SeBa} \citep{RN236,RN193}. Our star-forming regions are evolved over a time period of 5 Myr, covering the age estimates for NGC 2264 used for this analysis and we take snapshots every 0.1 Myr. Our initial radii are set at 1 pc (resulting in a higher initial stellar density, $\sim$150--10 000M$_{\sun}$ pc$^{-3}$) and 5 pc (resulting in a more moderate initial stellar density, $\sim$10--70\,M$_{\sun}$ pc$^{-3}$), we have not applied any external tidal field, i.e. a Galactic potential.

\begin{table*}
	\centering
	\caption{Ejected RW and WW stars from $N$-body simulations within the search radius of 100 pc at different times during the simulations. For all our initial condition combinations, we show averages from all 20 simulations and the maximum from a single simulation in the format average $\pm$ uncertainty / maximum. We count ejected binary systems as one star when calculating averages and maxima. The uncertainties in our averages are the standard deviations. We show ejected stars with masses from 0.3--8 M$_{\sun}$.}
	\label{tab:RW_WW_pred}
	\begin{tabular}{lcccccccc} 
		\hline
		Mass $m$ (M$_{\sun}$) & & & & Simulation ID \\
		\hline
		RW stars\\
		\hline
		&  16-03-1 & 30-03-1 & 16-05-1 & 30-05-1 &  16-03-5 & 20-03-5 & 16-05-5 & 20-05-5\\
		0.3 $\leq$ \textit{m} < 8.0 \\
		- after 1 Myr & 4.1$\,\pm$1.9 / 9 &  0.3$\,\pm$0.4  / 1 &  3.9$\,\pm$2.3  / 9&  0.3$\,\pm$0.5  / 2&  2.0$\,\pm$1.2  / 4&  0.2$\,\pm$0.5  / 2&  1.2$\,\pm$0.9  / 3&  0.3$\,\pm$0.4  / 1 \\ 
		- after 2 Myr &  2.9$\,\pm$1.6 / 6 &  0.4$\,\pm$0.6  / 2&  3.2$\,\pm$2.0  / 9&  0.6$\,\pm$0.9  / 4 &  1.5$\,\pm$1.1  / 4&  0.2$\,\pm$0.5  / 2&  1.0$\,\pm$0.9  / 3&  0.2$\,\pm$0.4  / 1 \\ 
		- after 3 Myr &  1.3$\,\pm$0.8 / 3  &  0.4$\,\pm$0.6  / 2&  1.6$\,\pm$1.2  / 3 &  0.3$\,\pm$0.7  / 3&  0.7$\,\pm$0.7  / 2 & 
		0  / 0&  0.4$\,\pm$0.6  / 1&  0.1$\,\pm$0.2  / 1\\ 
		- after 4 Myr &  0.5$\,\pm$0.7  / 2  &  0.4$\,\pm$0.5  / 1&  0.7$\,\pm$0.8  / 3 &  0.2$\,\pm$0.5  / 2 &  0.2$\,\pm$0.4  / 1 &  0 /  0&  0.1$\,\pm$0.3  / 1&  0.1$\,\pm$0.2  / 1\\ 
		- after 5 Myr &  0.3$\,\pm$0.3  / 2  &  0.3$\,\pm$0.5  / 2 &  0.4$\,\pm$0.6  / 2 &  0.1$\,\pm$0.2  / 1&  0.1$\,\pm$0.2  / 1&  0/ 0&  0.2$\,\pm$0.4  / 1&  0 / 0\\ 
		\hline
		WW stars\\
		\hline
		&  16-03-1 & 30-03-1 & 16-05-1 & 30-05-1 &  16-03-5 & 20-03-5 & 16-05-5 & 20-05-5 \\
		0.3 $\leq$ \textit{m} < 8.0 \\
        - after 1 Myr &  27.6$\,\pm$7.0  / 43  &  2.0$\,\pm$1.4  / 5&  25.3$\,\pm$5.2  / 36&  1.1$\,\pm$1.1  / 4 &  9.0$\,\pm$3.1  / 17&  1.9$\,\pm$1.6  / 6&  9.0$\,\pm$2.0  / 14&  2.3$\,\pm$1.2  / 5\\ 
		- after 2 Myr &  29.2$\,\pm$6.9  / 46  & 2.5$\,\pm$1.4  / 5&  27.2$\,\pm$5.8  / 42&  1.5$\,\pm$1.3  / 5&  9.6$\,\pm$3.2  / 17&  3.1$\,\pm$1.7  / 6&  10.2$\,\pm$2.2  / 15&  2.5$\,\pm$1.3  / 5\\ 
		- after 3 Myr &  30.3$\,\pm$6.9  / 47  &  3.1$\,\pm$1.6  / 7&  29.9$\,\pm$6.2  / 45  & 1.8$\,\pm$1.4  / 5&  10.1$\,\pm$3.8  / 19&  3.4$\,\pm$1.8  /6&  10.3$\,\pm$2.4  / 16&  2.8$\,\pm$1.3  / 5\\ 
		- after 4 Myr &  29.6$\,\pm$6.8  / 44 &  4.3$\,\pm$1.9  / 9&  28.3$\,\pm$5.7  / 41&  2.3$\,\pm$1.7  / 6&  10.3$\,\pm$3.6 / 19&  3.5$\,\pm$1.8 / 7&  10.1$\,\pm$2.7  / 16&  3.1$\,\pm$1.3  / 5\\ 
		- after 5 Myr &  28.1$\,\pm$6.4  / 42  &  4.4$\,\pm$2.1  / 9&  27.0$\,\pm$5.6  / 38&  2.4$\,\pm$1.9  / 7&  10.0$\,\pm$3.8  / 19&  3.8$\,\pm$1.9  / 7&  10.0$\,\pm$3.1  / 16&  3.2$\,\pm$1.6  / 8\\ 
		\hline
	\end{tabular}
\end{table*}

\begin{table*}
	\caption{Ejected RW and WW stars from $N$-body simulations within the search radius of 100 pc at different times during the simulations for the initial condition ID 16-05-1 and the two alternative binary settings described in section~\ref{sec:alt_set-up}. For these initial condition combinations, we show averages from all 20 simulations and the maximum from a single simulation in the format average $\pm$ uncertainty / maximum. We count ejected binary systems as one star when calculating averages and maxima. The uncertainties in our averages are the standard deviations. We show ejected stars with masses from 0.3--8 M$_{\sun}$.}
	\label{tab:RW_WW_pred_bin}
	\begin{tabular}{lcccccccc} 
		\hline
		Mass $m$ (M$_{\sun}$) & &Simulation ID \\
		\hline
		RW stars\\
		\hline
		&  16-05-1& 16-05-1-100-bin & 16-05-1-eq-mass\\
		0.3 $\leq$ \textit{m} < 8.0 \\
		- after 1 Myr & 3.9$\,\pm$2.3 / 9& 1.7$\,\pm$1.5 / 6 &  5.7$\,\pm$2.4  / 12 \\  
		- after 2 Myr & 3.2$\,\pm$2.0 / 9& 1.2$\,\pm$1.2 / 5 &  4.2$\,\pm$1.9  / 8 \\ 
		- after 3 Myr & 1.6$\,\pm$1.2 / 3& 0.5$\,\pm$0.8 / 2  &  1.6$\,\pm$1.1  / 4 \\ 
		- after 4 Myr & 0.7$\,\pm$0.8 / 3& 0.3$\,\pm$0.6  / 2  &  0.4$\,\pm$0.5  / 1\\ 
		- after 5 Myr & 0.4$\,\pm$0.6 / 2& 0.3$\,\pm$0.5  / 2  &  0.2$\,\pm$0.4  / 1 \\ 
		\hline
		WW stars\\
		\hline
		&   16-05-1& 16-05-1-100-bin & 16-05-1-eq-mass \\
		0.3 $\leq$ \textit{m} < 8.0 \\
        - after 1 Myr & 25.3$\,\pm$5.2  / 36 & 26.3$\,\pm$5.9  / 43  &  27.1$\,\pm$7.0  / 43\\ 
		- after 2 Myr & 27.2$\,\pm$5.8  / 42& 28.3$\,\pm$6.3  / 44  & 30.1$\,\pm$7.2  / 43\\ 
		- after 3 Myr & 29.9$\,\pm$6.2  / 45& 28.8$\,\pm$6.3  / 44  &  30.4$\,\pm$7.5  / 45\\ 
		- after 4 Myr & 28.3$\,\pm$5.7  / 41& 29.8$\,\pm$6.1  / 44 &  30.0$\,\pm$7.6  / 45\\ 
		- after 5 Myr & 27.0$\,\pm$5.6  / 38& 28.2$\,\pm$5.9  / 42  &  27.6$\,\pm$6.6  / 41\\ 
		\hline
	\end{tabular}
\end{table*}

The stellar systems in our simulations undergo both stellar and binary evolution and we see several supernovae after 4 Myr. These supernovae are the result of the stellar evolution of the highest mass stars in our simulations, all of which leave a black hole as the supernova remnant during the analysis time of 5 Myr.

\subsection{Numbers from the simulations}

From our $N$-body simulations, we have predicted numbers for RW and WW stars across the full mass range (0.1--50 M$_{\sun}$), which we show in the appendix Table~\ref{tab:RW_WW_pred_full_mass}. In the results from the \textit{Gaia} DR2 observations we find the star (\textit{Gaia} DR2 3326637442758920960) with the smallest mass amongst the RW/WWs has a mass of $\sim$0.3 M$_{\sun}$ and is an M3-star \citep{2014A&A...570A..82V}. 

Even though this star does not trace back to NGC 2264 in 3D, it gives us an indication of the lowest mass we are able to identify within our data set. Our \textit{Gaia} DR2 results show that there are no high-mass (> 8 M$_{\sun}$) RWs or WWs from this region. When comparing our simulations to the \textit{Gaia} DR2 observations, we therefore reduce the mass range to 0.3--8 M$_{\sun}$. 
The number of ejected stars in the two velocity ranges from our simulations are shown in Table~\ref{tab:RW_WW_pred} for all our initial conditions within a radius of 100 pc. Table~\ref{tab:RW_WW_pred_full_mass} shows the numbers for the complete mass range separated into low/intermediate mass stars (0.1--8 M$_{\sun}$) and high-mass stars (> 8 M$_{\sun}$), also within 100 pc.

\citet{RN309} showed that we find more and faster RW/WW stars from simulations where the initial conditions are more spatially substructured and/or subvirial. We find the same trend in our simulations. Simulations with more initial spatial substructure produce a higher number of RW and WW stars (in particular simulations with IDs 16-03-1 and 16-05-1). At this high level of initial substructure (fractal dimension $D$ = 1.6), a change in the initial virial ratio, which sets the global bulk motion of the stars in our simulations, has very little effect on the number of RWs. Due to the initial spatial substructure, even globally initially virialised simulations are subvirial on local scales and undergo local collapse \citep{RN4,RN1}, which causes ejection of stars at early times in the simulations.

The maximum number of RWs from a single simulation is ten, which we achieve in 16-03-1 ($D$ = 1.6, $\alpha_{\text{vir}}$ = 0.3, radius 1pc) between 0.6--0.8 Myr and in 16-05-1 ($D$ = 1.6, $\alpha_{\text{vir}}$ = 0.5, radius 1pc) between 1.5--1.9 Myr. These high numbers of RWs are only achieved at early ages and reduce quickly to a maximum of only two RWs after 5 Myr. High numbers of RWs occur only in initially highly substructured regions with an initial radius of 1 pc, regardless of initial virial ratio. For regions with an initial radius of 5 pc, none of our simulations reach these numbers of RWs and the highest number is four RWs achieved at early ages (up to 2 Myr).

The highest velocities are also reached at early simulation times. In Simulation ID 16-03-1, the fastest velocity is reached after 0.8 Myr with a value of 216\,km\,s$^{-1}$ by a 0.8 M$_{\sun}$ star. This is not only the fastest velocity in this set of simulations, but across all of the simulations. At this velocity it takes this star only 0.4 Myr to leave the 100 pc region around its ejection site. The maximum velocities reached in the simulations are related to the initial conditions. Simulations that start more spatially substructured (lower fractal dimension $D$) and in addition have a higher initial stellar density produce RWs with higher velocities. For these simulations the maximum velocities are above 100\,km\,s$^{-1}$. If they are initially subvirial the maximum velocity reaches even higher values above 200\,km\,s$^{-1}$. These main results from the simulations in this analysis are in line with what is shown in \citet{RN309} for a simpler set of initial conditions without primordial binaries.

We see high numbers of walkaways (up to a maximum $\sim$40--50) in two initial condition settings, both initially highly substructured and with an initial radius of 1 pc. We find lower numbers of WWs (up to a maximum $\sim$15--20) from initially highly substructured simulations with an initial radius of 5 pc. Simulations with less or no initial substructure produce even fewer WWs (<10 WWs).

\subsubsection{Results from the alternative binary systems set-up}

The maximum and average numbers of ejected RW/WW stars from the two alternative binary system set-ups for initial condition ID 16-05-1 are shown in Table~\ref{tab:RW_WW_pred_bin} for the mass range 0.3--8\,M$_{\sun}$. Table~\ref{tab:RW_WW_pred_full_mass_bin} in the Appendix provides the numbers for the full mass range of 0.1-8\,M$_{\sun}$ and also for high-mass stars >8\,M$_{\sun}$.

Simulation ID 16-05-1-100-bin is set-up with an initial binary fraction of 100 per cent and a different separation distribution than the base case 16-05-1, but it does have the same flat mass ratio distribution. The simulations show smaller maximum and average numbers of RW stars at all ages. The maximum number of RWs is six (four fewer than in the base case) in the relevant mass range for our analysis of 0.3--8\,M$_{\sun}$. These six remain within 100 pc until an age of 1.9 Myr. Five RWs are present for another 0.1 Myr, after which the number drops to three RWs until an age of 2.8 Myr. After $\sim$3 Myr, this alternative set-up shows virtually the same number of RWs within 100 pc as the base case, making these two alternative binary system set-ups indistinguishable. The maximum number of WWs in this mass range is slightly higher throughout most of the simulation times covered here. The averages are similar and consistent with each other within the uncertainties.

Simulation ID 16-05-1-eq-mass has the same initial condition as our base set-up 16-05-1 apart from the mass ratio for stars with a mass >8\,M$_{\sun}$, which is set to be an equal mass ratio instead of being drawn from a flat distribution. These simulations show a larger maximum number of RWs up to an age of $\sim$3 Myr in the relevant mass range 0.3--8\,M$_{\sun}$. The average numbers of RWs at these early times are also larger than in the base case of 16-05-1, but are consistent within the uncertainties. The maximum number of RWs is 13 (3 more than in the base case), but this high number is only present within 100 pc of the centre between the ages of 0.5--0.7 Myr. After 3 Myr, the number or RWs quickly drop off below the numbers recorded for the base case. The maximum number of WWs is slightly higher than those in the base case, as are the averages. However, this difference is not significant.

\section{Discussion}\label{Discuss}

\subsection{3D-candidates tracing back to more than one subregion}\label{Discuss_double_trace_back}

In our results, we find several candidates that can be traced back to more than one subcluster as their velocity vector is oriented in such a way that an origin in more than one of the subclusters is possible. Five of the 3D RW-candidates that we trace back to S\,Mon can also be traced back to IRS\,1 and/or IRS\,2. Three 3D RW-candidates trace back to IRS\,2 but also to IRS\,1. Four 3D-WW traces back to S\,Mon but also to IRS\,1 and/or IRS\,2 and two 3D-WW trace back both to IRS\,1 and IRS\,2.

For these candidates, we check the flight times for each of these trajectories and consider the candidates to have originated from the subcluster which has the smallest flight time, i.e. time since ejection. Our reasoning for this classification is as follows: if a star gets ejected from one region and then passes through another region it is possible that interactions in the second region can alter its trajectory. Once this interaction happens, it would be difficult/impossible to trace this star back to its origin region. We consider the alignment of a star's trajectory to a region ''behind`` another region to be a chance-alignment and consider the first region a star traces back to as its origin region. However, it remains possible for these stars to have originated in the region further away from its current location, so in our results tables we will still consider them a possible origin region, but we note the more likely birth region.

For one of the 3D-WW trace-backs (\textit{Gaia} DR2 3326695991753074304) that traces back to IRS\,1 and IRS\,2, this approach fails. It is possibly still located within the overlap section of both subclusters and therefore could belong to either.

\subsection{S Mon}\label{Discuss_SMon}

When considering the approach for double trace-backs the number of successfully traced back 3D RW-candidates from S\,Mon reduces from nine to five stars. Of the 13 traced back 3D WW-stars, 11 stars remain after removing double trace-backs and the 3D-WW MS star with an age estimate of 50 Myr. Two of these eleven have flight times that are larger than their estimated age, suggesting that they might have not originated from S\,Mon. This leaves us with nine WWs after also excluding these two.

We have two 3D-WW candidates that only turn into WWs, when we consider their RVs. The 2D-velocity of one of these candidates (\textit{Gaia} DR2 3326739933562218496) is so small that based on this it would still be considered bound to the cluster; it is also still located within the cluster (when the upper distance estimate is used). For these type of candidates it is important to make sure that their higher RV does not originate from binary motion. For NGC~2264~404 (\textit{Gaia} DR2 3326739933562218496), there are only two RV measurements available on the Simbad database, a very old measurement from 1953 \citep{1953GCRV..C......0W} and the \textit{Gaia} DR2 measurement. The older measurement has no error provided but the quality indicator ``E'' on Simbad means that the error is likely large. We consider these two measurements as being consistent with each other, so it does not suggest a binary origin. This star appeared in \citet{2013ApJ...773...54K}, where it was also not identified as a binary, but as a YSO, so we keep it as part of our 3D-WW list.

\begin{figure}
    \centering
    \includegraphics[width=0.9\linewidth]{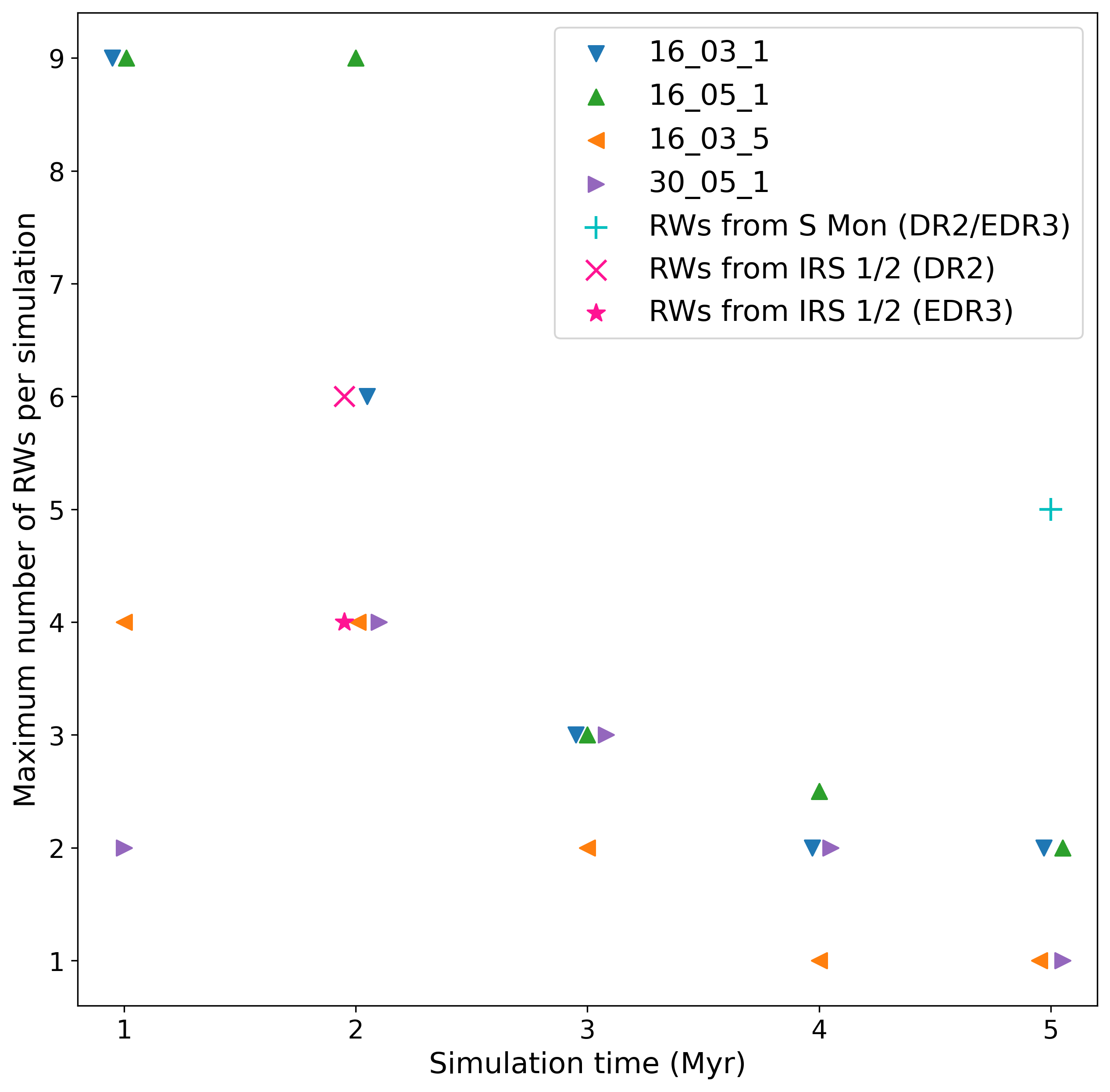}
    \caption{Maximum number of RWs for four selected initial conditions for simulation times from 1--5 Myr. The maximum number of RWs decreases at later times in the simulations and we see that the highest number of RWs is achieved in two different initial conditions (16-03-1 and 16-05-1). We also plot the observed numbers for S\,Mon (5 RWs) and IRS\,1/2 (6/4 RWs) at their maximum ages used for the PARSEC isochrones (5/2 Myr).}
    \label{fig:Sim_RW_num}
\end{figure}

We now compare the S\,Mon 3D-RW and WW stars from the \textit{Gaia} DR2 observations to our simulations. The five RW stars found exceed the averages and maxima of most of the initial conditions sets shown in Table~\ref{tab:RW_WW_pred}. This allows us to exclude several of these initial conditions as highly unlikely to be those of S\,Mon. However, we might have to revisit them again in the future, especially when a better restriction of the extent of the subcluster in the radial direction (i.e. depth) is available. Fig.~\ref{fig:Sim_RW_num} shows the maximum number of RWs at five different times for four selected initial conditions and we also plot the number of RWs we find in the observations. The maximum number of RWs decreases at later times in the simulations and we see that the highest number of RWs is achieved in two different initial condition sets (16-03-1 and 16-05-1). We also plot the observed numbers for S\,Mon (5 RWs) at its maximum age used for the PARSEC isochrones (5 Myr).

We find that only initial condition sets that feature a high amount of initial spatial substructure ($D$ = 1.6) can be fitted to the observations. Any simulation with a smooth initial spatial distribution, i.e. $D$ = 3.0, or a moderate amount of initial spatial substructure, i.e. $D$ = 2.0 produces too few RWs to be consistent with the observations. For Simulation ID 16-03-1 ($D$ = 1.6, $\alpha_{\text{vir}}$ = 0.3, radius 1pc), we find a maximum of five RWs within 100 pc in our simulations up to an age of 2.5 Myr. When we consider the average number, the upper age at which this number of RWs is still present in the vicinity of S\,Mon reduces to 1.7 Myr. For Simulation ID 16-05-1 ($D$ = 1.6, $\alpha_{\text{vir}}$ = 0.5, radius 1pc), we find five RWs up to a maximum age of 2.6 Myr and when considering the average number, the maximum age of our simulations where the RWs/WWs are consistent with the observations reduces to 2.1 Myr.

The fastest RW that can only be traced back to S\,Mon is \textit{Gaia} DR2 3326893624672681216 with a space velocity of $\sim$62 km\,s$^{-1}$. When we compare this velocity to the highest velocities reached in Simulation ID 16-03-1, the fastest star in our simulations has a similar velocity compared to this star between 2.5 Myr and 3.4 Myr. At 2 Myr, we find that the velocity of the fastest star within 100 pc in the simulations is double at $\sim$124 km\,s$^{-1}$. This $\sim$1.0 M$_{\sun}$ star leaves the 100 pc boundary shortly thereafter. After 3.4 Myr, a $\sim$1.2 M$_{\sun}$ star gets ejected at a velocity of $\sim$87 km\,s$^{-1}$ and becomes the fastest star until it also leaves the 100 pc radius. At this point, the highest velocity across the simulations drops considerably to $\sim$48 km\,s$^{-1}$. 

When comparing the fastest S\,Mon RW-star to the simulations with initial condition ID 16-05-1, we find that between 2.5--2.7 Myr, the velocity of the fastest star from all 20 simulations ($\sim$66 km\,s$^{-1}$) is comparable to the observations. At most other times below 5 Myr, we see maximum velocities above this value (74--126 km\,s$^{-1}$) due to fast stars being continuously ejected and replacing those that leave the 100 pc region. 

When we compare the averages of the highest velocities from all 20 simulations, we see much lower average-maximum velocities. For 16-03-1 at 1 Myr, the average of the highest velocities from all 20 simulations is $\sim$76 km\,s$^{-1}$, this average drops to $\sim$53 km\,s$^{-1}$ at 2 Myr and to $\sim$38 km\,s$^{-1}$ at 3 Myr. For 16-05-1 at 1 Myr, the average of the highest velocities is $\sim$55 km\,s$^{-1}$, after 2 Myr, this drops to $\sim$44 km\,s$^{-1}$ and to $\sim$35 km\,s$^{-1}$ after 3 Myr. Most of these velocities are below the velocity of the fastest star from S\,Mon.


Our upper age estimate for S\,Mon of less than 3 Myr is lower than that from most other studies. This lower age estimate is due to the high number of RWs that we can trace back to the region. We now briefly comment on the likelihood that all five RWs actually originated in S\,Mon. Two of the five candidates only trace back to the region when we consider the measurement errors. The first one is \textit{Gaia} DR2 3131997187129420672 and it only traces back on the sky when considering the errors in proper motion. In addition it only becomes a 3D-candidate when considering the errors both in distance (error is $\sim$25 pc) and RV. The second one is \textit{Gaia} DR2 3132474680112352128, which once again only traces back on the sky when considering its proper motion errors. In 3D, it only traces back when considering either its distance or its RV error.

To confirm these candidates, we check the latest data from \textit{Gaia} Early Data Release 3 (EDR3) with measurements covering a longer period than that in \textit{Gaia} DR2 \citep{2020arXiv201201533G}. This could possibly reduce the proper motion errors of these two stars and lead to them no longer tracing back to S\,Mon. In \textit{Gaia} EDR3, there has been no update to the radial velocities. We use \textit{Gaia} EDR3 to double-check all \textit{Gaia} DR2 3D-candidates from our analysis. Similar to our \textit{Gaia} DR2 analysis, we do not use the \textit{Gaia} DR2 parallaxes, but Bayesian estimated photogeometric distances from \citet{2021AJ....161..147B} providing distances for $\sim$1.35 billion stars in \textit{Gaia} EDR3. As \textit{Gaia} EDR3 is missing extinction and reddening values for all stars, we cannot utilise \textit{Gaia} EDR3 in the same way as DR2 to confirm stellar ages, so we have only used it to confirm the 3D trace-backs.

Both of the above mentioned error-only candidates still trace back to the region in 3D using \textit{Gaia} EDR3. Also, the remaining three 3D-RWs from S\,Mon continue to trace back to this region using the most recent kinematic data. So, after this check, we still have 5 RWs tracing back to S\,Mon, which fits to two sets of initial conditions in the simulations.

We have a fairly low number of traced back WWs from S\,Mon, considering that we have found five RWs. We compare these nine WWs to those predicted in simulations that fit the RW numbers (16-03-1 and 16-05-1). We find that the number of WWs is far below the average/maximum predicted by these simulations at all ages. However, of the 18 2D-WW candidates with RV, over 70 per cent trace back to S\,Mon in 3D. It is highly likely that when we measure RVs for the remaining 47 2D-candidates, we will be able to increase the number of traced back WWs.

Like for the RWs, we also have WW-candidates that only trace back on the sky given their velocity errors: \textit{Gaia} DR2 3331597816450524288 and \textit{Gaia} DR2 3327203588170236672. When considering \textit{Gaia} EDR3, we lose both of these stars, as well as \textit{Gaia} DR2 3326740693772293248 and \textit{Gaia} DR2 3326713442204844160, which would leave us with five WWs. However, we can expect to identify additional 2D and possibly 3D-WW candidates in \textit{Gaia} EDR3 that do not trace back using \textit{Gaia} DR2.

After comparing the \textit{Gaia} DR2 RW/WWs to the simulations, we find that S\,Mon appears to be consistent with initial conditions that show a high level of initial substructure and either an initially subvirial ratio or are in virial equilibrium. In these two initial condition set-ups the initial stellar surface density would have been fairly high with $\sim$3000 stars pc$^{-2}$ and an initial mass density of $\sim$10 000 M$_{\sun}$ pc$^{-3}$, which would be similar to the initial density suggested for the ONC \citep[see][and references therein]{RN8}.

\subsubsection{Comparing RWs from the alternative binary system set-ups}

We incorporated two alternative set-ups for the binary systems in our simulations for the initial condition with ID 16-05-1 (fractal dimension $D$ = 1.6, $\alpha_{\text{vir}}$ = 0.3, radius 1 pc). For the simulations with an initial binary fraction of 100 per cent and a different separation distribution, we find at least five RWs up to an age of 2.0 Myr, compared to 2.6 Myr for the base case. For the simulations with an equal-mass ratio for high-mass stars instead of a flat distribution, we find five RWs up to an age of 2.7 Myr, which is virtually the same age as in the base case. Here, we can also match the average number of RWs up to an age of 2.2 Myr, which is virtually the same age as in the base case (2.1 Myr). The change in the set-up of the binaries does not cause a significant change in our results and does not change our conclusions about the likely initial conditions

Even though we cannot match the average number from one of these two sets of simulations with the observations, both remain consistent with the maxima from the base case. Therefore, S\,Mon remains consistent with the same initial spatial and kinematic substructure conditions. In addition, the differences in the maximum number of RWs within 100 pc stemming from the alternative binary set-ups virtually disappear after 3 Myr, when the fastest RWs ejected at early times have left this region.

\subsection{IRS 1 and IRS 2}

While the centres of IRS\,1 and IRS\,2 are located around 10 pc from each other \citep{RN264} in radial distance, the estimated ages for these subclusters are the same and our chosen boundaries of the regions overlap considerably. For the comparison to our simulations, we consider these two subclusters as one region. In our simulations, we find several examples of a single star-forming region having evolved into two subclusters with centres located several pc from each other. We find a total of seven 3D-RWs that we can trace-back to either IRS\,1 or IRS\,2 and five 3D-WWs, after we apply the approach for double trace-back described in section~\ref{Discuss_double_trace_back}. The number of 3D-RWs further reduces to six 3D-RWs, when excluding the MS 3D-RW, which has been suggested to have an age of 159 Myr \citep{2012AstL...38..694G}.

Most of the RWs and WWs that trace back to both regions show very similar flight times, which is not surprising given their close proximity on the sky. This further highlights that the decision to treat IRS\,1 and 2 as one initial star-forming region appears to be valid. 

We use a PARSEC isochrone for IRS\,1/2 with an age several Myr younger than S\,Mon (2 Myr instead of 5 Myr) and we trace back more RWs to IRS\,1/2 than to S\,Mon. This result is fully in line with the predictions from simulations, if these regions started from the same initial conditions. Due to the higher velocity of RWs, they leave our 100 pc search region much more quickly than WWs, so the older a region is the fewer RWs we can expect to find within a 100 pc radius.

Our six RWs in the observations are only predicted in two of our initial conditions sets: ID 16-03-1 and 16-05-1. Both of these simulations start with a high amount of initial spatial substructure ($D$ = 1.6) and an initial radius of 1 pc. These regions can either be subvirial or in virial equilibrium.

For the initial conditions represented in 16-03-1 ($D$ = 1.6, $\alpha_{\text{vir}}$ = 0.3, radius 1pc), the maximum number of observed RWs is consistent with one of our simulations up to an age of 2.1 Myr. The average matches up to an age of 1.2 Myr. In the 16-05-1 ($D$ = 1.6, $\alpha_{\text{vir}}$ = 0.5, radius 1pc) simulations, the maximum number of RWs from the observations is consistent with one simulation up to an age of 2.4 Myr, which is above the age we choose for our isochrone (2 Myr). We can match the averages up to an age of 1.1 Myr. 

The upper age estimates for these two initial conditions are higher than the upper end of the age estimates we find in the literature and any reduction in RWs would further increase these age estimates. However, we can only match the RWs to a single simulation in each of the two initial condition sets at these high ages and also have individual simulations predicting this high number of RWs at minimum ages of 0.1 Myr for 16-03-1 and 0.2 Myr for 16-05-1.

The fastest star from IRS\,1/2 is \textit{Gaia} DR2 3132688088448708736 with a space velocity of $\sim$88 km\,s$^{-1}$. When comparing this velocity to the maximum reached in either of the simulations represented by 16-03-1 and 16-05-1 at early ages (< 2Myr), the simulations predict stars of even higher velocities within 100 pc. When comparing the average of the maximum velocities for all 20 simulations, for 16-03-1 we reach the highest value of $\sim$86 km\,s$^{-1}$ at simulation time 0.8 Myr. For 16-05-1, this average never reaches velocities over 80 km\,s$^{-1}$. 

We now double-check the 3D-RWs we find in \textit{Gaia} DR2 against EDR3. Two of the 3D-RW stars trace back to IRS\,1/2 on the sky only when considering the proper motion errors. These two are \textit{Gaia} DR2 3134140405870541568 (trace-back to IRS\,1) and \textit{Gaia} DR2 3132688088448708736 (trace-back to IRS\,2). We lose the former of these stars when tracing back the stars with \textit{Gaia} EDR3, but retain the latter. We also lose one further star \textit{Gaia} DR2 3134176728413264896, reducing the total number of 3D-RWs from six to four. In this \textit{Gaia} EDR3 check, we also lose \textit{Gaia} DR2 3326630811329448576, which is the MS star with an age estimate of 159 Myr we previously already excluded.

Reducing the number of RWs from six to four increases the possible ages in Simulation IDs 16-03-1 and 16-05-1. For initial condition set 16-03-1, the maximum upper age from the simulation increases from 2.1 to 2.7 Myr and we comparing the average we go from 1.2 Myr to 2.2 Myr. The lowest age, where we see a maximum of 4 RWs in these simulations remains at 0.1 Myr.

For initial condition ID 16-05-1, the upper age from the maximum increases from 2.4 to 2.9 Myr and the average from 1.1 Myr to 2.4 Myr. The lowest age, where we see a maximum of 4 RWs in these simulations is 0.1 Myr. This reduction of RWs to four also opens up two other possible initial conditions, however, only when fitting it to the maximum number of ejected stars as none of the averages fit to the observations. 

Simulation ID 30-05-1 ($D$ = 3.0, $\alpha_{\text{vir}}$ = 0.5, radius 1 pc) has one simulation with four RWs at ages of 1.1--2.1 Myr, whereas the average number of RWs in simulations is too low at any age. A second additional option for the initial conditions of IRS\,1/2 is found in Simulation ID 16-03-5 ($D$ = 1.6, $\alpha_{\text{vir}}$ = 0.3, radius 5 pc). Once again we cannot match the average number of RWs to the observations. But we find that the maximum number from one of the simulations is consistent up to an age of 2.1 Myr and we find further simulations with this number of RWs down to an age of 0.1 Myr.

Fig.~\ref{fig:Sim_RW_num} shows the maximum number of RWs at five different times for these four possible initial conditions and we also plot the number of RWs we find in the observations. The maximum number of RWs decreases at later times in the simulations and we see that the highest number of RWs is achieved in two different initial condition sets (16-03-1 and 16-05-1). We also plot the observed numbers for IRS\,1/2 (6 RWs) at their maximum age used for the PARSEC isochrones (2 Myr) and include the number of RWs after double-checking our results with \textit{Gaia} EDR3 (4 RWs).

When considering the average age estimates with four RWs from 16-03-1 and 16-05-1, we find that these upper age estimates are far above those that other literature sources predict for these subclusters, However, we also have simulations predicting much lower ages, so both of these initial conditions are viable options. We now turn to the 3D-WWs to evaluate if these provide any further insights.

\begin{figure*}
    \centering
    \begin{minipage}[t]{1.0\columnwidth}
        \centering
        \vspace{0pt}
    	\includegraphics[width=1.0\columnwidth]{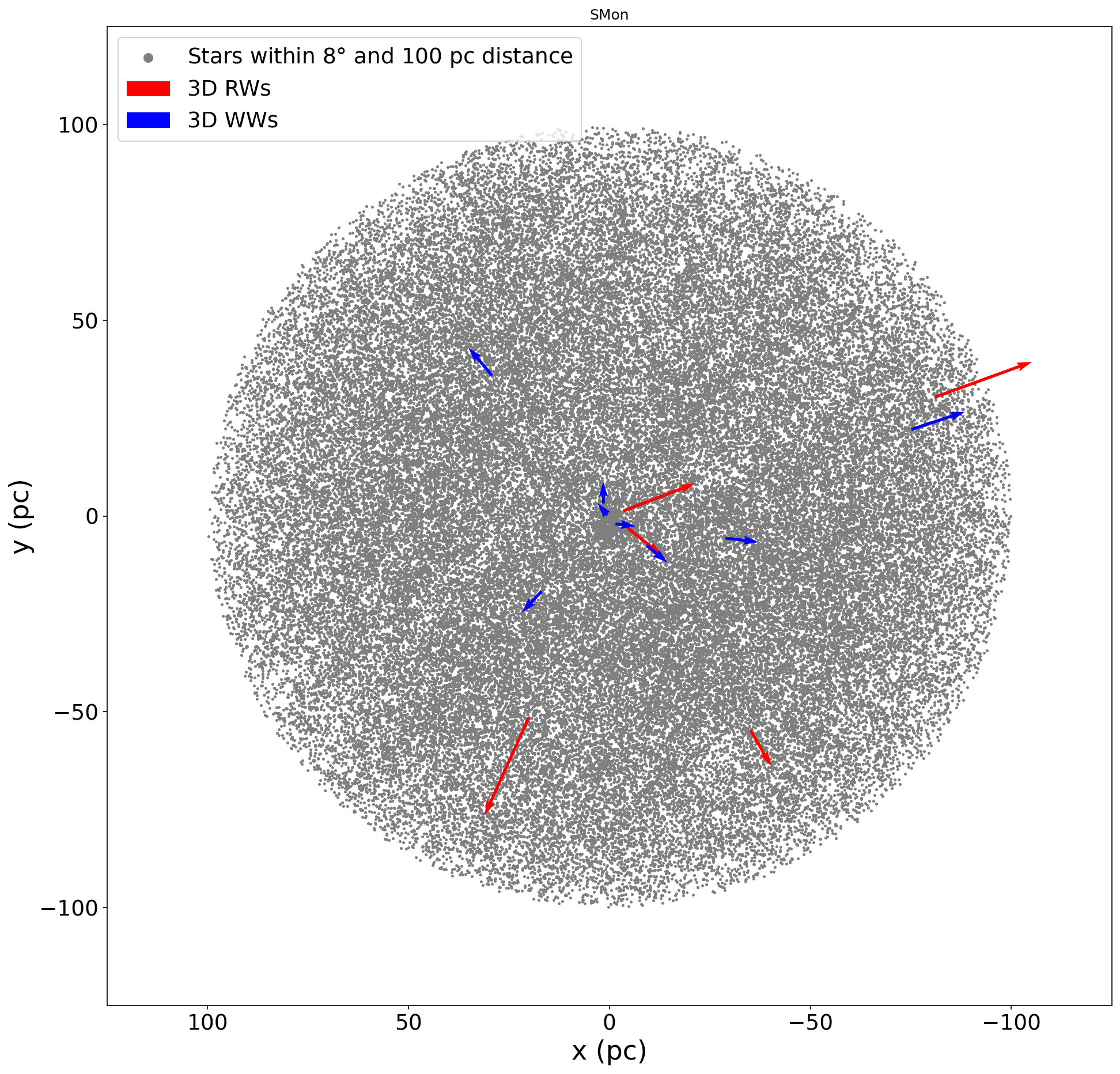}
    \end{minipage}
    \hfill{}
    \begin{minipage}[t]{1.0\columnwidth}
        \centering
        \vspace{0pt}
        \includegraphics[width=1.0\columnwidth]{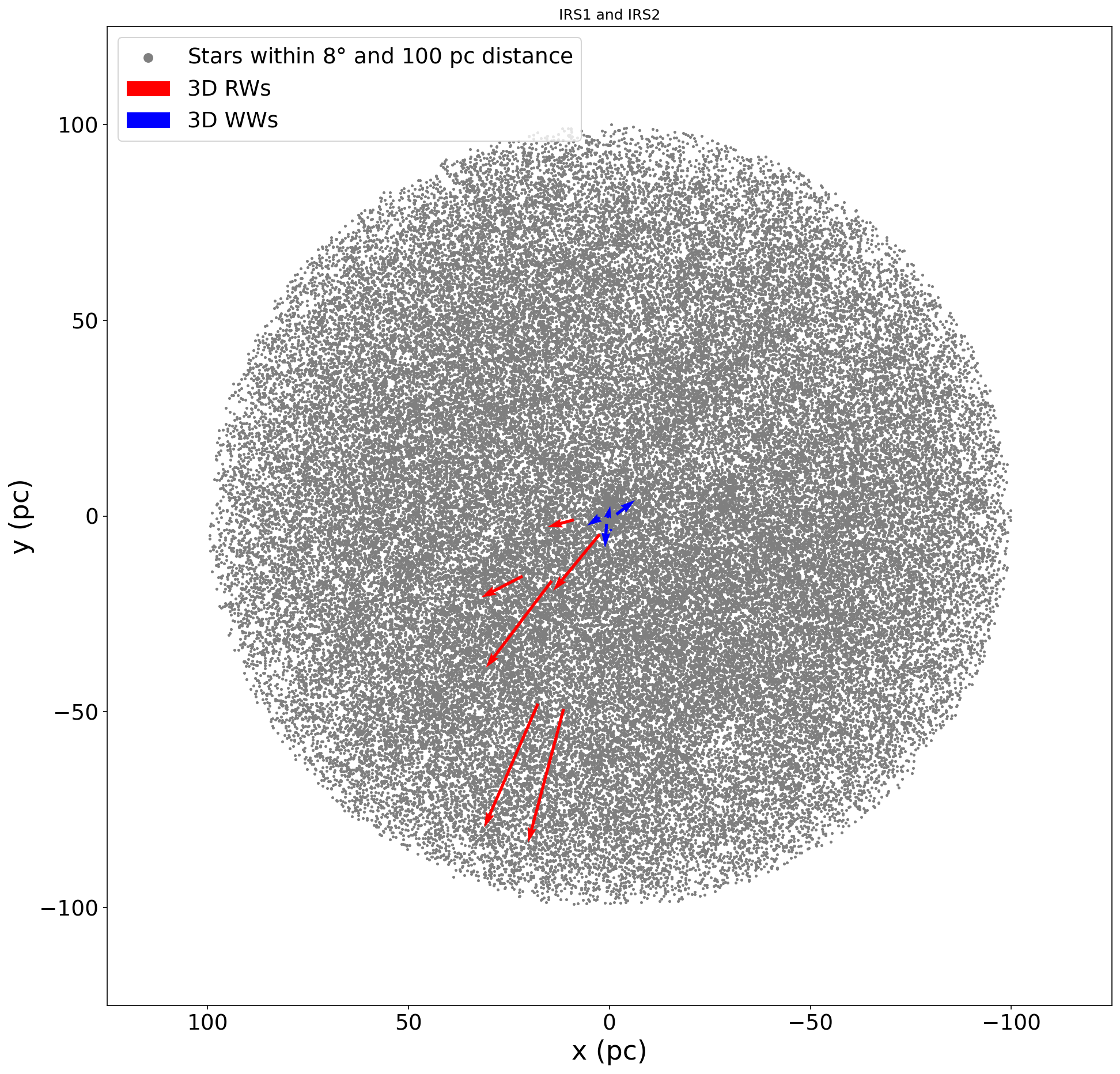}
    \end{minipage}
    \caption{Location of the identified 3D RW and WW stars. Left: S\,Mon, Right: IRS\,1 and 2. The subclusters extend to a radius of $\sim$2 pc and are located at the centres of the plot. The x-axis is inverted to replicate the orientation on the sky (i.e. decreasing right ascension from left to right). The RW stars are plotted in ``red'', with the length of the arrows indicating their 2D-velocity in the respective rest frame, the WW stars are plotted in ``blue''. Note: one 3D-WW in each figure is not visible due to a very small 2D-velocity.}
    \label{fig:Cand_location}
\end{figure*}



We find five 3D WWs originating from IRS\,1/2, which is much lower than the predicted numbers from simulations 16-03-1 and 16-05-1 at any age in our simulations. There are still a few 2D-candidates that are missing RVs, however, these are only 19 further candidates. If all of these 2D-candidates turned into 3D-candidates (i.e. 24 3D-WWs), we would still not reach the maximum number of WWs predicted by these two initial conditions. However, we would be able to match the average number of WWs in both initial set-ups at all ages. So the number of WWs do not enable us to further constrain the initial conditions.

Using \textit{Gaia} EDR3 to also check the 3D-WWs, we lose one of these (\textit{Gaia} DR2 3326693857153492736) taking us to four 3D-WWs. We now match this number to the two additional initial conditions that open up after the \textit{Gaia} EDR3 check of the 3D-RWs. For 30-05-1, this number of WWs only fits the maximum number at most ages, however, but does not match any of the averages. Any further WW discoveries in \textit{Gaia} EDR3 or other sources would make this initial condition unsuitable for IRS\,1/2. With initial condition 16-03-5, this number of WWs is still well below the maximum and below the averages, however, this gap would be much easier to close than for 16-03-1 and 16-05-1. 

\subsubsection{Comparing RWs from the alternative binary system set-ups}

We now compare the observation results for IRS\,1/2 to the two sets of simulations with alternative set-ups for the binary systems. For the simulations with an initial binary fraction of 100 per cent and a different separation distribution, we find at least six RWs up to an age of 1.9 Myr, compared to 2.4 Myr for the base case. For the simulations with an equal-mass ratio for high-mass stars instead of a flat distribution, we find a maximum of six RWs up to an age of 2.3 Myr, which is virtually the same age as in the base case. Here, we can also match the average number of RWs up to an age of 2.0 Myr, which is an increase compared to the base case (1.1 Myr).

Neither change in the set-up of the binaries causes a significant change in our results or our interpretation of them. Even though we cannot match the average number for one of these sets of simulations with the observations, they remain consistent with the maximum from the base case. Therefore, IRS\,1/2 remains consistent with the same initial conditions.

\subsection{Implications of the results for both subregions}

The number of ejected stars from S\,Mon and IRS\,1/2 are consistent with both regions having formed from initial conditions represented by Simulation IDs 16-03-1 or 16-05-1, regardless of the settings used for the set-up of the binary systems (tested for Simulation ID 16-05-1). The simulation differ only by the initial virial ratio. If 16-05-1 was the true initial condition, IRS\,1/2 could be a binary cluster \citep{RN278,RN288,RN287,RN290,RN291, 2021MNRAS.tmp..697B}, which is a common outcome in simulations with these initial conditions \citep{RN210}. If 16-03-5 remained as a set of initial conditions consistent with the observations, then IRS\,1/2 would have formed with a slightly different initial set-up than S\,Mon. Both regions would share their initially highly spatial substructure and possibly their initial subvirial radius with a different initial stellar density. The simulations with initial conditions represented by IDs 16-03-1 and 16-05-1 show initial stellar densities as high as $\sim$3000 stars pc$^{-2}$ and a mass density of $\sim$10 000 M$_{\sun}$ pc$^{-3}$, whereas simulations with initial condition ID 16-03-5 start with a lower initial stellar density of $\sim$400 stars pc$^{-2}$ and a mass density of $\sim$150 M$_{\sun}$ pc$^{-3}$.

In past studies \citep{2006ApJ...648.1090F, 2015AJ....149..119T,2018MNRAS.476.3160C}, these authors found evidence of a clumpy (initial) spatial substructure and a correlation with the kinematic substructure. From our results presented here, we suggest that our results confirm these previous results using a different approach and provide avenues for further study. A study focused on the possible initial spatial substructure using alternative approaches such as mass segregation and the $\mathcal{Q}$-parameter would allow us to analyse this further. We present this study in a companion paper \citep[][]{Parker+Schoettler2021}.

Fig.~\ref{fig:Cand_location} shows the location of the identified five 3D-RWs and nine 3D-WWs from S\,Mon on the left and the six 3D-RWs and five 3D-WWs from IRS\,1/2, that we found based on our analysis of the \textit{Gaia} DR2 data. We see clearly that the WWs from S\,Mon, which is older, have reached farther distances from the cluster than those ejected from IRS\,1/2, which are still in close proximity of the cluster. Also there appear to be no ejected stars from IRS\,1/2 that are on a North-West-trajectory.

Future \textit{Gaia} data releases and further RV measurements should provide us with more clarity thereby allowing us to better restrict the initial conditions of S\,Mon and IRS\,1/2, however, this \textit{Gaia} DR2 analysis has provided a framework of possible initial conditions for the subclusters that can be used for future analyses of new observational data.

\section{Conclusions}

In this paper, we combine \textit{Gaia} DR2 and EDR3 observations with predictions from $N$-body simulations to search for RW and WW stars from the three subclusters S\,Mon, IRS\,1 and IRS\,2 in NGC 2264 within a distance of 100 pc to constrain the initial conditions of these regions. The conclusions from our simulations and the search in \textit{Gaia} DR2/EDR3 are summarised as follows:

\renewcommand{\labelenumi}{(\roman{enumi})}
\begin{enumerate}
  \item We find five 3D-RWs and nine 3D-WWs within 100 pc of S\,Mon that we can trace back to this subcluster using \textit{Gaia} DR2. All of these RWs and WWs are either low- or intermediate-mass stars. We find two 3D-WWs, which have already reached the MS, all others are still pre-MS stars. When comparing the number of ejected stars to those predicted in $N$-body simulations, we find that S\,Mon appears to have evolved from initial conditions with an initially highly substructured distribution with a high stellar density ($\sim$10 000 M$_{\sun}$ pc$^{-3}$) and either initially subvirial or virialised velocities.
  \item While we have searched separately for ejected stars from IRS\,1 and IRS\,2 in \textit{Gaia} DR2, we treat these subclusters as one when comparing them to simulations, due to their similar age and considerable overlapping boundaries. For IRS\,1/2, we trace-back six 3D-RWs and five 3D-WWs still within 100 pc of these subclusters. As for S\,Mon, all of these stars are either low- or intermediate-mass stars. The higher number of 3D-RWs suggests that this region is younger than S\,Mon, as the number of RWs drops for the same initial condition with older ages. We find the same initial conditions fit to IRS\,1/2 than to S\,Mon.
  \item We use \textit{Gaia} DR2 for our analysis, as this release provides extinction and reddening values, which are not available in the more recent \textit{Gaia} EDR3. This allowed us to use extinction/reddening corrected CAMDs to predict ages for our ejected stars. However, we use the more accurate \textit{Gaia} EDR3 astrometry to check if our ejected stars still trace back given the more updated data. For S\,Mon, the number of 3D-RWs does not change, however, two 3D-RWs no longer trace back to IRS\,1/2 in \textit{Gaia} EDR3. With this lower number of four 3D-RWs, two further sets of initial conditions become viable options for IRS\,1/2. One set has smooth spatial substructure, low to moderate initial stellar density ($\sim$150 M$_{\sun}$ pc$^{-3}$) and is initially virialised; 
the other is initially highly substructured, subvirial and with a low initial stellar density ($\sim$70 M$_{\sun}$ pc$^{-3}$). 
  \item For all subclusters, we only find a very low number of 3D-WWs of nine or less, when some of the possible initial conditions predict 3--4 times that number. This is likely due to the lack of RVs available for our 2D-trace-backs and the number should increase with further RV availability.
\end{enumerate}

\section*{Acknowledgements}

CS acknowledges PhD funding from the 4IR Science and Technology Facilities Council (STFC) Centre for Doctoral Training in Data Intensive Science. RJP acknowledges support from the Royal Society in the form of a Dorothy Hodgkin Fellowship.

This work has made use of data from the European Space Agency (ESA) mission {\it Gaia} (\url{https://www.cosmos.esa.int/gaia}), processed by the {\it Gaia} Data Processing and Analysis Consortium (DPAC, \url{https://www.cosmos.esa.int/web/gaia/dpac/consortium}). Funding for the DPAC has been provided by national institutions, in particular the institutions participating in the {\it Gaia} Multilateral Agreement. 

This research has made use of the SIMBAD database, operated at CDS and the VizieR catalogue access tool, CDS, Strasbourg, France.

\section*{Data Availability Statement}
The data underlying this article were accessed from the \textit{Gaia} archive, \url{https://gea.esac.esa.int/archive/}. The derived data generated in this research will be shared on reasonable request to the corresponding author.




\bibliographystyle{mnras}
\bibliography{Main_document} 

\begin{thebibliography}{}
\makeatletter
\relax
\def\mn@urlcharsother{\let\do\@makeother \do\$\do\&\do\#\do\^\do\_\do\%\do\~}
\def\mn@doi{\begingroup\mn@urlcharsother \@ifnextchar [ {\mn@doi@}
  {\mn@doi@[]}}
\def\mn@doi@[#1]#2{\def\@tempa{#1}\ifx\@tempa\@empty \href
  {http://dx.doi.org/#2} {doi:#2}\else \href {http://dx.doi.org/#2} {#1}\fi
  \endgroup}
\def\mn@eprint#1#2{\mn@eprint@#1:#2::\@nil}
\def\mn@eprint@arXiv#1{\href {http://arxiv.org/abs/#1} {{\tt arXiv:#1}}}
\def\mn@eprint@dblp#1{\href {http://dblp.uni-trier.de/rec/bibtex/#1.xml}
  {dblp:#1}}
\def\mn@eprint@#1:#2:#3:#4\@nil{\def\@tempa {#1}\def\@tempb {#2}\def\@tempc
  {#3}\ifx \@tempc \@empty \let \@tempc \@tempb \let \@tempb \@tempa \fi \ifx
  \@tempb \@empty \def\@tempb {arXiv}\fi \@ifundefined
  {mn@eprint@\@tempb}{\@tempb:\@tempc}{\expandafter \expandafter \csname
  mn@eprint@\@tempb\endcsname \expandafter{\@tempc}}}

\bibitem[\protect\citeauthoryear{{Adams}, {Proszkow}, {Fatuzzo}  \&
  {Myers}}{{Adams} et~al.}{2006}]{RN272}
{Adams} F.~C.,  {Proszkow} E.~M.,  {Fatuzzo} M.,   {Myers} P.~C.,  2006,
  \mn@doi [\apj] {10.1086/500393}, 641, 504

\bibitem[\protect\citeauthoryear{{Allen}}{{Allen}}{1972}]{1972ApJ...172L..55A}
{Allen} D.~A.,  1972, \mn@doi [\apjl] {10.1086/180890}, 172, L55

\bibitem[\protect\citeauthoryear{{Allison}, {Goodwin}, {Parker}, {Portegies
  Zwart}, {de Grijs}  \& {Kouwenhoven}}{{Allison} et~al.}{2009}]{RN15}
{Allison} R.~J.,  {Goodwin} S.~P.,  {Parker} R.~J.,  {Portegies Zwart} S.~F.,
  {de Grijs} R.,   {Kouwenhoven} M.~B.~N.,  2009, \mn@doi [\mnras]
  {10.1111/j.1365-2966.2009.14508.x}, 395, 1449

\bibitem[\protect\citeauthoryear{{Allison}, {Goodwin}, {Parker}, {Portegies
  Zwart}  \& {de Grijs}}{{Allison} et~al.}{2010}]{RN4}
{Allison} R.~J.,  {Goodwin} S.~P.,  {Parker} R.~J.,  {Portegies Zwart} S.~F.,
  {de Grijs} R.,  2010, \mn@doi [\mnras] {10.1111/j.1365-2966.2010.16939.x},
  407, 1098

\bibitem[\protect\citeauthoryear{{Anders} et~al.,}{{Anders}
  et~al.}{2019}]{2019A&A...628A..94A}
{Anders} F.,  et~al., 2019, \mn@doi [\aap] {10.1051/0004-6361/201935765}, 628,
  A94

\bibitem[\protect\citeauthoryear{Andrae et~al.,}{Andrae et~al.}{2018}]{RN303}
Andrae R.,  et~al., 2018, \mn@doi [\aap] {10.1051/0004-6361/201732516}, 616, A8

\bibitem[\protect\citeauthoryear{{Arnold}, {Goodwin}, {Griffiths}  \&
  {Parker}}{{Arnold} et~al.}{2017}]{RN210}
{Arnold} B.,  {Goodwin} S.~P.,  {Griffiths} D.~W.,   {Parker} R.~J.,  2017,
  \mn@doi [\mnras] {10.1093/mnras/stx1719}, 471, 2498

\bibitem[\protect\citeauthoryear{Bailer-Jones, Rybizki, Fouesneau, Mantelet  \&
  Andrae}{Bailer-Jones et~al.}{2018}]{RN305}
Bailer-Jones C. A.~L.,  Rybizki J.,  Fouesneau M.,  Mantelet G.,   Andrae R.,
  2018, \mn@doi [\aj] {10.3847/1538-3881/aacb21}, 156, 58

\bibitem[\protect\citeauthoryear{{Bailer-Jones}, {Rybizki}, {Fouesneau},
  {Demleitner}  \& {Andrae}}{{Bailer-Jones} et~al.}{2021}]{2021AJ....161..147B}
{Bailer-Jones} C.~A.~L.,  {Rybizki} J.,  {Fouesneau} M.,  {Demleitner} M.,
  {Andrae} R.,  2021, \mn@doi [\aj] {10.3847/1538-3881/abd806}, 161, 147

\bibitem[\protect\citeauthoryear{{Baratella} et~al.,}{{Baratella}
  et~al.}{2020}]{2020A&A...634A..34B}
{Baratella} M.,  et~al., 2020, \mn@doi [\aap] {10.1051/0004-6361/201937055},
  634, A34

\bibitem[\protect\citeauthoryear{{Baxter}, {Covey}, {Muench},
  {F{\H{u}}r{\'e}sz}, {Rebull}  \& {Szentgyorgyi}}{{Baxter}
  et~al.}{2009}]{2009AJ....138..963B}
{Baxter} E.~J.,  {Covey} K.~R.,  {Muench} A.~A.,  {F{\H{u}}r{\'e}sz} G.,
  {Rebull} L.,   {Szentgyorgyi} A.~H.,  2009, \mn@doi [\aj]
  {10.1088/0004-6256/138/3/963}, 138, 963

\bibitem[\protect\citeauthoryear{{Bestenlehner} et~al.,}{{Bestenlehner}
  et~al.}{2011}]{2011A&A...530L..14B}
{Bestenlehner} J.~M.,  et~al., 2011, \mn@doi [\aap]
  {10.1051/0004-6361/201117043}, 530, L14

\bibitem[\protect\citeauthoryear{{Bischoff} et~al.,}{{Bischoff}
  et~al.}{2020}]{2020AN....341..908B}
{Bischoff} R.,  et~al., 2020, \mn@doi [Astronomische Nachrichten]
  {10.1002/asna.202013793}, 341, 908

\bibitem[\protect\citeauthoryear{{Bisht}, {Zhu}, {Yadav}, {Ganesh}, {Rangwal},
  {Durgapal}, {Sariya}  \& {Jiang}}{{Bisht} et~al.}{2021}]{2021MNRAS.tmp..697B}
{Bisht} D.,  {Zhu} Q.,  {Yadav} R.~K.~S.,  {Ganesh} S.,  {Rangwal} G.,
  {Durgapal} A.,  {Sariya} D.~P.,   {Jiang} I.-G.,  2021, \mn@doi [\mnras]
  {10.1093/mnras/stab691}

\bibitem[\protect\citeauthoryear{{Blaauw}}{{Blaauw}}{1956}]{RN255}
{Blaauw} A.,  1956, \mn@doi [\pasp] {10.1086/126983}, 68, 495

\bibitem[\protect\citeauthoryear{{Blaauw}}{{Blaauw}}{1961}]{RN67}
{Blaauw} A.,  1961, \bain, \href
  {https://ui.adsabs.harvard.edu/\#abs/1961BAN....15..265B} {15, 265}

\bibitem[\protect\citeauthoryear{{Blaauw} \& {Morgan}}{{Blaauw} \&
  {Morgan}}{1954}]{RN325}
{Blaauw} A.,  {Morgan} W.~W.,  1954, \mn@doi [\apj] {10.1086/145866}, 119, 625

\bibitem[\protect\citeauthoryear{{Bonnell}, {Smith}, {Davies}  \&
  {Horne}}{{Bonnell} et~al.}{2001}]{RN271}
{Bonnell} I.~A.,  {Smith} K.~W.,  {Davies} M.~B.,   {Horne} K.,  2001, \mn@doi
  [\mnras] {10.1046/j.1365-8711.2001.04171.x}, 322, 859

\bibitem[\protect\citeauthoryear{Boubert, Guillochon, Hawkins, Ginsburg, Evans
  \& Strader}{Boubert et~al.}{2018}]{RN316}
Boubert D.,  Guillochon J.,  Hawkins K.,  Ginsburg I.,  Evans N.~W.,   Strader
  J.,  2018, \mn@doi [\mnras] {10.1093/mnras/sty1601}, 479, 2789

\bibitem[\protect\citeauthoryear{Bressan, Marigo, Girardi, Salasnich, Dal~Cero,
  Rubele  \& Nanni}{Bressan et~al.}{2012}]{RN225}
Bressan A.,  Marigo P.,  Girardi L.,  Salasnich B.,  Dal~Cero C.,  Rubele S.,
  Nanni A.,  2012, \mn@doi [\mnras] {10.1111/j.1365-2966.2012.21948.x}, 427,
  127

\bibitem[\protect\citeauthoryear{{Bressert} et~al.,}{{Bressert}
  et~al.}{2010}]{RN59}
{Bressert} E.,  et~al., 2010, \mn@doi [\mnras]
  {10.1111/j.1745-3933.2010.00946.x}, 409, L54

\bibitem[\protect\citeauthoryear{Bromley, Kenyon, Brown  \& Geller}{Bromley
  et~al.}{2018}]{RN320}
Bromley B.~C.,  Kenyon S.~J.,  Brown W.~R.,   Geller M.~J.,  2018, \mn@doi
  [\apj] {10.3847/1538-4357/aae83e}, 868, 25

\bibitem[\protect\citeauthoryear{{Brown}, {Lattanzi}, {Kenyon}  \&
  {Geller}}{{Brown} et~al.}{2018}]{2018ApJ...866...39B}
{Brown} W.~R.,  {Lattanzi} M.~G.,  {Kenyon} S.~J.,   {Geller} M.~J.,  2018,
  \mn@doi [\apj] {10.3847/1538-4357/aadb8e}, 866, 39

\bibitem[\protect\citeauthoryear{{Buckner} et~al.,}{{Buckner}
  et~al.}{2020}]{2020A&A...636A..80B}
{Buckner} A. S.~M.,  et~al., 2020, \mn@doi [\aap]
  {10.1051/0004-6361/201936935}, 636, A80

\bibitem[\protect\citeauthoryear{{Busso} et~al.,}{{Busso}
  et~al.}{2018}]{2018gDR2.reptE...5B}
{Busso} G.,  et~al., 2018, Technical report, {Gaia DR2 documentation Chapter 5:
  Photometry}

\bibitem[\protect\citeauthoryear{{Cannon} \& {Pickering}}{{Cannon} \&
  {Pickering}}{1993}]{1993yCat.3135....0C}
{Cannon} A.~J.,  {Pickering} E.~C.,  1993, VizieR Online Data Catalog, \href
  {https://ui.adsabs.harvard.edu/abs/1993yCat.3135....0C} {p. III/135A}

\bibitem[\protect\citeauthoryear{{Cantat-Gaudin} \& {Anders}}{{Cantat-Gaudin}
  \& {Anders}}{2020}]{2020A&A...633A..99C}
{Cantat-Gaudin} T.,  {Anders} F.,  2020, \mn@doi [\aap]
  {10.1051/0004-6361/201936691}, 633, A99

\bibitem[\protect\citeauthoryear{{Cantat-Gaudin} et~al.,}{{Cantat-Gaudin}
  et~al.}{2018}]{2018A&A...618A..93C}
{Cantat-Gaudin} T.,  et~al., 2018, \mn@doi [\aap]
  {10.1051/0004-6361/201833476}, 618, A93

\bibitem[\protect\citeauthoryear{{Castelaz} \& {Grasdalen}}{{Castelaz} \&
  {Grasdalen}}{1988}]{1988ApJ...335..150C}
{Castelaz} M.~W.,  {Grasdalen} G.,  1988, \mn@doi [\apj] {10.1086/166915}, 335,
  150

\bibitem[\protect\citeauthoryear{{Chabrier}}{{Chabrier}}{2001}]{2001ApJ...554.1274C}
{Chabrier} G.,  2001, \mn@doi [\apj] {10.1086/321401}, 554, 1274

\bibitem[\protect\citeauthoryear{{Chabrier}}{{Chabrier}}{2005}]{RN200}
{Chabrier} G.,  2005, in {Corbelli} E.,  {Palla} F.,   {Zinnecker} H.,  eds,
  Astrophysics and Space Science Library Vol. 327, The Initial Mass Function 50
  Years Later. p.~41, \mn@doi{10.1007/978-1-4020-3407-7_5}

\bibitem[\protect\citeauthoryear{{Cieza} \& {Baliber}}{{Cieza} \&
  {Baliber}}{2007}]{2007ApJ...671..605C}
{Cieza} L.,  {Baliber} N.,  2007, \mn@doi [\apj] {10.1086/522080}, 671, 605

\bibitem[\protect\citeauthoryear{{Costado} \& {Alfaro}}{{Costado} \&
  {Alfaro}}{2018}]{2018MNRAS.476.3160C}
{Costado} M.~T.,  {Alfaro} E.~J.,  2018, \mn@doi [\mnras]
  {10.1093/mnras/sty447}, 476, 3160

\bibitem[\protect\citeauthoryear{{Cruzal{\`e}bes} et~al.,}{{Cruzal{\`e}bes}
  et~al.}{2019}]{2019MNRAS.490.3158C}
{Cruzal{\`e}bes} P.,  et~al., 2019, \mn@doi [\mnras] {10.1093/mnras/stz2803},
  490, 3158

\bibitem[\protect\citeauthoryear{{Dahm}}{{Dahm}}{2008}]{2008hsf1.book..966D}
{Dahm} S.~E.,  2008, in {Reipurth} B.,  ed., ~ Vol. 4, Handbook of Star Forming
  Regions, Volume I. p.~966

\bibitem[\protect\citeauthoryear{{Dahm}, {Simon}, {Proszkow}  \&
  {Patten}}{{Dahm} et~al.}{2007}]{2007AJ....134..999D}
{Dahm} S.~E.,  {Simon} T.,  {Proszkow} E.~M.,   {Patten} B.~M.,  2007, \mn@doi
  [\aj] {10.1086/519954}, 134, 999

\bibitem[\protect\citeauthoryear{{Dias}, {Alessi}, {Moitinho}  \&
  {L{\'e}pine}}{{Dias} et~al.}{2002}]{2002A&A...389..871D}
{Dias} W.~S.,  {Alessi} B.~S.,  {Moitinho} A.,   {L{\'e}pine} J.~R.~D.,  2002,
  \mn@doi [\aap] {10.1051/0004-6361:20020668}, 389, 871

\bibitem[\protect\citeauthoryear{{Dias}, {Alessi}, {Moitinho}  \&
  {Lepine}}{{Dias} et~al.}{2012}]{2012yCat....102022D}
{Dias} W.~S.,  {Alessi} B.~S.,  {Moitinho} A.,   {Lepine} J.~R.~D.,  2012,
  VizieR Online Data Catalog, \href
  {https://ui.adsabs.harvard.edu/abs/2012yCat....102022D} {p. B/ocl}

\bibitem[\protect\citeauthoryear{{Drew}, {Herrero}, {Mohr-Smith},
  {Mongui{\'o}}, {Wright}, {Kupfer}  \& {Napiwotzki}}{{Drew}
  et~al.}{2018}]{RN293}
{Drew} J.~E.,  {Herrero} A.,  {Mohr-Smith} M.,  {Mongui{\'o}} M.,  {Wright}
  N.~J.,  {Kupfer} T.,   {Napiwotzki} R.,  2018, \mn@doi [\mnras]
  {10.1093/mnras/sty1905}, 480, 2109

\bibitem[\protect\citeauthoryear{{Drew}, {Mongui{\'o}}  \& {Wright}}{{Drew}
  et~al.}{2021}]{2021MNRAS.508.4952D}
{Drew} J.~E.,  {Mongui{\'o}} M.,   {Wright} N.~J.,  2021, \mn@doi [\mnras]
  {10.1093/mnras/stab2905}, 508, 4952

\bibitem[\protect\citeauthoryear{{Duch{\^e}ne} \& {Kraus}}{{Duch{\^e}ne} \&
  {Kraus}}{2013}]{2013ARA&A..51..269D}
{Duch{\^e}ne} G.,  {Kraus} A.,  2013, \mn@doi [\araa]
  {10.1146/annurev-astro-081710-102602}, 51, 269

\bibitem[\protect\citeauthoryear{{Duch{\^e}ne}, {Lacour}, {Moraux}, {Goodwin}
  \& {Bouvier}}{{Duch{\^e}ne} et~al.}{2018}]{RN257}
{Duch{\^e}ne} G.,  {Lacour} S.,  {Moraux} E.,  {Goodwin} S.,   {Bouvier} J.,
  2018, \mn@doi [\mnras] {10.1093/mnras/sty1180}, 478, 1825

\bibitem[\protect\citeauthoryear{{Duflot}, {Figon}  \& {Meyssonnier}}{{Duflot}
  et~al.}{1995}]{1995A&AS..114..269D}
{Duflot} M.,  {Figon} P.,   {Meyssonnier} N.,  1995, \aaps, \href
  {https://ui.adsabs.harvard.edu/abs/1995A&AS..114..269D} {114, 269}

\bibitem[\protect\citeauthoryear{{Eldridge}, {Langer}  \& {Tout}}{{Eldridge}
  et~al.}{2011}]{RN137}
{Eldridge} J.~J.,  {Langer} N.,   {Tout} C.~A.,  2011, \mn@doi [\mnras]
  {10.1111/j.1365-2966.2011.18650.x}, 414, 3501

\bibitem[\protect\citeauthoryear{{Evans} et~al.,}{{Evans}
  et~al.}{2018}]{2018A&A...616A...4E}
{Evans} D.~W.,  et~al., 2018, \mn@doi [\aap] {10.1051/0004-6361/201832756},
  616, A4

\bibitem[\protect\citeauthoryear{{F{\H{u}}r{\'e}sz} et~al.,}{{F{\H{u}}r{\'e}sz}
  et~al.}{2006}]{2006ApJ...648.1090F}
{F{\H{u}}r{\'e}sz} G.,  et~al., 2006, \mn@doi [\apj] {10.1086/506140}, 648,
  1090

\bibitem[\protect\citeauthoryear{Farias, Tan  \& Chatterjee}{Farias
  et~al.}{2019}]{RN312}
Farias J.~P.,  Tan J.~C.,   Chatterjee S.,  2019, \mn@doi [\mnras]
  {10.1093/mnras/sty3470}, 483, 4999

\bibitem[\protect\citeauthoryear{{Farias}, {Tan}  \& {Eyer}}{{Farias}
  et~al.}{2020}]{2020ApJ...900...14F}
{Farias} J.~P.,  {Tan} J.~C.,   {Eyer} L.,  2020, \mn@doi [\apj]
  {10.3847/1538-4357/aba699}, 900, 14

\bibitem[\protect\citeauthoryear{{Fehrenbach}, {Burnage}  \&
  {Figuiere}}{{Fehrenbach} et~al.}{1992}]{1992A&AS...95..541F}
{Fehrenbach} C.,  {Burnage} R.,   {Figuiere} J.,  1992, \aaps, \href
  {https://ui.adsabs.harvard.edu/abs/1992A&AS...95..541F} {95, 541}

\bibitem[\protect\citeauthoryear{{\VAN{Fuente Marcos}{De la}{de la} Fuente
  Marcos} \& {De La Fuente Marcos}}{{\VAN{Fuente Marcos}{De la}{de la} Fuente
  Marcos} \& {De La Fuente Marcos}}{2009}]{RN287}
{\VAN{Fuente Marcos}{De la}{de la} Fuente Marcos} R.,  {De La Fuente Marcos}
  C.,  2009, \mn@doi [\aap] {10.1051/0004-6361/200912297}, 500, L13

\bibitem[\protect\citeauthoryear{{\VAN{Fuente Marcos}{de la}{de la} Fuente
  Marcos} \& {de la Fuente Marcos}}{{\VAN{Fuente Marcos}{de la}{de la} Fuente
  Marcos} \& {de la Fuente Marcos}}{2018}]{RN252}
{\VAN{Fuente Marcos}{de la}{de la} Fuente Marcos} R.,  {de la Fuente Marcos}
  C.,  2018, \mn@doi [Research Notes of the American Astronomical Society]
  {10.3847/2515-5172/aac5d7}, 2, 35

\bibitem[\protect\citeauthoryear{{Gaia Collaboration} et~al.,}{{Gaia
  Collaboration} et~al.}{2016}]{RN308}
{Gaia Collaboration} et~al., 2016, \mn@doi [\aap]
  {10.1051/0004-6361/201629272}, 595, A1

\bibitem[\protect\citeauthoryear{{Gaia Collaboration} et~al.,}{{Gaia
  Collaboration} et~al.}{2018a}]{RN238}
{Gaia Collaboration} et~al., 2018a, \mn@doi [\aap]
  {10.1051/0004-6361/201833051}, 616, A1

\bibitem[\protect\citeauthoryear{{Gaia Collaboration} et~al.,}{{Gaia
  Collaboration} et~al.}{2018b}]{2018A&A...616A..10G}
{Gaia Collaboration} et~al., 2018b, \mn@doi [\aap]
  {10.1051/0004-6361/201832843}, 616, A10

\bibitem[\protect\citeauthoryear{{Gaia Collaboration} et~al.,}{{Gaia
  Collaboration} et~al.}{2021}]{2020arXiv201201533G}
{Gaia Collaboration} et~al., 2021, \mn@doi [\aap]
  {10.1051/0004-6361/202039657}, 649, A1

\bibitem[\protect\citeauthoryear{{Getman} et~al.,}{{Getman}
  et~al.}{2014}]{2014ApJ...787..108G}
{Getman} K.~V.,  et~al., 2014, \mn@doi [\apj] {10.1088/0004-637X/787/2/108},
  787, 108

\bibitem[\protect\citeauthoryear{{Getman}, {Feigelson}, {Kuhn}  \&
  {Garmire}}{{Getman} et~al.}{2019}]{2019MNRAS.487.2977G}
{Getman} K.~V.,  {Feigelson} E.~D.,  {Kuhn} M.~A.,   {Garmire} G.~P.,  2019,
  \mn@doi [\mnras] {10.1093/mnras/stz1457}, 487, 2977

\bibitem[\protect\citeauthoryear{{Gontcharov}}{{Gontcharov}}{2012}]{2012AstL...38..694G}
{Gontcharov} G.~A.,  2012, \mn@doi [Astronomy Letters]
  {10.1134/S1063773712110035}, 38, 694

\bibitem[\protect\citeauthoryear{{Goodwin} \& {Whitworth}}{{Goodwin} \&
  {Whitworth}}{2004}]{RN14}
{Goodwin} S.~P.,  {Whitworth} A.~P.,  2004, \mn@doi [\aap]
  {10.1051/0004-6361:20031529}, 413, 929

\bibitem[\protect\citeauthoryear{{Hambly} et~al.,}{{Hambly}
  et~al.}{2018}]{2018gDR2.reptE..14H}
{Hambly} N.,  et~al., 2018, Technical report, {Gaia DR2 documentation Chapter
  14: Datamodel description}

\bibitem[\protect\citeauthoryear{{Heiter}, {Soubiran}, {Netopil}  \&
  {Paunzen}}{{Heiter} et~al.}{2014}]{2014A&A...561A..93H}
{Heiter} U.,  {Soubiran} C.,  {Netopil} M.,   {Paunzen} E.,  2014, \mn@doi
  [\aap] {10.1051/0004-6361/201322559}, 561, A93

\bibitem[\protect\citeauthoryear{{Herbig}}{{Herbig}}{1954}]{1954ApJ...119..483H}
{Herbig} G.~H.,  1954, \mn@doi [\apj] {10.1086/145854}, 119, 483

\bibitem[\protect\citeauthoryear{{Hetem} \& {Gregorio-Hetem}}{{Hetem} \&
  {Gregorio-Hetem}}{2019}]{2019MNRAS.490.2521H}
{Hetem} A.,  {Gregorio-Hetem} J.,  2019, \mn@doi [\mnras]
  {10.1093/mnras/stz2698}, 490, 2521

\bibitem[\protect\citeauthoryear{{Hillenbrand}, {Strom}, {Calvet}, {Merrill},
  {Gatley}, {Makidon}, {Meyer}  \& {Skrutskie}}{{Hillenbrand}
  et~al.}{1998}]{1998AJ....116.1816H}
{Hillenbrand} L.~A.,  {Strom} S.~E.,  {Calvet} N.,  {Merrill} K.~M.,  {Gatley}
  I.,  {Makidon} R.~B.,  {Meyer} M.~R.,   {Skrutskie} M.~F.,  1998, \mn@doi
  [\aj] {10.1086/300536}, 116, 1816

\bibitem[\protect\citeauthoryear{{Hoogerwerf}, {de Bruijne}  \& {de
  Zeeuw}}{{Hoogerwerf} et~al.}{2001}]{RN50}
{Hoogerwerf} R.,  {de Bruijne} J.~H.~J.,   {de Zeeuw} P.~T.,  2001, \mn@doi
  [\aap] {10.1051/0004-6361:20000014}, 365, 49

\bibitem[\protect\citeauthoryear{Irrgang, Kreuzer  \& Heber}{Irrgang
  et~al.}{2018}]{RN315}
Irrgang A.,  Kreuzer S.,   Heber U.,  2018, \mn@doi [\aap]
  {10.1051/0004-6361/201833874}, 620, A48

\bibitem[\protect\citeauthoryear{{Jackson} et~al.,}{{Jackson}
  et~al.}{2016}]{2016A&A...586A..52J}
{Jackson} R.~J.,  et~al., 2016, \mn@doi [\aap] {10.1051/0004-6361/201527507},
  586, A52

\bibitem[\protect\citeauthoryear{{Jackson} et~al.,}{{Jackson}
  et~al.}{2020}]{2020MNRAS.496.4701J}
{Jackson} R.~J.,  et~al., 2020, \mn@doi [\mnras] {10.1093/mnras/staa1749}, 496,
  4701

\bibitem[\protect\citeauthoryear{{Jaehnig}, {Da Rio}  \& {Tan}}{{Jaehnig}
  et~al.}{2015}]{RN248}
{Jaehnig} K.~O.,  {Da Rio} N.,   {Tan} J.~C.,  2015, \mn@doi [\apj]
  {10.1088/0004-637X/798/2/126}, 798, 126

\bibitem[\protect\citeauthoryear{{Janson}, {Jayawardhana}, {Girard},
  {Lafreni{\`e}re}, {Bonavita}, {Gizis}  \& {Brandeker}}{{Janson}
  et~al.}{2012}]{2012ApJ...758L...2J}
{Janson} M.,  {Jayawardhana} R.,  {Girard} J.~H.,  {Lafreni{\`e}re} D.,
  {Bonavita} M.,  {Gizis} J.,   {Brandeker} A.,  2012, \mn@doi [\apjl]
  {10.1088/2041-8205/758/1/L2}, 758, L2

\bibitem[\protect\citeauthoryear{{Karlsson}}{{Karlsson}}{1972}]{1972A&AS....7...35K}
{Karlsson} B.,  1972, \aaps, \href
  {https://ui.adsabs.harvard.edu/abs/1972A&AS....7...35K} {7, 35}

\bibitem[\protect\citeauthoryear{{Kharchenko}, {Piskunov}, {R{\"o}ser},
  {Schilbach}  \& {Scholz}}{{Kharchenko} et~al.}{2005}]{2005A&A...438.1163K}
{Kharchenko} N.~V.,  {Piskunov} A.~E.,  {R{\"o}ser} S.,  {Schilbach} E.,
  {Scholz} R.~D.,  2005, \mn@doi [\aap] {10.1051/0004-6361:20042523}, 438, 1163

\bibitem[\protect\citeauthoryear{Kim, Lu, Konopacky, Chu, Toller, Anderson,
  Theissen  \& Morris}{Kim et~al.}{2019}]{RN322}
Kim D.,  Lu J.~R.,  Konopacky Q.,  Chu L.,  Toller E.,  Anderson J.,  Theissen
  C.~A.,   Morris M.~R.,  2019, \mn@doi [\aj] {10.3847/1538-3881/aafb09}, 157,
  109

\bibitem[\protect\citeauthoryear{{King}, {Soderblom}, {Fischer}  \&
  {Jones}}{{King} et~al.}{2000}]{2000ApJ...533..944K}
{King} J.~R.,  {Soderblom} D.~R.,  {Fischer} D.,   {Jones} B.~F.,  2000,
  \mn@doi [\apj] {10.1086/308695}, 533, 944

\bibitem[\protect\citeauthoryear{{Klagyivik} et~al.,}{{Klagyivik}
  et~al.}{2013}]{2013ApJ...773...54K}
{Klagyivik} P.,  et~al., 2013, \mn@doi [\apj] {10.1088/0004-637X/773/1/54},
  773, 54

\bibitem[\protect\citeauthoryear{{K{\"o}hler}, {Petr-Gotzens}, {McCaughrean},
  {Bouvier}, {Duch{\^e}ne}, {Quirrenbach}  \& {Zinnecker}}{{K{\"o}hler}
  et~al.}{2006}]{2006A&A...458..461K}
{K{\"o}hler} R.,  {Petr-Gotzens} M.~G.,  {McCaughrean} M.~J.,  {Bouvier} J.,
  {Duch{\^e}ne} G.,  {Quirrenbach} A.,   {Zinnecker} H.,  2006, \mn@doi [\aap]
  {10.1051/0004-6361:20054561}, 458, 461

\bibitem[\protect\citeauthoryear{{Kounkel}, {Hartmann}, {Tobin}, {Mateo},
  {Bailey}  \& {Spencer}}{{Kounkel} et~al.}{2016}]{2016ApJ...821....8K}
{Kounkel} M.,  {Hartmann} L.,  {Tobin} J.~J.,  {Mateo} M.,  {Bailey} John~I.
  I.,   {Spencer} M.,  2016, \mn@doi [\apj] {10.3847/0004-637X/821/1/8}, 821, 8

\bibitem[\protect\citeauthoryear{{Kounkel} et~al.,}{{Kounkel}
  et~al.}{2019}]{2019AJ....157..196K}
{Kounkel} M.,  et~al., 2019, \mn@doi [\aj] {10.3847/1538-3881/ab13b1}, 157, 196

\bibitem[\protect\citeauthoryear{{Kouwenhoven}, {Brown}, {Portegies Zwart}  \&
  {Kaper}}{{Kouwenhoven} et~al.}{2007}]{2007AA...474...77K}
{Kouwenhoven} M.~B.~N.,  {Brown} A.~G.~A.,  {Portegies Zwart} S.~F.,   {Kaper}
  L.,  2007, \mn@doi [\aap] {10.1051/0004-6361:20077719}, 474, 77

\bibitem[\protect\citeauthoryear{{Kroupa}}{{Kroupa}}{1995}]{RN261}
{Kroupa} P.,  1995, \mn@doi [\mnras] {10.1093/mnras/277.4.1491}, 277, 1491

\bibitem[\protect\citeauthoryear{{Kroupa} \& {Petr-Gotzens}}{{Kroupa} \&
  {Petr-Gotzens}}{2011}]{2011A&A...529A..92K}
{Kroupa} P.,  {Petr-Gotzens} M.~G.,  2011, \mn@doi [\aap]
  {10.1051/0004-6361/201015989}, 529, A92

\bibitem[\protect\citeauthoryear{Kuhn, Hillenbrand, Sills, Feigelson  \&
  Getman}{Kuhn et~al.}{2019}]{RN264}
Kuhn M.~A.,  Hillenbrand L.~A.,  Sills A.,  Feigelson E.~D.,   Getman K.~V.,
  2019, \mn@doi [\apj] {10.3847/1538-4357/aaef8c}, 870, 32

\bibitem[\protect\citeauthoryear{{Lada} \& {Lada}}{{Lada} \&
  {Lada}}{2003}]{RN25}
{Lada} C.~J.,  {Lada} E.~A.,  2003, \mn@doi [\araa]
  {10.1146/annurev.astro.41.011802.094844}, 41, 57

\bibitem[\protect\citeauthoryear{{Lada}, {Young}  \& {Greene}}{{Lada}
  et~al.}{1993}]{1993ApJ...408..471L}
{Lada} C.~J.,  {Young} E.~T.,   {Greene} T.~P.,  1993, \mn@doi [\apj]
  {10.1086/172605}, 408, 471

\bibitem[\protect\citeauthoryear{{Lamm}, {Bailer-Jones}, {Mundt}, {Herbst}  \&
  {Scholz}}{{Lamm} et~al.}{2004}]{2004A&A...417..557L}
{Lamm} M.~H.,  {Bailer-Jones} C.~A.~L.,  {Mundt} R.,  {Herbst} W.,   {Scholz}
  A.,  2004, \mn@doi [\aap] {10.1051/0004-6361:20035588}, 417, 557

\bibitem[\protect\citeauthoryear{{Larson}}{{Larson}}{1981}]{RN27}
{Larson} R.~B.,  1981, \mn@doi [\mnras] {10.1093/mnras/194.4.809}, 194, 809

\bibitem[\protect\citeauthoryear{Lennon et~al.,}{Lennon et~al.}{2018}]{RN313}
Lennon D.~J.,  et~al., 2018, \mn@doi [\aap] {10.1051/0004-6361/201833465}, 619,
  A78

\bibitem[\protect\citeauthoryear{{Li} et~al.,}{{Li}
  et~al.}{2021}]{2021ApJS..252....3L}
{Li} Y.-B.,  et~al., 2021, \mn@doi [\apjs] {10.3847/1538-4365/abc16e}, 252, 3

\bibitem[\protect\citeauthoryear{Lindegren}{Lindegren}{2018}]{LL:LL-124}
Lindegren L.,  2018, {R}e-normalising the astrometric chi-square in {G}aia
  {D}{R}2, GAIA-C3-TN-LU-LL-124, \url
  {http://www.rssd.esa.int/doc_fetch.php?id=3757412}

\bibitem[\protect\citeauthoryear{Lindegren et~al.,}{Lindegren
  et~al.}{2018}]{RN307}
Lindegren L.,  et~al., 2018, \mn@doi [\aap] {10.1051/0004-6361/201832727}, 616,
  A2

\bibitem[\protect\citeauthoryear{{Lucy}}{{Lucy}}{2006}]{2006A&A...457..629L}
{Lucy} L.~B.,  2006, \mn@doi [\aap] {10.1051/0004-6361:20065746}, 457, 629

\bibitem[\protect\citeauthoryear{{Luo}, {Zhao}, {Zhao}  \& {et al.}}{{Luo}
  et~al.}{2019}]{2019yCat.5164....0L}
{Luo} A.~L.,  {Zhao} Y.~H.,  {Zhao} G.,   {et al.} 2019, VizieR Online Data
  Catalog, \href {https://ui.adsabs.harvard.edu/abs/2019yCat.5164....0L} {p.
  V/164}

\bibitem[\protect\citeauthoryear{{Ma{\'\i}z Apell{\'a}niz}}{{Ma{\'\i}z
  Apell{\'a}niz}}{2019}]{2019A&A...630A.119M}
{Ma{\'\i}z Apell{\'a}niz} J.,  2019, \mn@doi [\aap]
  {10.1051/0004-6361/201935885}, 630, A119

\bibitem[\protect\citeauthoryear{{Ma{\'\i}z Apell{\'a}niz} \&
  {Weiler}}{{Ma{\'\i}z Apell{\'a}niz} \& {Weiler}}{2018}]{2018A&A...619A.180M}
{Ma{\'\i}z Apell{\'a}niz} J.,  {Weiler} M.,  2018, \mn@doi [\aap]
  {10.1051/0004-6361/201834051}, 619, A180

\bibitem[\protect\citeauthoryear{{Ma{\'\i}z Apell{\'a}niz}, {Crespo Bellido},
  {Barb{\'a}}, {Fern{\'a}ndez Aranda}  \& {Sota}}{{Ma{\'\i}z Apell{\'a}niz}
  et~al.}{2020}]{2020A&A...643A.138M}
{Ma{\'\i}z Apell{\'a}niz} J.,  {Crespo Bellido} P.,  {Barb{\'a}} R.~H.,
  {Fern{\'a}ndez Aranda} R.,   {Sota} A.,  2020, \mn@doi [\aap]
  {10.1051/0004-6361/202038228}, 643, A138

\bibitem[\protect\citeauthoryear{Marchetti, Rossi  \& Brown}{Marchetti
  et~al.}{2019}]{RN319}
Marchetti T.,  Rossi E.~M.,   Brown A. G.~A.,  2019, \mn@doi [\mnras]
  {10.1093/mnras/sty2592}, 490, 157

\bibitem[\protect\citeauthoryear{{Margulis}, {Lada}  \& {Young}}{{Margulis}
  et~al.}{1989}]{1989ApJ...345..906M}
{Margulis} M.,  {Lada} C.~J.,   {Young} E.~T.,  1989, \mn@doi [\apj]
  {10.1086/167960}, 345, 906

\bibitem[\protect\citeauthoryear{{Mari{\~n}as}, {Lada}, {Teixeira}  \&
  {Lada}}{{Mari{\~n}as} et~al.}{2013}]{2013ApJ...772...81M}
{Mari{\~n}as} N.,  {Lada} E.~A.,  {Teixeira} P.~S.,   {Lada} C.~J.,  2013,
  \mn@doi [\apj] {10.1088/0004-637X/772/2/81}, 772, 81

\bibitem[\protect\citeauthoryear{{Marks} \& {Kroupa}}{{Marks} \&
  {Kroupa}}{2012}]{RN277}
{Marks} M.,  {Kroupa} P.,  2012, \mn@doi [\aap] {10.1051/0004-6361/201118231},
  543, A8

\bibitem[\protect\citeauthoryear{{Maschberger}}{{Maschberger}}{2013}]{RN203}
{Maschberger} T.,  2013, \mn@doi [\mnras] {10.1093/mnras/sts479}, 429, 1725

\bibitem[\protect\citeauthoryear{{Mason}, {Gies}, {Hartkopf}, {Bagnuolo}, {ten
  Brummelaar}  \& {McAlister}}{{Mason} et~al.}{1998}]{1998AJ....115..821M}
{Mason} B.~D.,  {Gies} D.~R.,  {Hartkopf} W.~I.,  {Bagnuolo} William~G. J.,
  {ten Brummelaar} T.,   {McAlister} H.~A.,  1998, \mn@doi [\aj]
  {10.1086/300234}, 115, 821

\bibitem[\protect\citeauthoryear{{Mayne} \& {Naylor}}{{Mayne} \&
  {Naylor}}{2008}]{2008MNRAS.386..261M}
{Mayne} N.~J.,  {Naylor} T.,  2008, \mn@doi [\mnras]
  {10.1111/j.1365-2966.2008.13025.x}, 386, 261

\bibitem[\protect\citeauthoryear{{Mayor}, {Duquennoy}, {Halbwachs}  \&
  {Mermilliod}}{{Mayor} et~al.}{1992}]{1992ASPC...32...73M}
{Mayor} M.,  {Duquennoy} A.,  {Halbwachs} J.~L.,   {Mermilliod} J.~C.,  1992,
  in {McAlister} H.~A.,  {Hartkopf} W.~I.,  eds,  Astronomical Society of the
  Pacific Conference Series Vol. 32, IAU Colloq. 135: Complementary Approaches
  to Double and Multiple Star Research. p.~73

\bibitem[\protect\citeauthoryear{{McBride} \& {Kounkel}}{{McBride} \&
  {Kounkel}}{2019}]{2019ApJ...884....6M}
{McBride} A.,  {Kounkel} M.,  2019, \mn@doi [\apj] {10.3847/1538-4357/ab3df9},
  884, 6

\bibitem[\protect\citeauthoryear{{McCuskey}}{{McCuskey}}{1959}]{1959ApJS....4...23M}
{McCuskey} S.~W.,  1959, \mn@doi [\apjs] {10.1086/190043}, \href
  {https://ui.adsabs.harvard.edu/abs/1959ApJS....4...23M} {4, 23}

\bibitem[\protect\citeauthoryear{{McGinnis}, {Dougados}, {Alencar}, {Bouvier}
  \& {Cabrit}}{{McGinnis} et~al.}{2018}]{2018A&A...620A..87M}
{McGinnis} P.,  {Dougados} C.,  {Alencar} S.~H.~P.,  {Bouvier} J.,   {Cabrit}
  S.,  2018, \mn@doi [\aap] {10.1051/0004-6361/201731629}, 620, A87

\bibitem[\protect\citeauthoryear{{McMillan}, {Vesperini}  \& {Portegies
  Zwart}}{{McMillan} et~al.}{2007}]{2007ApJ...655L..45M}
{McMillan} S. L.~W.,  {Vesperini} E.,   {Portegies Zwart} S.~F.,  2007, \mn@doi
  [\apjl] {10.1086/511763}, 655, L45

\bibitem[\protect\citeauthoryear{{\VAN{Mink}{De}{de} Mink}, {Sana}, {Langer},
  {Izzard}  \& {Schneider}}{{\VAN{Mink}{De}{de} Mink} et~al.}{2014}]{RN136}
{\VAN{Mink}{De}{de} Mink} S.~E.,  {Sana} H.,  {Langer} N.,  {Izzard} R.~G.,
  {Schneider} F.~R.~N.,  2014, \mn@doi [\apj] {10.1088/0004-637X/782/1/7}, 782,
  7

\bibitem[\protect\citeauthoryear{{Moeckel} \& {Bonnell}}{{Moeckel} \&
  {Bonnell}}{2009a}]{2009MNRAS.396.1864M}
{Moeckel} N.,  {Bonnell} I.~A.,  2009a, \mn@doi [\mnras]
  {10.1111/j.1365-2966.2009.14813.x}, 396, 1864

\bibitem[\protect\citeauthoryear{{Moeckel} \& {Bonnell}}{{Moeckel} \&
  {Bonnell}}{2009b}]{2009MNRAS.400..657M}
{Moeckel} N.,  {Bonnell} I.~A.,  2009b, \mn@doi [\mnras]
  {10.1111/j.1365-2966.2009.15499.x}, 400, 657

\bibitem[\protect\citeauthoryear{{Muench}, {Lada}, {Luhman}, {Muzerolle}  \&
  {Young}}{{Muench} et~al.}{2007}]{2007AJ....134..411M}
{Muench} A.~A.,  {Lada} C.~J.,  {Luhman} K.~L.,  {Muzerolle} J.,   {Young} E.,
  2007, \mn@doi [\aj] {10.1086/518560}, 134, 411

\bibitem[\protect\citeauthoryear{{Naylor}}{{Naylor}}{2009}]{2009MNRAS.399..432N}
{Naylor} T.,  2009, \mn@doi [\mnras] {10.1111/j.1365-2966.2009.15295.x}, 399,
  432

\bibitem[\protect\citeauthoryear{{Netopil}, {Paunzen}, {Heiter}  \&
  {Soubiran}}{{Netopil} et~al.}{2016}]{2016A&A...585A.150N}
{Netopil} M.,  {Paunzen} E.,  {Heiter} U.,   {Soubiran} C.,  2016, \mn@doi
  [\aap] {10.1051/0004-6361/201526370}, 585, A150

\bibitem[\protect\citeauthoryear{{Nicholson}, {Parker}, {Church}, {Davies},
  {Fearon}  \& {Walton}}{{Nicholson} et~al.}{2019}]{2019MNRAS.485.4893N}
{Nicholson} R.~B.,  {Parker} R.~J.,  {Church} R.~P.,  {Davies} M.~B.,  {Fearon}
  N.~M.,   {Walton} S. R.~J.,  2019, \mn@doi [\mnras] {10.1093/mnras/stz606},
  485, 4893

\bibitem[\protect\citeauthoryear{{Ochsenbein}, {Bauer}  \&
  {Marcout}}{{Ochsenbein} et~al.}{2000}]{2000A&AS..143...23O}
{Ochsenbein} F.,  {Bauer} P.,   {Marcout} J.,  2000, \mn@doi [\aaps]
  {10.1051/aas:2000169}, 143, 23

\bibitem[\protect\citeauthoryear{{Oh} \& {Kroupa}}{{Oh} \&
  {Kroupa}}{2016}]{RN241}
{Oh} S.,  {Kroupa} P.,  2016, \mn@doi [\aap] {10.1051/0004-6361/201628233},
  590, A107

\bibitem[\protect\citeauthoryear{{{\"O}pik}}{{{\"O}pik}}{1924}]{1924PTarO..25f...1O}
{{\"O}pik} E.,  1924, Publications of the Tartu Astrofizica Observatory, \href
  {https://ui.adsabs.harvard.edu/abs/1924PTarO..25f...1O} {25, 1}

\bibitem[\protect\citeauthoryear{{Park}, {Goodwin}  \& {Kim}}{{Park}
  et~al.}{2018}]{RN258}
{Park} S.-M.,  {Goodwin} S.~P.,   {Kim} S.~S.,  2018, \mn@doi [\mnras]
  {10.1093/mnras/sty1083}, 478, 183

\bibitem[\protect\citeauthoryear{{Parker}}{{Parker}}{2014}]{RN8}
{Parker} R.~J.,  2014, \mn@doi [\mnras] {10.1093/mnras/stu2054}, 445, 4037

\bibitem[\protect\citeauthoryear{{Parker} \& {Meyer}}{{Parker} \&
  {Meyer}}{2014}]{2014MNRAS.442.3722P}
{Parker} R.~J.,  {Meyer} M.~R.,  2014, \mn@doi [\mnras]
  {10.1093/mnras/stu1101}, 442, 3722

\bibitem[\protect\citeauthoryear{{Parker} \& {Quanz}}{{Parker} \&
  {Quanz}}{2012}]{RN269}
{Parker} R.~J.,  {Quanz} S.~P.,  2012, \mn@doi [\mnras]
  {10.1111/j.1365-2966.2011.19911.x}, 419, 2448

\bibitem[\protect\citeauthoryear{{Parker} \& {Reggiani}}{{Parker} \&
  {Reggiani}}{2013}]{2013MNRAS.432.2378P}
{Parker} R.~J.,  {Reggiani} M.~M.,  2013, \mn@doi [\mnras]
  {10.1093/mnras/stt600}, 432, 2378

\bibitem[\protect\citeauthoryear{{Parker} \& {Schoettler}}{{Parker} \&
  {Schoettler}}{2021}]{Parker+Schoettler2021}
{Parker} R.~J.,  {Schoettler} C.,  2021, \mnras

\bibitem[\protect\citeauthoryear{{Parker} \& {Wright}}{{Parker} \&
  {Wright}}{2016}]{RN1}
{Parker} R.~J.,  {Wright} N.~J.,  2016, \mn@doi [\mnras]
  {10.1093/mnras/stw087}, 457, 3430

\bibitem[\protect\citeauthoryear{{Parker}, {Wright}, {Goodwin}  \&
  {Meyer}}{{Parker} et~al.}{2014}]{RN5}
{Parker} R.~J.,  {Wright} N.~J.,  {Goodwin} S.~P.,   {Meyer} M.~R.,  2014,
  \mn@doi [\mnras] {10.1093/mnras/stt2231}, 438, 620

\bibitem[\protect\citeauthoryear{{Paunzen}, {Duffee}, {Heiter}, {Kuschnig}  \&
  {Weiss}}{{Paunzen} et~al.}{2001}]{2001A&A...373..625P}
{Paunzen} E.,  {Duffee} B.,  {Heiter} U.,  {Kuschnig} R.,   {Weiss} W.~W.,
  2001, \mn@doi [\aap] {10.1051/0004-6361:20010630}, 373, 625

\bibitem[\protect\citeauthoryear{{Peretto}, {Andr{\'e}}  \&
  {Belloche}}{{Peretto} et~al.}{2006}]{2006A&A...445..979P}
{Peretto} N.,  {Andr{\'e}} P.,   {Belloche} A.,  2006, \mn@doi [\aap]
  {10.1051/0004-6361:20053324}, 445, 979

\bibitem[\protect\citeauthoryear{{Peretto}, {Hennebelle}  \&
  {Andr{\'e}}}{{Peretto} et~al.}{2007}]{2007A&A...464..983P}
{Peretto} N.,  {Hennebelle} P.,   {Andr{\'e}} P.,  2007, \mn@doi [\aap]
  {10.1051/0004-6361:20065653}, 464, 983

\bibitem[\protect\citeauthoryear{{Peter}, {Feldt}, {Henning}  \&
  {Hormuth}}{{Peter} et~al.}{2012}]{2012A&A...538A..74P}
{Peter} D.,  {Feldt} M.,  {Henning} T.,   {Hormuth} F.,  2012, \mn@doi [\aap]
  {10.1051/0004-6361/201015027}, 538, A74

\bibitem[\protect\citeauthoryear{{Pietrzynski} \& {Udalski}}{{Pietrzynski} \&
  {Udalski}}{2000}]{RN288}
{Pietrzynski} G.,  {Udalski} A.,  2000, \actaa, \href
  {https://ui.adsabs.harvard.edu/\#abs/2000AcA....50..355P} {50, 355}

\bibitem[\protect\citeauthoryear{{Pinsonneault} \& {Stanek}}{{Pinsonneault} \&
  {Stanek}}{2006}]{2006ApJ...639L..67P}
{Pinsonneault} M.~H.,  {Stanek} K.~Z.,  2006, \mn@doi [\apjl] {10.1086/502799},
  639, L67

\bibitem[\protect\citeauthoryear{{Platais} et~al.,}{{Platais}
  et~al.}{2020}]{2020AJ....159..272P}
{Platais} I.,  et~al., 2020, \mn@doi [\aj] {10.3847/1538-3881/ab8d42}, 159, 272

\bibitem[\protect\citeauthoryear{{Portegies Zwart}, {Makino}, {McMillan}  \&
  {Hut}}{{Portegies Zwart} et~al.}{1999}]{RN236}
{Portegies Zwart} S.~F.,  {Makino} J.,  {McMillan} S.~L.~W.,   {Hut} P.,  1999,
  \aap, 348, 117

\bibitem[\protect\citeauthoryear{{Portegies Zwart}, {McMillan}, {Hut}  \&
  {Makino}}{{Portegies Zwart} et~al.}{2001}]{RN193}
{Portegies Zwart} S.~F.,  {McMillan} S. L.~W.,  {Hut} P.,   {Makino} J.,  2001,
  \mn@doi [\mnras] {10.1046/j.1365-8711.2001.03976.x}, 321, 199

\bibitem[\protect\citeauthoryear{{Poveda}, {Ruiz}  \& {Allen}}{{Poveda}
  et~al.}{1967}]{RN189}
{Poveda} A.,  {Ruiz} J.,   {Allen} C.,  1967, Boletin de los Observatorios
  Tonantzintla y Tacubaya, \href
  {https://ui.adsabs.harvard.edu/\#abs/1967BOTT....4...86P} {4, 86}

\bibitem[\protect\citeauthoryear{{Priyatikanto}, {Kouwenhoven}, {Arifyanto},
  {Wulandari}  \& {Siregar}}{{Priyatikanto} et~al.}{2016}]{RN290}
{Priyatikanto} R.,  {Kouwenhoven} M.~B.~N.,  {Arifyanto} M.~I.,  {Wulandari}
  H.~R.~T.,   {Siregar} S.,  2016, \mn@doi [\mnras] {10.1093/mnras/stw060},
  457, 1339

\bibitem[\protect\citeauthoryear{Raddi, Hollands, Gänsicke, Townsley, Hermes,
  Gentile~Fusillo  \& Koester}{Raddi et~al.}{2018}]{RN323}
Raddi R.,  Hollands M.~A.,  Gänsicke B.~T.,  Townsley D.~M.,  Hermes J.~J.,
  Gentile~Fusillo N.~P.,   Koester D.,  2018, \mn@doi [\mnras]
  {10.1093/mnrasl/sly103}, 479, L96

\bibitem[\protect\citeauthoryear{{Raghavan}, {Farrington}, {ten Brummelaar},
  {McAlister}, {Ridgway}, {Sturmann}, {Sturmann}  \& {Turner}}{{Raghavan}
  et~al.}{2012}]{2012ApJ...745...24R}
{Raghavan} D.,  {Farrington} C.~D.,  {ten Brummelaar} T.~A.,  {McAlister}
  H.~A.,  {Ridgway} S.~T.,  {Sturmann} L.,  {Sturmann} J.,   {Turner} N.~H.,
  2012, \mn@doi [\apj] {10.1088/0004-637X/745/1/24}, 745, 24

\bibitem[\protect\citeauthoryear{{Rapson}, {Pipher}, {Gutermuth}, {Megeath},
  {Allen}, {Myers}  \& {Allen}}{{Rapson} et~al.}{2014}]{2014ApJ...794..124R}
{Rapson} V.~A.,  {Pipher} J.~L.,  {Gutermuth} R.~A.,  {Megeath} S.~T.,  {Allen}
  T.~S.,  {Myers} P.~C.,   {Allen} L.~E.,  2014, \mn@doi [\apj]
  {10.1088/0004-637X/794/2/124}, 794, 124

\bibitem[\protect\citeauthoryear{{Rate} \& {Crowther}}{{Rate} \&
  {Crowther}}{2020}]{2020MNRAS.493.1512R}
{Rate} G.,  {Crowther} P.~A.,  2020, \mn@doi [\mnras] {10.1093/mnras/stz3614},
  493, 1512

\bibitem[\protect\citeauthoryear{{Rebull} et~al.,}{{Rebull}
  et~al.}{2002}]{2002AJ....123.1528R}
{Rebull} L.~M.,  et~al., 2002, \mn@doi [\aj] {10.1086/338904}, 123, 1528

\bibitem[\protect\citeauthoryear{{Rebull}, {Stauffer}, {Megeath}, {Hora}  \&
  {Hartmann}}{{Rebull} et~al.}{2006}]{2006ApJ...646..297R}
{Rebull} L.~M.,  {Stauffer} J.~R.,  {Megeath} S.~T.,  {Hora} J.~L.,
  {Hartmann} L.,  2006, \mn@doi [\apj] {10.1086/504865}, 646, 297

\bibitem[\protect\citeauthoryear{{Reggiani} \& {Meyer}}{{Reggiani} \&
  {Meyer}}{2011}]{2011ApJ...738...60R}
{Reggiani} M.~M.,  {Meyer} M.~R.,  2011, \mn@doi [\apj]
  {10.1088/0004-637X/738/1/60}, 738, 60

\bibitem[\protect\citeauthoryear{{Reggiani} \& {Meyer}}{{Reggiani} \&
  {Meyer}}{2013}]{2013A&A...553A.124R}
{Reggiani} M.,  {Meyer} M.~R.,  2013, \mn@doi [\aap]
  {10.1051/0004-6361/201321631}, 553, A124

\bibitem[\protect\citeauthoryear{{Reipurth}, {Guimar{\~a}es}, {Connelley}  \&
  {Bally}}{{Reipurth} et~al.}{2007}]{2007AJ....134.2272R}
{Reipurth} B.,  {Guimar{\~a}es} M.~M.,  {Connelley} M.~S.,   {Bally} J.,  2007,
  \mn@doi [\aj] {10.1086/523596}, 134, 2272

\bibitem[\protect\citeauthoryear{{\VAN{Rosa}{De}{de} Rosa}
  et~al.,}{{\VAN{Rosa}{De}{de} Rosa} et~al.}{2012}]{2012MNRAS.422.2765D}
{\VAN{Rosa}{De}{de} Rosa} R.~J.,  et~al., 2012, \mn@doi [\mnras]
  {10.1111/j.1365-2966.2011.20397.x}, 422, 2765

\bibitem[\protect\citeauthoryear{{\VAN{Rosa}{De}{de} Rosa}
  et~al.,}{{\VAN{Rosa}{De}{de} Rosa} et~al.}{2014}]{2014MNRAS.437.1216D}
{\VAN{Rosa}{De}{de} Rosa} R.~J.,  et~al., 2014, \mn@doi [\mnras]
  {10.1093/mnras/stt1932}, 437, 1216

\bibitem[\protect\citeauthoryear{{Rozhavskii}, {Kuz'mina}  \&
  {Vasilevskii}}{{Rozhavskii} et~al.}{1976}]{RN278}
{Rozhavskii} F.~G.,  {Kuz'mina} V.~A.,   {Vasilevskii} A.~E.,  1976, \mn@doi
  [Astrophysics] {10.1007/BF01002037}, 12, 204

\bibitem[\protect\citeauthoryear{{Salpeter}}{{Salpeter}}{1955}]{RN204}
{Salpeter} E.~E.,  1955, \mn@doi [\apj] {10.1086/145971}, 121, 161

\bibitem[\protect\citeauthoryear{{Sana} et~al.,}{{Sana}
  et~al.}{2012}]{2012Sci...337..444S}
{Sana} H.,  et~al., 2012, \mn@doi [Science] {10.1126/science.1223344}, 337, 444

\bibitem[\protect\citeauthoryear{{Sana} et~al.,}{{Sana}
  et~al.}{2013}]{2013AA...550A.107S}
{Sana} H.,  et~al., 2013, \mn@doi [\aap] {10.1051/0004-6361/201219621}, 550,
  A107

\bibitem[\protect\citeauthoryear{{Schoettler} \& {Parker}}{{Schoettler} \&
  {Parker}}{2021}]{2021MNRAS.501L..12S}
{Schoettler} C.,  {Parker} R.~J.,  2021, \mn@doi [\mnras]
  {10.1093/mnrasl/slaa182}, 501, L12

\bibitem[\protect\citeauthoryear{Schoettler, Parker, Arnold, Grimmett, de
  Bruijne  \& Wright}{Schoettler et~al.}{2019}]{RN309}
Schoettler C.,  Parker R.~J.,  Arnold B.,  Grimmett L.~P.,  de Bruijne J.,
  Wright N.~J.,  2019, \mn@doi [\mnras] {10.1093/mnras/stz1487}, 487, 4615

\bibitem[\protect\citeauthoryear{{Schoettler}, {de Bruijne}, {Vaher}  \&
  {Parker}}{{Schoettler} et~al.}{2020}]{2020MNRAS.495.3104S}
{Schoettler} C.,  {de Bruijne} J.,  {Vaher} E.,   {Parker} R.~J.,  2020,
  \mn@doi [\mnras] {10.1093/mnras/staa1228}, 495, 3104

\bibitem[\protect\citeauthoryear{{Sch{\"o}nrich}, {Binney}  \&
  {Dehnen}}{{Sch{\"o}nrich} et~al.}{2010}]{2010MNRAS.403.1829S}
{Sch{\"o}nrich} R.,  {Binney} J.,   {Dehnen} W.,  2010, \mn@doi [\mnras]
  {10.1111/j.1365-2966.2010.16253.x}, 403, 1829

\bibitem[\protect\citeauthoryear{{Spina} et~al.,}{{Spina}
  et~al.}{2017}]{2017A&A...601A..70S}
{Spina} L.,  et~al., 2017, \mn@doi [\aap] {10.1051/0004-6361/201630078}, 601,
  A70

\bibitem[\protect\citeauthoryear{{Stassun} et~al.,}{{Stassun}
  et~al.}{2019}]{2019AJ....158..138S}
{Stassun} K.~G.,  et~al., 2019, \mn@doi [\aj] {10.3847/1538-3881/ab3467}, 158,
  138

\bibitem[\protect\citeauthoryear{{Stone}}{{Stone}}{1991}]{RN276}
{Stone} R.~C.,  1991, \mn@doi [\aj] {10.1086/115880}, 102, 333

\bibitem[\protect\citeauthoryear{{Sung}, {Stauffer}  \& {Bessell}}{{Sung}
  et~al.}{2009}]{2009AJ....138.1116S}
{Sung} H.,  {Stauffer} J.~R.,   {Bessell} M.~S.,  2009, \mn@doi [\aj]
  {10.1088/0004-6256/138/4/1116}, 138, 1116

\bibitem[\protect\citeauthoryear{{Teixeira} et~al.,}{{Teixeira}
  et~al.}{2006}]{2006ApJ...636L..45T}
{Teixeira} P.~S.,  et~al., 2006, \mn@doi [\apjl] {10.1086/500009}, 636, L45

\bibitem[\protect\citeauthoryear{{Teixeira}, {Lada}, {Marengo}  \&
  {Lada}}{{Teixeira} et~al.}{2012}]{2012A&A...540A..83T}
{Teixeira} P.~S.,  {Lada} C.~J.,  {Marengo} M.,   {Lada} E.~A.,  2012, \mn@doi
  [\aap] {10.1051/0004-6361/201015326}, 540, A83

\bibitem[\protect\citeauthoryear{{Tetzlaff}, {Neuh{\"a}user}  \&
  {Hohle}}{{Tetzlaff} et~al.}{2011}]{RN263}
{Tetzlaff} N.,  {Neuh{\"a}user} R.,   {Hohle} M.~M.,  2011, \mn@doi [\mnras]
  {10.1111/j.1365-2966.2010.17434.x}, 410, 190

\bibitem[\protect\citeauthoryear{{Thompson}, {Corbin}, {Young}  \&
  {Schneider}}{{Thompson} et~al.}{1998}]{1998ApJ...492L.177T}
{Thompson} R.~I.,  {Corbin} M.~R.,  {Young} E.,   {Schneider} G.,  1998,
  \mn@doi [\apjl] {10.1086/311096}, 492, L177

\bibitem[\protect\citeauthoryear{{Tobin}, {Hartmann}, {F{\H{u}}r{\'e}sz}, {Hsu}
   \& {Mateo}}{{Tobin} et~al.}{2015}]{2015AJ....149..119T}
{Tobin} J.~J.,  {Hartmann} L.,  {F{\H{u}}r{\'e}sz} G.,  {Hsu} W.-H.,   {Mateo}
  M.,  2015, \mn@doi [\aj] {10.1088/0004-6256/149/4/119}, 149, 119

\bibitem[\protect\citeauthoryear{{Tokovinin}}{{Tokovinin}}{2018}]{2018ApJS..235....6T}
{Tokovinin} A.,  2018, \mn@doi [\apjs] {10.3847/1538-4365/aaa1a5}, 235, 6

\bibitem[\protect\citeauthoryear{{Vaher}}{{Vaher}}{2020}]{2020RNAAS...4..116V}
{Vaher} E.,  2020, \mn@doi [Research Notes of the American Astronomical
  Society] {10.3847/2515-5172/aba952}, 4, 116

\bibitem[\protect\citeauthoryear{{Venuti} et~al.,}{{Venuti}
  et~al.}{2014}]{2014A&A...570A..82V}
{Venuti} L.,  et~al., 2014, \mn@doi [\aap] {10.1051/0004-6361/201423776}, 570,
  A82

\bibitem[\protect\citeauthoryear{{Venuti} et~al.,}{{Venuti}
  et~al.}{2015}]{2015A&A...581A..66V}
{Venuti} L.,  et~al., 2015, \mn@doi [\aap] {10.1051/0004-6361/201526164}, 581,
  A66

\bibitem[\protect\citeauthoryear{{Venuti} et~al.,}{{Venuti}
  et~al.}{2017}]{2017A&A...599A..23V}
{Venuti} L.,  et~al., 2017, \mn@doi [\aap] {10.1051/0004-6361/201629537}, 599,
  A23

\bibitem[\protect\citeauthoryear{{Venuti} et~al.,}{{Venuti}
  et~al.}{2018}]{2018A&A...609A..10V}
{Venuti} L.,  et~al., 2018, \mn@doi [\aap] {10.1051/0004-6361/201731103}, 609,
  A10

\bibitem[\protect\citeauthoryear{{Vincke} \& {Pfalzner}}{{Vincke} \&
  {Pfalzner}}{2016}]{2016ApJ...828...48V}
{Vincke} K.,  {Pfalzner} S.,  2016, \mn@doi [\apj]
  {10.3847/0004-637X/828/1/48}, 828, 48

\bibitem[\protect\citeauthoryear{{Voroshilov}, {Guseva}, {Kalandadze},
  {Kolesnik}, {Kuznetsov}, {Metreveli}  \& {Shapovalov}}{{Voroshilov}
  et~al.}{1985}]{1985cbvm.book.....V}
{Voroshilov} V.~I.,  {Guseva} N.~G.,  {Kalandadze} N.~B.,  {Kolesnik} L.~N.,
  {Kuznetsov} V.~I.,  {Metreveli} M.~D.,   {Shapovalov} A.~N.,  1985,
  {Catalogue of BV magnitudes and spectral classes for 6000 stars. Ukrainian
  Acad. Nauk, Kiev, 1-140.}

\bibitem[\protect\citeauthoryear{{Weiler}}{{Weiler}}{2018}]{2018A&A...617A.138W}
{Weiler} M.,  2018, \mn@doi [\aap] {10.1051/0004-6361/201833462}, 617, A138

\bibitem[\protect\citeauthoryear{{Wenger} et~al.,}{{Wenger}
  et~al.}{2000}]{2000A&AS..143....9W}
{Wenger} M.,  et~al., 2000, \mn@doi [\aaps] {10.1051/aas:2000332}, 143, 9

\bibitem[\protect\citeauthoryear{{Wilson}}{{Wilson}}{1953}]{1953GCRV..C......0W}
{Wilson} R.~E.,  1953, Carnegie Institute Washington D.C. Publication, \href
  {https://ui.adsabs.harvard.edu/abs/1953GCRV..C......0W} {p.~0}

\bibitem[\protect\citeauthoryear{{\VAN{Wit}{De}{de} Wit}, {Testi}, {Palla}  \&
  {Zinnecker}}{{\VAN{Wit}{De}{de} Wit} et~al.}{2005}]{RN190}
{\VAN{Wit}{De}{de} Wit} W.~J.,  {Testi} L.,  {Palla} F.,   {Zinnecker} H.,
  2005, \mn@doi [\aap] {10.1051/0004-6361:20042489}, 437, 247

\bibitem[\protect\citeauthoryear{{Wootton} \& {Parker}}{{Wootton} \&
  {Parker}}{2019}]{2019MNRAS.485L..48W}
{Wootton} B.~A.,  {Parker} R.~J.,  2019, \mn@doi [\mnras]
  {10.1093/mnrasl/sly238}, 485, L48

\bibitem[\protect\citeauthoryear{{Wright} \& {Mamajek}}{{Wright} \&
  {Mamajek}}{2018}]{2018MNRAS.476..381W}
{Wright} N.~J.,  {Mamajek} E.~E.,  2018, \mn@doi [\mnras]
  {10.1093/mnras/sty207}, 476, 381

\bibitem[\protect\citeauthoryear{{Yeom}, {Lee}, {Koo}, {Beers}  \&
  {Kim}}{{Yeom} et~al.}{2019}]{2019JKAS...52...57Y}
{Yeom} B.-S.,  {Lee} Y.~S.,  {Koo} J.-R.,  {Beers} T.~C.,   {Kim} Y.~K.,  2019,
  \mn@doi [Journal of Korean Astronomical Society] {10.5303/JKAS.2019.52.3.57},
  52, 57

\bibitem[\protect\citeauthoryear{{Zari}, {Hashemi}, {Brown}, {Jardine}  \& {de
  Zeeuw}}{{Zari} et~al.}{2018}]{2018A&A...620A.172Z}
{Zari} E.,  {Hashemi} H.,  {Brown} A.~G.~A.,  {Jardine} K.,   {de Zeeuw} P.~T.,
   2018, \mn@doi [\aap] {10.1051/0004-6361/201834150}, 620, A172

\bibitem[\protect\citeauthoryear{{Zhao}, {Jiang}, {Gao}, {Li}  \& {Sun}}{{Zhao}
  et~al.}{2018}]{2018ApJ...855...12Z}
{Zhao} H.,  {Jiang} B.,  {Gao} S.,  {Li} J.,   {Sun} M.,  2018, \mn@doi [\apj]
  {10.3847/1538-4357/aaacd0}, 855, 12

\bibitem[\protect\citeauthoryear{{Zhong}, {Chen}, {Kouwenhoven}, {Li}, {Shao}
  \& {Hou}}{{Zhong} et~al.}{2019}]{RN291}
{Zhong} J.,  {Chen} L.,  {Kouwenhoven} M.~B.~N.,  {Li} L.,  {Shao} Z.,   {Hou}
  J.,  2019, \mn@doi [\aap] {10.1051/0004-6361/201834334}, 624, A34

\bibitem[\protect\citeauthoryear{{Zwintz} et~al.,}{{Zwintz}
  et~al.}{2017}]{2017A&A...601A.101Z}
{Zwintz} K.,  et~al., 2017, \mn@doi [\aap] {10.1051/0004-6361/201630327}, 601,
  A101

\makeatother
\end{thebibliography}



\appendix

\section{RW/WW predictions from simulations for the full mass range}

Tables~\ref{tab:RW_WW_pred_full_mass} and \ref{tab:RW_WW_pred_full_mass_bin} (for alternative binary system set-ups) provide predicted numbers for the amount of RW/WW stars that we should expect to find around young star-forming regions based on simulations with different initial conditions as shown in Table~\ref{tab:Init_cond}.

\begin{table*}
	\centering
	\caption{Ejected RW and WW stars from $N$-body simulations within the search radius of 100 pc at different times during the simulations. For all our initial condition combinations, we show averages from all 20 simulations and the maximum from a single simulation in the format average $\pm$ uncertainty / maximum. We count ejected binary systems as one star when calculating averages and maxima. The uncertainties in our averages are the standard deviations. We show ejected stars with masses from 0.1--50 M$_{\sun}$.}
	\label{tab:RW_WW_pred_full_mass}
	\begin{tabular}{lcccccccc} 
		\hline
		Mass $m$ (M$_{\sun}$) & & & & Simulation ID \\
		\hline
		RW stars\\
		\hline
		&  16-03-1 & 30-03-1 & 16-05-1 & 30-05-1 &  16-03-5 & 20-03-5 & 16-05-5 & 20-05-5\\
		0.1 $\leq$ \textit{m} < 8.0 \\
		- after 1 Myr & 7.8$\,\pm$2.3 /10 &  0.6$\,\pm$0.7  / 2 &  8.6$\,\pm$4.4  / 19&  0.4$\,\pm$0.7  / 2&  3.9$\,\pm$1.6  / 7&  0.4$\,\pm$0.8  / 3&  3.4$\,\pm$1.6  / 6&  0.4$\,\pm$0.7  / 2 \\ 
		- after 2 Myr &  5.4$\,\pm$1.9 / 9 &  0.6$\,\pm$0.9  / 4&  6.6$\,\pm$2.9  / 12&  0.8$\,\pm$1.1  / 4 &  2.8$\,\pm$1.4  / 6&  0.4$\,\pm$0.7  / 2&  2.1$\,\pm$1.2  / 4&  0.3$\,\pm$0.4  / 1 \\ 
		- after 3 Myr &  2.0$\,\pm$1.2 / 5  &  0.5$\,\pm$1.1  / 5&  2.8$\,\pm$1.9  / 8 &  0.5$\,\pm$0.9  / 3&  0.8$\,\pm$0.9  / 3&  0.2$\,\pm$0.4  / 1&  0.8$\,\pm$1.0  / 3&  0.1$\,\pm$0.3  / 1\\ 
		- after 4 Myr &  0.8$\,\pm$1.1  / 4  &  0.4$\,\pm$0.6  / 2&  0.9$\,\pm$0.8  / 3 &  0.3$\,\pm$0.5  / 2 &  0.2$\,\pm$0.4  / 1&  0 / 0&  0.1$\,\pm$0.3  / 1&  0.2$\,\pm$0.4  / 1\\ 
		- after 5 Myr &  0.3$\,\pm$0.6  / 2  &  0.3$\,\pm$0.6  / 2 &  0.4$\,\pm$0.6  / 2 &  0.2$\,\pm$0.4  / 1&  0.1$\,\pm$0.2  / 1&  0/ 0&  0.2$\,\pm$0.4  / 1&  0.1$\,\pm$0.3  / 1\\ 
		\textit{m} $\geq$ 8.0  \\
	    - after 1 Myr &  0.1$\,\pm$0.2 / 1 &  0 / 0 &  0.1$\,\pm$0.1  / 1 &  0 / 0 &  0.1$\,\pm$0.2  / 1&  0.1$\,\pm$0.2  / 1&  0.1$\,\pm$0.2  / 1&  0 / 0\\ 
		- after 2 Myr &  0.1$\,\pm$0.3  / 1 &  0 / 0 &   0 / 0&  0 / 0 &  0 / 0 &  0 / 0&  0.1$\,\pm$0.2  / 1&  0 / 0\\ 
		- after 3 Myr &  0.1$\,\pm$0.2  / 1  &  0 / 0&   0 / 0 &  0 / 0 &  0 / 0 &  0 / 0&  0  / 0&  0 / 0\\ 
		- after 4 Myr &  0.2$\,\pm$0.5 / 2  &  0.1$\,\pm$0.2  / 1&  0.1$\,\pm$0.1  / 1 &  0 / 0 &  0 / 0 &  0 / 0&  0  / 0&  0 / 0\\ 
		- after 5 Myr &  0.1$\,\pm$0.2  / 1  &  0.1$\,\pm$0.2  / 1&  0.1$\,\pm$0.1  / 1 &  0.1$\,\pm$0.4  / 2&  0 / 0 &  0 / 0&  0/ 0&  0 / 0\\ 		\hline
		WW stars\\
		\hline
		&  16-03-1 & 30-03-1 & 16-05-1 & 30-05-1 &  16-03-5 & 20-03-5 & 16-05-5 & 20-05-5 \\
		0.1 $\leq$ \textit{m} < 8.0 \\
        - after 1 Myr &  53.6$\,\pm$9.3  / 79  &  4.1$\,\pm$2.6  / 10&  51.7$\,\pm$8.5  / 65&  2.0$\,\pm$1.6  / 6 &  20.4$\,\pm$4.3  / 28&  6.1$\,\pm$2.8  / 13&  18.9$\,\pm$4.9  / 31&  4.9$\,\pm$2.1  / 10\\ 
		- after 2 Myr &  55.6$\,\pm$10.2  / 72  &  5.0$\,\pm$2.8  / 10&  54.1$\,\pm$9.1  / 72&  2.9$\,\pm$2.0  / 7&  21.7$\,\pm$4.2  / 28&  6.7$\,\pm$2.7  / 12&  20.9$\,\pm$5.4  / 34&  5.5$\,\pm$1.9  / 10\\ 
		- after 3 Myr &  57.4$\,\pm$10.1  / 73  &  5.9$\,\pm$2.9  / 13&  55.8$\,\pm$9.9  / 75  &3.5$\,\pm$2.5  / 9&  22.5$\,\pm$4.3  / 30&  7.4$\,\pm$2.9  / 12&  21.4$\,\pm$5.8  / 34&  5.9$\,\pm$2.0  / 10\\ 
		- after 4 Myr &  55.5$\,\pm$9.5  / 70 &  7.4$\,\pm$2.9  / 14&  54.2$\,\pm$9.3  / 71&  4.1$\,\pm$2.8  / 9&  22.1$\,\pm$4.2  / 29&  7.5$\,\pm$2.37 / 12&  20.6$\,\pm$5.7  / 32&  6.3$\,\pm$2.1  / 11\\ 
		- after 5 Myr &  51.7$\,\pm$9.3  / 67  &  7.7$\,\pm$2.9  / 13&  50.9$\,\pm$8.7  / 66&  4.3$\,\pm$2.8  / 9&  21.1$\,\pm$4.0  / 29&  7.7$\,\pm$2.7  / 14&  19.8$\,\pm$5.6  / 30&  6.5$\,\pm$2.5  / 13\\ 
		\textit{m} $\geq$ 8.0  \\
        - after 1 Myr &  1.0$\,\pm$0.7  / 3  &  0 / 0 &  0.7$\,\pm$1.0  / 3&  0 / 0 &  0.2$\,\pm$0.3  / 1&  0.2$\,\pm$0.6  / 2&  0.3$\,\pm$0.6  / 2&  0.2$\,\pm$0.6  / 2\\ 
		- after 2 Myr &  1.2$\,\pm$0.8  / 3  &  0.1$\,\pm$0.3  / 2&  0.9$\,\pm$1.0  / 3&  0 / 0 &  0.1$\,\pm$0.3  / 1&  0.2$\,\pm$0.6  / 2&  0.2$\,\pm$0.5  / 2&  0.2$\,\pm$0.6  / 2\\ 
		- after 3 Myr &  1.3$\,\pm$0.7  / 3  &  0.1$\,\pm$0.4  / 2 &  1.2$\,\pm$1.0  / 3&  0 / 0 &  0.2$\,\pm$0.4  / 1&  0.2$\,\pm$0.6  / 2&  0.2$\,\pm$0.5  / 2&  0.2$\,\pm$0.6  / 2\\ 
		- after 4 Myr &  1.5$\,\pm$0.9  / 4  &  0.4$\,\pm$0.5  / 2&  1.2$\,\pm$0.9  / 3&  0.1$\,\pm$0.2  / 1&  0.2$\,\pm$0.4  / 1&  0.2$\,\pm$0.6  / 2&  0.3$\,\pm$0.5  /2&  0.2$\,\pm$0.6  / 2\\ 
		- after 5 Myr &  1.6$\,\pm$0.9  / 4  &  0.4$\,\pm$0.5  / 2&  1.3$\,\pm$1.1  / 4&  0.3$\,\pm$0.5  / 2&  0.3$\,\pm$0.6  / 2&  0.2$\,\pm$0.6  / 2&  0.4$\,\pm$0.8  / 3&  0.2$\,\pm$0.6  / 2\\ 
		\hline
	\end{tabular}
\end{table*}

\begin{table*}
	\centering
	\caption{Ejected RW and WW stars from $N$-body simulations within the search radius of 100 pc at different times during the simulations. We show the results for the two alternative binary system set-ups in comparison to the original. For all our initial condition combinations, we show averages from all 20 simulations and the maximum from a single simulation in the format average $\pm$ uncertainty / maximum. We count ejected binary systems as one star when calculating averages and maxima. The uncertainties in our averages are the standard deviations. We show ejected stars with masses from 0.1--50 M$_{\sun}$.}
	\label{tab:RW_WW_pred_full_mass_bin}
	\begin{tabular}{lcccccccc} 
		\hline
		Mass $m$ (M$_{\sun}$) & & Simulation ID \\
		\hline
		RW stars\\
		\hline
		&  16-05-1& 16-05-1-100-bin & 16-05-1-eq-mass \\
		0.1 $\leq$ \textit{m} < 8.0 \\
		- after 1 Myr &8.6$\,\pm$4.4 /19  & 3.1$\,\pm$2.0 /8 &  11.3$\,\pm$4.2  / 23 \\ 
		- after 2 Myr & 6.6$\,\pm$2.9 /12 & 2.3$\,\pm$2.0 / 7 &  7.2$\,\pm$3.1  / 13\\ 
		- after 3 Myr & 2.8$\,\pm$1.9 /8 & 0.9$\,\pm$1.2 / 5  &  2.4$\,\pm$1.7  / 5\\ 
		- after 4 Myr & 0.9$\,\pm$0.8 /3 & 0.5$\,\pm$0.6  / 2  &  0.6$\,\pm$0.8  / 3\\ 
		- after 5 Myr & 0.4$\,\pm$0.6 /2 & 0.3$\,\pm$0.5  / 2  &  0.4$\,\pm$0.7  / 2 \\ 
		\textit{m} $\geq$ 8.0  \\
	    - after 1 Myr &  0.1$\,\pm$0.1  / 1  & 0 / 0 &   0.1$\,\pm$0.2 / 1 \\ 
		- after 2 Myr & 0 / 0 & 0.1$\,\pm$0.1  / 1 &  0.1$\,\pm$0.2 / 1 \\ 
		- after 3 Myr & 0 / 0  & 0.1$\,\pm$0.4  / 2  &  0 / 0\\ 
		- after 4 Myr & 0.1$\,\pm$0.1  / 1 & 0/ 0  &  0 / 0\\ 
		- after 5 Myr & 0.1$\,\pm$0.1  / 1 & 0.1$\,\pm$0.3  / 1  &  0.1$\,\pm$0.2  / 1\\
		\hline
		WW stars\\
		\hline
		& 16-05-1 & 16-05-1-100-bin & 16-05-1-eq-mass \\
		0.1 $\leq$ \textit{m} < 8.0 \\
        - after 1 Myr &51.7$\,\pm$8.5 /65 & 50.0$\,\pm$12.6  / 83 &  53.7$\,\pm$13.0  / 83\\ 
		- after 2 Myr &54.1$\,\pm$9.1  / 72 & 53.0$\,\pm$12.7  / 87 &  57.7$\,\pm$12.8  / 83\\ 
		- after 3 Myr &55.8$\,\pm$9.9  / 75 & 54.2$\,\pm$9.9  / 91  &  58.8$\,\pm$13.3 / 87\\ 
		- after 4 Myr & 54.2$\,\pm$9.3  / 71 & 55.3$\,\pm$13.4  / 93 &  56.6$\,\pm$13.0  / 83\\ 
		- after 5 Myr &50.9$\,\pm$8.7  / 66 & 53.1$\,\pm$13.1  / 90  &  52.4$\,\pm$11.4  / 76\\
		\textit{m} $\geq$ 8.0  \\
        - after 1 Myr &0.7$\,\pm$1.0  / 3 & 0.3$\,\pm$0.4  / 1  &  1.0$\,\pm$1.1 / 4 \\ 
		- after 2 Myr & 0.9$\,\pm$1.0  / 3& 0.2$\,\pm$0.4  / 1  &  1.3$\,\pm$1.3  / 5\\ 
		- after 3 Myr & 1.2$\,\pm$1.0  / 3& 0.3$\,\pm$0.4  / 1  &  1.6$\,\pm$1.6  / 5 \\ 
		- after 4 Myr & 1.2$\,\pm$0.9  / 3& 0.3$\,\pm$0.6  / 2  &  1.5$\,\pm$1.3  / 4\\ 
		- after 5 Myr &1.3$\,\pm$1.1  / 4 & 0.3$\,\pm$0.6  / 2  &  1.6$\,\pm$1.7  / 6\\ 
		\hline
	\end{tabular}
\end{table*}

\section{2D-RW and WW candidates from S Mon and IRS 1/2}\label{2D-app}

Tables~\ref{tab:RWC_2D_SMon_app}, \ref{tab:WWC_2D_SMon_app}, \ref{tab:RWC_2D_IRS_app} and \ref{tab:WWC_2D_IRS_app} provide information on the 2D-RW and WW candidates that can be traced back to either S\,Mon or IRS\,1/2. The tables also include the non-3D trace-backs to both regions.

\begin{table*}
    \renewcommand\arraystretch{1.2}
	\centering
	\caption{S\,Mon RW star 2D candidates sorted by decreasing 2D-velocity. Column 2+3: velocity in S\,Mon rest frame [rf]; Column 3: RV sources - $^{a}$\textit{Gaia} DR2, $^{b}$\citet{2016A&A...586A..52J}, $^{c}$\citet{2019yCat.5164....0L}, $^{d}$\citet{2019AJ....157..196K}, $^{e}$\citet{1992A&AS...95..541F}, 
	$^{f}$\citet{1995A&AS..114..269D}; Column 4: indication of 3D-candidate status; Column 5: minimum flight time since ejection (crossing of search boundary); Column 6: age from PARSEC isochrones \citep{RN225}; Column 7--8: from literature sources -
	$^{1}$\citet{2019yCat.5164....0L}, $^{2}$\citet{2015A&A...581A..66V}, $^{3}$\citet{2017A&A...599A..23V}, $^{4}$\citet{2014A&A...570A..82V}, $^{5}$\citet{2019AJ....157..196K}, $^{6}$\citet{2004A&A...417..557L},
	$^{7}$\citet{1972A&AS....7...35K},
	$^{8}$\citet{1993yCat.3135....0C}, 
	$^{9}$\citet{1985cbvm.book.....V}, 
	$^{10}$\citet{2001A&A...373..625P}, 
	$^{11}$\citet{2002AJ....123.1528R}, 
	$^{12}$\citet{2019AJ....158..138S},
	$^{13}$\citet{2019A&A...628A..94A}, 
	$^{14}$\citet{2018A&A...609A..10V}.}
	\label{tab:RWC_2D_SMon_app}
	\begin{tabular}{lccccccccc} 
		\hline
		\textit{Gaia} DR2 source-id & 2D-velocity rf & Radial velocity rf & 3D-cand. & Flight time & Iso. age &  Mass & Spectral type  \\
		 & (km\,s$^{-1}$) & (km\,s$^{-1}$) & & (Myr) & (Myr) &  (M$_{\sun}$) & \\
		\hline
    3134372235323909248 & 74.9  $\pm$0.5   & -58.2 $\pm$4.7$^{c}$   & no    &  -     & 2.3\,$^{+7.7}_{-1.5}$      & 0.9--1.0$^{12,13}$ & G5$^{1}$ \\
    3326642764222483328 & 72.8  $\pm$0.8   &-  & -     & 0.1   & 9.0\,$^{+41.0}_{-7.0}$  & 0.6--0.8$^{12,13}$    & - \\
    3353807577672866432 & 68.4  $\pm$0.6   & 86.2  $\pm$3.7$^{a}$   & no    &   -    &  1.5\,$^{+1.5}_{-1.0}$  &   1.1$^{13}$ & K1$^{1}$ \\
    3326673142526903296 & 58.7  $\pm$0.6   &-  & -     & 0.1   & 20.0\,$^{+15.0}_{-18.0}$    &  1.0$^{12,13}$ & - \\
    3157444730917606656 & 46.3  $\pm$2.6   &-  & -     & 0.9   & 20.0\,$^{+12.0}_{-15.0}$  & 0.6$^{12}$    & - \\
    3326639882300319744 & 46.1  $\pm$1.0   & -17.1 $\pm$2.9$^{b}$   & no    &    -   & 4.0\,$^{+16.0}_{-3.0}$ &0.9--1.1$^{12,13}$     &-  \\
    3331426086479602944 & 41.9  $\pm$0.5   &-  & -     & 1.1   &   8.0\,$\pm$5.0    & 0.9--1.0$^{12,13}$& -\\
    3134341650859166208 & 39.2  $\pm$0.9   &-  & -     & 0.4   &   6.0\,$^{+25.0}_{-4.0}$     &  0.9$^{12,13}$& -\\
    3326519142179922688 & 37.7  $\pm$0.8   &-  & -     & 0.3   &    12.0\,$^{+28.0}_{-11.0}$    &  0.9--1.0$^{12,13}$&-\\
    3326992103978048512 & 36.8  $\pm$0.5   & -16.3 $\pm$3.9$^{a}$   & no    &   -    &    5.0\,$^{+5.0}_{-3.0}$    & 1.3--1.7$^{12,13}$ & -\\
    3328110440447873024 & 36.6  $\pm$8.8   & -   & -    &    0.8   &   100.0\,$^{+100.0}_{-98.5}$ & - & - & - \\
    3326736570603687168 & 33.6  $\pm$0.6   & 105.7 $\pm$2.9$^{b}$   & no    &     -  &   30.0\,$^{+20.0}_{-26.0}$    &  0.8--1.0$^{12,13}$ & G7$^{6}$\\
    3326595352079413888 & 32.8  $\pm$0.5   & -73.0 $\pm$10.0$^{a}$   & no    &-       &   10.0\,$^{+5.0}_{-8.5}$      &1.1--1.2$^{12,13}$ & F2$^{1}$ \\
    3357507846618517248 & 31.8  $\pm$0.8   &-  & -     & 1.9   &    7.0\,$\pm$3.0    &0.6--0.9$^{12,13}$ &-  \\
    3355894515164624000 & 30.4  $\pm$4.1   &-  & -     & 1.8   &   15.0\,$^{+20.0}_{-12.0}$     & 0.4$^{12}$ & - \\
    3318797443817943040 & 27.8  $\pm$0.6   & -28.0 $\pm$3.2$^{a}$   & no    & -      & 1.5\,$^{+2.5}_{-1.0}$      & 1.2$^{13}$ & -\\
    3132210251867043200 & 23.9  $\pm$0.7   & -38.9 $\pm$7.0$^{c}$   & no    &  -     &    5.0\,$^{+7.0}_{-3.0}$   & 1.0$^{13}$ & G5$^{1}$ \\
    3133279939241456000 & 21.6  $\pm$0.6   & -50.7 $\pm$5.3$^{a}$   & no    &   -    &   9.0\,$^{+6.0}_{-7.0}$     & 1.0$^{12,13}$ &-  \\
    3328003654677416192 & 20.9  $\pm$0.6   & 29.0  $\pm$3.2$^{a}$   & no    &    -   &  0.2\,$^{+0.8}_{-0.1}$       &  1.1$^{13}$ &-\\
    3131085447170896128 & 13.0  $\pm$0.6   & 34.9  $\pm$3.4$^{a}$   & no    &   -    &  5.0\,$\pm$2.0     & 1.1--1.5$^{12,13}$- &  \\
	\hline
	\end{tabular}
\end{table*}

\begin{table*}
    \renewcommand\arraystretch{1.2}
	\centering
	\caption{S\,Mon WW star 2D candidates. Column 2+3: velocity in S\,Mon rest frame [rf]; Column 3: RV sources - $^{a}$\textit{Gaia} DR2, $^{b}$\citet{2016A&A...586A..52J}, $^{c}$\citet{2019yCat.5164....0L}, $^{d}$\citet{2019AJ....157..196K}, $^{e}$\citet{1992A&AS...95..541F}, 
	$^{f}$\citet{1995A&AS..114..269D}; Column 4: indication of 3D-candidate status; Column 5: minimum flight time since ejection (crossing of search boundary); Column 6: age from PARSEC isochrones \citep{RN225}; Column 7--8: from literature sources - $^{1}$\citet{2019yCat.5164....0L}, $^{2}$\citet{2015A&A...581A..66V}, $^{3}$\citet{2017A&A...599A..23V}, $^{4}$\citet{2014A&A...570A..82V}, $^{5}$\citet{2019AJ....157..196K}, $^{6}$\citet{2004A&A...417..557L},
	$^{7}$\citet{1972A&AS....7...35K},
	$^{8}$\citet{1993yCat.3135....0C}, 
	$^{9}$\citet{1985cbvm.book.....V}, 
	$^{10}$\citet{2001A&A...373..625P}, 
	$^{11}$\citet{2002AJ....123.1528R}, 
	$^{12}$\citet{2019AJ....158..138S},
	$^{13}$\citet{2019A&A...628A..94A},
	$^{14}$\citet{2018A&A...609A..10V}.}
	\label{tab:WWC_2D_SMon_app}
	\begin{tabular}{lccccccc} 
		\hline
		\textit{Gaia} DR2 source-id & 2D-velocity rf & Radial velocity rf & 3D-cand. & Flight time & Iso. age & Mass & Spectral type  \\
		 & (km\,s$^{-1}$) & (km\,s$^{-1}$) & & (Myr) & (Myr) &  (M$_{\sun}$) &\\
		\hline
    3326731554082039296 & 27.3  $\pm$0.6   & -   &  -     & 0.2&  20.0\,$^{+10.0}_{-16.0}$      & 1.0$^{12,13}$ & -\\
    3326610298565496320 & 27.1  $\pm$1.3   & -   &  -     & 0.2&  20.0\,$^{+25.0}_{-17.0}$       &  0.7--0.9$^{12,13}$ &-\\
    3326670389452200832 & 23.0  $\pm$0.9   & -   &   -    & 0.2& 50.0\,$^{+20.0}_{-46.0}$       &  0.7--0.8$^{12,13}$  &-\\
    3331136468244152832 & 22.2  $\pm$1.5   & -   &  -     & 1.9&   5.0\,$^{+25.0}_{-3.0}$    &  0.7--1.0$^{12,13}$  &-\\
    3130519576640345344 & 21.5  $\pm$1.5   & -   &  -     & 3.7&   1.5\,$^{+2.5}_{-1.0}$    &  - &- \\
    3326628230053022080 & 21.0  $\pm$2.9   & -   &  -     & 0.4&   50.0\,$^{+50.0}_{-46.0}$      & 0.6$^{12}$  & -\\
    3158160611766998400 & 20.7  $\pm$0.7   & -   &   -    & 1.9& 4.8\,$^{+2.0}_{-1.5}$      & 1.6--1.7$^{12,13}$  & F0$^{1}$ \\
    3327753756299891328 & 20.0  $\pm$0.6   & -   &    -   & 0.7&   20.0\,$^{+5.0}_{-15.0}$      & 0.7--0.9$^{12,13}$  & -\\
    3161262853763313024 & 18.9  $\pm$0.9   & 22.6  $\pm$6.7$^{a}$   & no    & -     &  6.0\,$\pm$2.0       & 1.4--1.6$^{12,13}$  & -\\
    3134469026706163456 & 18.6  $\pm$0.6   & -   &     -  & 0.6& 10.0\,$^{+20.0}_{-7.5}$      & 1.1$^{13}$  & -\\
    3326641462848241408 & 17.6  $\pm$0.6   & -   &  -     & 0.4&  20.0\,$^{+5.0}_{-16.0}$       &  1.0--1.1$^{12,13}$  &-\\
    3326535012082791424 & 17.5  $\pm$1.1   & -   &   -    & 0.5&  12.0\,$^{+50.0}_{-10.0}$     &  0.5--0.7$^{12,13}$  &-\\
    3326641462848241792 & 17.0  $\pm$0.6   & -   &    -   & 0.3&  7.0\,$^{+25.0}_{-5.5}$      &  0.8--1.0$^{12,13}$ & -\\
    3350754336961977984 & 15.1  $\pm$0.6   & -   &     -  & 0.4&   25.0\,$^{+10.0}_{-11.0}$     &  0.9--1.0$^{12,13}$  &-\\
    3326170730136908032 & 15.1  $\pm$1.8   & -   &  -     & 1.2&   2.0\,$\pm$0.2     &  2.2$^{13}$  & A0/1$^{7}$  \\
    3350679806397436928 & 14.7  $\pm$0.6   & -   &   -    & 0.9&  9.0\,$^{+3.0}_{-6.0}$      &  1.0--1.2$^{12,13}$  &-\\
    3356837591202942976 & 14.3  $\pm$0.6   & -   &    -   & 4.9&   5.0\,$^{+5.0}_{-2.5}$     & 0.7--0.9$^{12,13}$ & -\\
    3350847554932397056 & 13.9  $\pm$0.9   & -   &     -  & 1.5&   2.0\,$^{+2.0}_{-1.0}$    & 0.5$^{12}$  &- \\
    3350758013453747328 & 13.8  $\pm$0.5   & -   &      - & 0.8&  20.0\,$^{+10.0}_{-16.0}$     & 1.0$^{12,13}$  & -\\
    3326951215889632128 & 13.7  $\pm$0.5   & -   &-       & 0.2& 15.0\,$^{+5.0}_{-12.0}$     &  1.2--1.5$^{12,13}$  &-\\
    3326626344563404672 & 12.9  $\pm$0.7   & -   &   -    & 0.6&  32.0\,$^{+8.0}_{-30.0}$     & 0.8$^{12,13}$  & -\\
    3326737189078970880 & 12.6  $\pm$1.5   & -   &    -   & -0.1&  2.5\,$^{+17.5}_{-2.0}$      & - &-\\
    3327689606667717248 & 11.5  $\pm$2.2   & -   &     -  & 2.3&  2.0\,$^{+5.0}_{-1.8}$     &  0.7$^{12}$  &-\\
    3327310343876193536 & 11.5  $\pm$0.7   & -   & -      & 2.7&  4.0\,$^{+6.0}_{-3.6}$     & 1.2--1.7$^{12,13}$  & -\\
    3326644138611656832 & 11.1  $\pm$0.5   & 18.6  $\pm$3.1$^{a}$   & no    & -     &  5.0\,$^{+6.0}_{-4.6}$      & 1.1$^{13}$  &  -\\
    3352137969269081472 & 10.6  $\pm$0.6   & 8.8   $\pm$5.8$^{c}$   & no    & -     &   3.0\,$^{+3.0}_{-1.5}$     & 0.9$^{13}$ & K7$^{1}$\\
    3350774471768004736 & 10.0  $\pm$0.5   & - & -       & 0.7 &1.0\,$^{+8.0}_{-0.6}$ &1.0$^{13}$  & -\\
    3159113274168209152 & 9.9   $\pm$0.7   & - & -  & 3.9 & 4.0$\pm$0.5 & 2.0--2.1$^{12,13}$  & A2$^{8}$ \\
    3327841717229545856 & 9.9   $\pm$0.5   & 16.9  $\pm$6.4$^{a}$   & no    & -     &   7.0\,$\pm$3.0    & 1.5--1.9$^{12,13}$  & F0$^{9}$\\
    3355872628012737152 & 9.7   $\pm$2.9   & - & -      & 4.8 & 20.0\,$^{+25.0}_{-16.5}$  & 0.5$^{12}$ & -\\
    3326637442758920960 & 9.7   $\pm$0.6   & -1.4  $\pm$2.9$^{b}$   & no    & - & 2.0\,$\pm$0.5 & 0.3$^{4}$ & M3$^{2,4}$\\ 
    3326495017346142592 & 8.2   $\pm$1.5   & - & -      & 1.7 &0.4\,$^{+1.5}_{-0.2}$& - & -\\
    3133869616772191232 & 7.8   $\pm$0.9   & - & -      & 2.8 &2.7\,$^{+2.4}_{-1.4}$& 0.6$^{12}$ & -\\
    3351891609942897664 & 7.7   $\pm$0.6   & - & -       & 2.6* & 1.5\,$^{+0.8}_{-0.5}$& 1.0$^{13}$ & -\\
    3327882841540580224 & 7.6   $\pm$1.9   & - & -      & 4.4 & 0.5\,$^{+1.6}_{-0.4}$& - & -\\
	3326894591041117696 & 6.8   $\pm$1.1   & - & -       & 0.7 & 2.0\,$^{+6.0}_{-1.3}$& 1.8$^{13}$ & -\\
    3326994406080500608 & 6.8   $\pm$0.6   & - & -       & 1.3 & 3.0\,$^{+18.0}_{-1.3}$& 0.9$^{13}$ & -\\
    3326492028048885120 & 6.6   $\pm$2.2   & - & -       & 1.9 & 2.0\,$^{+2.0}_{-1.5}$& 0.3$^{12}$ & -\\
    3326589025592180864 & 6.2   $\pm$1.7   & - & -   & 0.9 &1.8\,$^{+6.2}_{-1.6}$& - & -\\
	3327008935952524032 & 6.1   $\pm$0.6   & - & -  & 0.9 & 3.5\,$^{+3.5}_{-1.9}$ & 1.5--1.8$^{12,13}$ & -\\
    3326909670670846976 & 6.0   $\pm$1.9   & - & -       & 0.1 &30.0\,$^{+120.0}_{-26.0}$& 0.6$^{12}$ & - \\
    \hline
\end{tabular}
\end{table*}   


\begin{table*} 
    \renewcommand\arraystretch{1.2}
	\begin{tabular}{lccccccc} 
	    \textbf{Table B2} - continued\\
		\hline
		\textit{Gaia} DR2 source-id & 2D-velocity rf & Radial velocity rf & 3D-cand. & Flight time & Iso. age & Mass & Spectral type  \\
		 & (km\,s$^{-1}$) & (km\,s$^{-1}$) & & (Myr) & (Myr) &  (M$_{\sun}$) &\\
		\hline
        3326707154372788480 & 5.9   $\pm$1.9   & - & -      & 0.3 &3.5\,$^{+26.0}_{-3.0}$& 0.4$^{12}$ & -\\
        3351060477934502016 & 5.8   $\pm$1.8   & - & -      & 1.7 &2.0\,$^{+2.0}_{-1.6}$& - & -\\
        3134320416539722368 & 5.8   $\pm$1.7   & - & -      & 2.7 &2.0\,$^{+2.0}_{-1.6}$& 0.7$^{12}$ & -\\
        3326525975471926656 & 5.8   $\pm$1.8   & - & -       & 1.7 &3.5\,$^{+17.0}_{-2.1}$& 0.4$^{12}$ & -\\
        3326629814897031040 & 5.6   $\pm$3.6   & - & -      & 1.4 &1.2\,$^{+8.8}_{-1.1}$& - & -\\
        3326581191571739648 & 5.5   $\pm$1.0   & - & -       & 1.5 &1.0\,$^{+4.3}_{-0.7}$& - & -\\
        3327852815425166592 & 5.4   $\pm$0.7   & - & -       & 3.2 &9.0\,$^{+11.0}_{-7.0}$& - & -\\
        3326690661698141056 & 5.4   $\pm$3.3   & - & -      & 0.4 &15.0\,$^{+85.0}_{-14.0}$& 0.8$^{12}$ & -\\
        3134443016384346496 & 5.2   $\pm$1.4   & - & -       & 2.9 &2.0\,$^{+4.0}_{-1.0}$& 0.5$^{12}$ & -\\
        3326641428488508928 & 5.1   $\pm$1.1   & - & -       & 1.2 &7.5\,$^{+22.5}_{-4.5}$& 0.2--0.5$^{4,12}$ & M4$^{4}$\\
        3326703688334117120 & 5.0   $\pm$2.0   & - & -   & 0.3 &20.0\,$^{+75.0}_{-18.0}$& 0.6 & -\\
        \hline
	\end{tabular}
\end{table*}

\begin{table*}
    \renewcommand\arraystretch{1.2}
	\centering
	\caption{IRS\,1/2 RW star 2D candidates sorted by decreasing 2D-velocity. Column 2+3: velocity in respective IRS rest frame [rf]; Column 3: RV sources - $^{a}$\textit{Gaia} DR2, $^{b}$\citet{2016A&A...586A..52J}, $^{c}$\citet{2019yCat.5164....0L}, $^{d}$\citet{2019AJ....157..196K}, $^{e}$\citet{1992A&AS...95..541F}, 	$^{f}$\citet{1995A&AS..114..269D}; Column 4: indication of 3D-candidate status; Column 5: minimum flight time since ejection (crossing of search boundary); Column 6: age from PARSEC isochrones \citep{RN225}; Column 7: Subcluster identification; Column 8--9: from literature sources - $^{1}$\citet{2019yCat.5164....0L}, $^{2}$\citet{2015A&A...581A..66V}, $^{3}$\citet{2017A&A...599A..23V}, $^{4}$\citet{2014A&A...570A..82V}, $^{5}$\citet{2019AJ....157..196K}, $^{6}$\citet{2004A&A...417..557L}, $^{7}$\citet{1972A&AS....7...35K},
	$^{8}$\citet{1993yCat.3135....0C}, $^{9}$\citet{1985cbvm.book.....V}, 
	$^{10}$\citet{2001A&A...373..625P}, $^{11}$\citet{2002AJ....123.1528R}, 
	$^{12}$\citet{2019AJ....158..138S}, $^{13}$\citet{2019A&A...628A..94A}, 
	$^{14}$\citet{2018A&A...609A..10V}.}
	\label{tab:RWC_2D_IRS_app}
	\begin{tabular}{lcccccccc} 
		\hline
		\textit{Gaia} DR2 source-id & 2D-velocity rf & Radial velocity rf & 3D-cand. & Flight time & Iso. age & Subcluster & Mass &  Spectral type  \\
		 & (km\,s$^{-1}$) & (km\,s$^{-1}$) & & (Myr) & (Myr) &  & (M$_{\sun}$) & \\
		\hline
        3326689188524905344  & 64.2  $\pm$0.8    & -1.3   $\pm$3.1$^{a}$   &   no    & -   &  4.0\,$^{+8.0}_{-3.5}$      & IRS\,1  &      1.0$^{13}$&       G7$^{1}$ \\
        3326689188524905344  & 63.4  $\pm$0.7    & -1.3   $\pm$3.1$^{a}$   & no    &-   &     4.0\,$^{+8.0}_{-3.5}$  & IRS\,2  &     1.0$^{13}$&           G$^{1}$ \\
         3326639882300319744  & 47.2  $\pm$1.1    & -16.8  $\pm$2.9$^{b}$   & no    & -   &   4.0\,$^{+16.0}_{-3.0}$     & IRS\,2  &      0.9--1.1$^{12,13}$&    -    \\
        3134341650859166208  & 40.5  $\pm$1.1    & -     & -     & 0.3   &  6.0\,$^{+25.0}_{-4.0}$      & IRS\,2  &       0.9$^{13}$&     -    \\
        3134341650859166208  & 39.1  $\pm$1.2    & -     & -     & 0.3   &  6.0\,$^{+25.0}_{-4.0}$      & IRS\,1  &            0.9$^{13}$& -   \\
        3326519142179922688  & 39.0    $\pm$1.0    & -     & -     & 0.2   &  12.0\,$^{+28.0}_{-11.0}$      & IRS\,2  &         0.9--1.0$^{12,13}$&    -   \\
        3326626928678945408  & 38.1  $\pm$0.8    & -     & -     & 0.1   &  3.8\,$^{+26.2}_{-3.0}$      & IRS\,2  &     0.7--0.9$^{12,13}$&     -     \\
        3326519142179922688  & 37.6  $\pm$1.1    & -     & -     & 0.1   & 12.0\,$^{+28.0}_{-11.0}$       & IRS\,1  &        0.9--1.0$^{12,13}$&     -   \\
        3355468591847181440  & 37.1  $\pm$0.8    & -     & -     & 1.3   &   3.5\,$^{+2.5}_{-1.5}$     & IRS\,2  &    1.0$^{13}$ &  -          \\
        3326626928678945408  & 36.9  $\pm$0.8    & -     & -     & 0.1   &  3.8\,$^{+26.2}_{-3.0}$     & IRS\,1  &       0.7--0.9$^{12,13}$&   -     \\
        3134452465312268032  & 32.9  $\pm$0.8    & -    & -     & 0.4   & 20.0\,$^{+10.0}_{-18.0}$    & IRS\,2  &          1.1$^{12,13}$ &   -  \\
        3355801022316951552  & 32.7  $\pm$3.3    & -     & -     & 1.7   &  100.0\,$^{+100.0}_{-98.0}$       & IRS\,1  &         0.6$^{12}$&    -   \\
        3355801022316951552  & 31.6  $\pm$3.2    & -     & -     & 1.7   &   100.0\,$^{+100.0}_{-98.0}$     & IRS\,2  &          0.6$^{12}$ &    -  \\
        3352821109587150592  & 26.4  $\pm$0.8    & 37.5   $\pm$7.0$^{a}$   & no    & 1.8   &     4.5\,$^{+7.5}_{-2.5}$    & IRS\,2  &        0.9--1.0$^{12,13}$ &   -     \\
        3133279939241456000  & 23.8  $\pm$0.7    & -50.6  $\pm$5.3$^{a}$   & no    & -   &     9.0\,$^{+6.0}_{-7.0}$   & IRS\,2  &      1.0$^{12,13}$ &   -      \\
        3133279939241456000  & 22.6  $\pm$0.9   & -50.5  $\pm$5.3$^{a}$   &   no    & -   &  9.0\,$^{+6.0}_{-7.0}$      & IRS\,1  &         1.0$^{12,13}$ &   -    \\
        3328003654677416192  & 20.2  $\pm$0.9    & 29.1   $\pm$3.2$^{a}$   &  no     & -   &   0.2\,$^{+0.8}_{-0.1}$    & IRS\,1  &        1.1$^{13}$ &   -     \\
        3328003654677416192  & 18.9  $\pm$0.7    & 29.1   $\pm$3.2$^{a}$   & no    & -   &    0.2\,$^{+0.8}_{-0.1}$   & IRS\,2  &         1.1$^{13}$ &      - \\
        3326698289559319936  & 6.1   $\pm$0.7    & -31.0  $\pm$11.7$^{a}$  & no    & -  &     10.0\,$^{+10.0}_{-9.0}$  & IRS\,2  &     1.0--1.7$^{4,12,13}$&   G5$^{2,4}$     \\ 
    \hline
	\end{tabular}
\end{table*}

\begin{table*}
    \renewcommand\arraystretch{1.2}
	\centering
	\caption{IRS\,1/2 WW star 2D candidates. Column 2+3: velocity in respective IRS rest frame [rf]; Column 3: RV sources - $^{a}$\textit{Gaia} DR2, $^{b}$\citet{2016A&A...586A..52J}, $^{c}$\citet{2019yCat.5164....0L}, $^{d}$\citet{2019AJ....157..196K}, $^{e}$\citet{1992A&AS...95..541F}, 
	$^{f}$\citet{1995A&AS..114..269D}; Column 4: indication of 3D-candidate status; Column 5: minimum flight time since ejection (crossing of search boundary); Column 6: age from PARSEC isochrones \citep{RN225}; Column 7: Subcluster identification; Column 8--9: from literature sources - $^{1}$\citet{2019yCat.5164....0L}, $^{2}$\citet{2015A&A...581A..66V}, $^{3}$\citet{2017A&A...599A..23V}, $^{4}$\citet{2014A&A...570A..82V}, $^{5}$\citet{2019AJ....157..196K}, $^{6}$\citet{2004A&A...417..557L},
	$^{7}$\citet{1972A&AS....7...35K},
	$^{8}$\citet{1993yCat.3135....0C}, 
	$^{9}$\citet{1985cbvm.book.....V}, 
	$^{10}$\citet{2001A&A...373..625P}, 
	$^{11}$\citet{2002AJ....123.1528R}, 
	$^{12}$\citet{2019AJ....158..138S},
	$^{13}$\citet{2019A&A...628A..94A},
	$^{14}$\citet{2018A&A...609A..10V}.}
	\label{tab:WWC_2D_IRS_app}
	\begin{tabular}{lccccccccr} 
		\hline
		\textit{Gaia} DR2 source-id & 2D-velocity rf & Radial velocity rf & 3D-cand. & Flight time & Iso. age & Subcluster  & Mass & Spectral type  \\
		 & (km\,s$^{-1}$) & (km\,s$^{-1}$) & & (Myr) & (Myr) & (M$_{\sun}$) & \\
		\hline
        3326691726849639168 & 23.8  $\pm$2.0   & - & -     & 0.0   &  6.0\,$^{+100.0}_{-5.0}$     & IRS\,2  &         0.8$^{13}$&-\\
        3326691726849639168 & 22.5  $\pm$2.0   & - & -     & 0.0   &  6.0\,$^{+100.0}_{-5.0}$     & IRS\,1  &      0.8$^{13}$ & -\\
        3326641462848241792 & 18.6  $\pm$0.7   & - & -     & 0.2   &  7.0\,$^{+25.0}_{-5.5}$      & IRS\,2  &  0.8--1.0$^{12,13}$    &  -\\
        3350980149161701888 & 17.1  $\pm$0.9   & -16.6 $\pm$6.4$^{a}$   & no    &  -      &   10.0\,$^{+5.0}_{-8.5}$    & IRS\,1  &   1.5$^{12,13}$    & F0$^{1}$ \\
        3350980149161701888 & 16.4  $\pm$0.7   & -16.6 $\pm$6.4$^{a}$   & no    &   -    &    10.0\,$^{+5.0}_{-8.5}$   & IRS\,2  &   1.5$^{12,13}$    & - \\
        3134369658343616640 & 15.4  $\pm$0.8   & - & -     & 1.2   &  4.0\,$^{+16.0}_{-3.5}$     & IRS\,2  &  1.0--1.1$^{12,13}$     &  -\\
        3134369658343616640 & 13.9  $\pm$0.9   & - & -     & 1.2   &  4.0\,$^{+16.0}_{-3.5}$    & IRS\,1  &   1.0--1.1$^{12,13}$    & - \\
        3326644138611656832 & 12.6  $\pm$0.7   & 18.7  $\pm$3.1$^{a}$   & no    &   -    &   5.0\,$^{+6.0}_{-4.6}$     & IRS\,2  &   1.1$^{13}$    & &- \\
        3326644138611656832 & 11.4  $\pm$0.8   & 18.8  $\pm$3.1$^{a}$   & no    & -      &   5.0\,$^{+6.0}_{-4.6}$     & IRS\,1  &   1.1$^{13}$    & &- \\
        3326637442758920960 & 10.9  $\pm$0.8   & -1.3  $\pm$2.9$^{b}$   & no    &  -     &   2.0\,$\pm$0.5    & IRS\,2  &  0.3$^{4}$ & M3$^{2,4}$\\
        3326637442758920960 & 9.8  $\pm$0.9   & -1.2  $\pm$2.9$^{b}$   & no    &  -   &   2.0\,$\pm$0.5    & IRS\,1  &  0.3$^{4}$ & M3$^{2,4}$\\ 
        3326704238089925120 & 9.3   $\pm$0.6   & -7.9  $\pm$2.9$^{d}$   & no   &   -   &   0.5\,$^{+3.}_{-0.4}$     & IRS\,2  &  0.6--1.0$^{4,13}$  &  M0$^{2,4}$ \\
        3326685512032888320 & 8.5   $\pm$1.0   & 13.4  $\pm$10.4$^{a}$   & no    &   -    &   3.0\,$^{+3.0}_{-1.5}$     & IRS\,1  &  1.2--2.1$^{4,12,13}$   & F5/G0$^{4,6}$ \\         
        3326480865429886720 & 8.5   $\pm$2.1   & -  & -     & 1.4& 2.0\,$^{+3.0}_{-1.6}$   &  & IRS\,2 &-&-\\
        3326480865429886720 & 7.7   $\pm$2.2   & -  & -     & 1.3& 2.0\,$^{+3.0}_{-1.6}$     & IRS\,1 &-&-\\
        3326495017346142592 & 6.9   $\pm$1.6   & -  & -     & 1.8& 0.4\,$^{+1.5}_{-0.2}$     & IRS\,1 & -&-\\
        3326634212943546496 & 6.7   $\pm$0.9   & -18.5 $\pm$5.0$^{b}$   & no    & -     & 2.0\,$^{+3.5}_{-1.3}$     & IRS\,2 & 1.8--2.5$^{12,13}$&-\\
        3327008935952524032 & 6.6   $\pm$0.8   & -  & -     & 1.7& 3.5\,$^{+3.5}_{-1.9}$    & IRS\,1 & 1.5--1.8$^{12,13}$ &-\\
        3326629814897031040 & 6.6   $\pm$3.7   & -  & -     & 0.5& 1.2\,$^{+8.8}_{-1.1}$     & IRS\,1 &-&-\\
        3134443016384346496 & 6.4   $\pm$1.5   & -  & -     & 1.6& 2.0\,$^{+4.0}_{-1.0}$     & IRS\,2 & 0.5$^{12}$&-\\
        3326589025592180864 & 6.0   $\pm$1.8   & -  & -     & 0.5& 1.8\,$^{+6.2}_{-1.6}$     & IRS\,2 & &-&- \\
        3326692620203367168 & 6.0   $\pm$2.9   & -  & -     & in cluster& 6.0\,$^{+46.0}_{-5.5}$     & IRS\,1 & 0.6$^{12}$ &-\\
        3326684549960233728 & 5.9   $\pm$1.0   & 12.7  $\pm$2.9$^{b}$   & no    & -     & 10.0\,$^{+25.0}_{-9.0}$      & IRS\,1 & 0.9--1.1$^{12,13}$ &-\\
        3326492028048885120 & 5.7   $\pm$2.2   & -  & -     & 1.5& 2.0\,$^{+2.0}_{-1.5}$     & IRS\,2 & 0.3$^{12}$&- \\
        3326690661698141056 & 5.7   $\pm$3.3   & -  & -     & in cluster& 15.0\,$^{+85.0}_{-14.0}$     & IRS\,2 & 0.8$^{12}$&-\\
        3350961560543556352 & 5.6   $\pm$1.7   & -  & -     & 1.7& 10.0\,$^{+38.0}_{-8.5}$     & IRS\,1 &-&-\\
        3350961560543556352 & 5.1   $\pm$1.6   & -  & -     & 1.6& 10.0\,$^{+38.0}_{-8.5}$      & IRS\,2 &-&-\\
        3326492028048885120 & 5.1   $\pm$2.3   & -  & -     & 2.0& 2.0\,$^{+2.0}_{-1.5}$     & IRS\,1 & 0.3$^{12}$\\
        3326589025592180864 & 5.1   $\pm$1.9   & -  & -     & 0.4& 1.8\,$^{+6.2}_{-1.6}$     & IRS\,1 &-&- \\
        3326678193408399232 & 3.6   $\pm$0.7   & -16.2 $\pm$4.5$^{a}$   & no   & 1.5   &    2.8\,$^{+2.2}_{-1.5}$    & IRS\,2  &    1.9   & F0$^{1}$ \\
        \hline
	\end{tabular}
\end{table*}

\section{Past visitors to NGC 2264}\label{App_visitors}

Table~\ref{tab:Past_visitors_list} provides information on sources that can be successfully traced back in 3D to NGC 2264 in the past 2/5 Myr (upper age limit for IRS\,1/2 and S\,Mon respectively) but their position on the CAMD identifies them as MS stars or older pre-MS stars.

\textit{Gaia} DR2 3327203588170236672 is a past visitor both in IRS\,1/2 but an ejected WW from S\,Mon, it is therefore older than IRS\,1/2, but young enough to have originated from S\,Mon. The same applies to \textit{Gaia} DR2 3134179335455713408, which is a past visitor to IRS\,1, but is shown to have been ejected from S\,Mon.

\begin{table*}
	\caption{Past visitors to NGC 2264 (3D trace-backs). Column 2+3: velocity in the respective NGC 2264 rest frame [rf]; Column 3: RV sources -- $^{a}$\textit{Gaia} DR2; Column 4: age from PARSEC isochrones \citep{RN225}; Column 5: from literature sources -- $^{1}$\citet{1985cbvm.book.....V}, $^{2}$\citet{1959ApJS....4...23M}, $^{3}$\citet{1993yCat.3135....0C},  $^{4}$\citet{2014A&A...570A..82V},
	$^{5}$\citet{1972A&AS....7...35K}; Column 6: Subcluster identification.}
	\label{tab:Past_visitors_list}
	\begin{tabular}{lcccccc} 
		\hline
		\textit{Gaia} DR2 source-id & 2D-velocity rf & Radial velocity rf & Iso. age & Spectral type &Subcluster \\
		 & (km\,s$^{-1}$) & (km\,s$^{-1}$) & (Myr)  & \\
		\hline
        Visitors at RW velocities\\		
		\hline
		S\,Mon \\
		\hline
        3128924155208294656 &	50.7	$\pm$0.7	&	-36.3	$\pm$4.7$^{a}$& $\sim$10--15 & - & -\\
        3326608919880182144 &	46.5	$\pm$0.6	&	34.5	$\pm$3.7$^{a}$& $\sim$6-12 &  B3/5$^{5}$ & -  \\
        3127637722304864000 &	44.7	$\pm$0.7	&	9.4	$\pm$4.2$^{a}$& $\sim$8-12 & - & -  \\
        3133406275699512320 &	37.8	$\pm$0.6	&	22.4	$\pm$3.4$^{a}$& $\sim$6--11 & - & -  \\
        3129614889029245696 &	35.0	$\pm$0.9	&	2.2	$\pm$6.3$^{a}$& $\sim$?8--12 & - & -  \\
        3129675293448445440 &	31.0	$\pm$0.7	&	-14.9	$\pm$5.3$^{a}$& $\sim$6--12 & - & -  \\
        3157158583015446016 &	30.8	$\pm$0.7	&	-16.7	$\pm$7.7$^{a}$& $\sim$13--14 & - & -  \\
        3131005736878997248 &	25.8	$\pm$0.5	&	-27.4	$\pm$3.0$^{a}$& $\sim$10--12 & - & -  \\
        3352025681642829952 &	23.2	$\pm$0.6	&	36.8	$\pm$3.4$^{a}$& $\sim$8-11 & - & -  \\
        3159544931266074880 &	21.2	$\pm$0.9	&	21.3	$\pm$6.3$^{a}$& $\sim$13--14 & - & -  \\
        3355756320296166528 &	12.9	$\pm$0.7	&	-29.7	$\pm$7.9$^{a}$& $\sim$7--10 & - & -  \\
		\hline
		IRS\,1 / IRS\,2\\
		\hline
	    3128924155208294656 & 52.0  $\pm$0.9   & -35.9 $\pm$4.7$^{a}$   &      $\sim$10--12&    - & IRS\,2       \\
        3128924155208294656 & 50.8  $\pm$1.1   & -35.7 $\pm$4.7$^{a}$   &  $\sim$10--12     &     - &  IRS\,1      \\
        3357475892062058752 & 46.1  $\pm$1.2   & -17.9 $\pm$6.3$^{a}$   &  $\sim$10--12     &       -    &IRS\,1      \\
        3357475892062058752 & 45.5  $\pm$1.0   & -17.9 $\pm$6.3$^{a}$   &      $\sim$10--12 &       - & IRS\,2      \\
        3127637722304864000 & 45.1  $\pm$0.9   & 9.7   $\pm$4.2$^{a}$   &     $\sim$8-12 &    -    &IRS\,2      \\
        3127637722304864000 & 44.1  $\pm$1.1   & 9.8   $\pm$4.2$^{a}$   & $\sim$8-12      &         - &IRS\,1        \\
        3133406275699512320 & 37.6  $\pm$0.9   & 22.6  $\pm$3.4$^{a}$   &      $\sim$6-11 &         - & IRS\,2    \\
        3132201314034932864 & 37.3  $\pm$1.0   & -34.2 $\pm$5.4$^{a}$   &   $\sim$3--11    &         - & IRS\,1       \\
        3133406275699512320 & 36.6  $\pm$1.1   & 22.7  $\pm$3.4$^{a}$   &  $\sim$6-11     &      - & IRS\,1        \\
        3132362529927264128 & 34.6  $\pm$0.9   & 1.0   $\pm$3.5$^{a}$   &      $\sim$6--10 &        - & IRS\,2     \\
        3132362529927264128 & 33.4  $\pm$1.1   & 1.1   $\pm$3.5$^{a}$   &  $\sim$6--10     &        - & IRS\,1        \\
        3129675293448445440 & 32.7  $\pm$0.9   & -14.7 $\pm$5.3$^{a}$   &      $\sim$6--11 &        - & IRS\,2      \\
        3129675293448445440 & 31.5  $\pm$1.1   & -14.6 $\pm$5.3$^{a}$   &   $\sim$6--11    &        - & IRS\,1        \\
        3129273692526220544 & 31.1  $\pm$0.9   & -50.0 $\pm$3.7$^{a}$   &     $\sim$10--13  &        - & IRS\,2     \\
        3352775922231300992 & 27.6  $\pm$1.1   & 19.3  $\pm$6.9$^{a}$   &     $\sim$8--12  &        - & IRS\,2     \\
        3352025681642829952 & 24.4  $\pm$1.1   & 36.8  $\pm$3.4$^{a}$   &  $\sim$8-11    &        - & IRS\,1       \\
        \hline
	    Visitors at WW velocities\\
		\hline
		S\,Mon \\
		\hline
        3326203616704192896	&	25.1	$\pm$0.6	&	13.3	$\pm$3.3$^{a}$& $\sim$5--11 & - & -\\
        3354418768701989248	&	18.8	$\pm$0.7	&   -5.6	$\pm$5.0$^{a}$& $\sim$8--12 & - & -\\
        3331208765431910528	&	18.6	$\pm$0.8	&	22.9	$\pm$5.5$^{a}$ & $\sim$5--12 & - & -\\
        3160409491002912512	&	16.2	$\pm$0.8	&	-9.5	$\pm$5.1$^{a}$& $\sim$11         & -  & -\\
        3158315471107784960	&	12.6	$\pm$0.5	&	-27.2	$\pm$3.2$^{a}$& $\sim$4--11 & -  & -\\
        3325585416291697408	&	11.0	$\pm$1.3	&	-8.3	$\pm$16.4$^{a}$& $\sim$8--12 & -  & -\\
		\hline
		IRS\,1 / IRS\,2 \\
		\hline
        3326616410302784256 & 20.5  $\pm$0.9   & 17.2  $\pm$6.5$^{a}$   &      $\sim$10--12 &         - & IRS\,2    &  \\
        3326616410302784256 & 19.5  $\pm$1.1   & 17.3  $\pm$6.5$^{a}$   &      $\sim$10--12 &        - & IRS\,1     &  \\
        3134179335455713408 & 13.3  $\pm$1.0   & 26.6  $\pm$6.6$^{a}$   &      $\sim$3--10 &        - &  IRS\,2    &  \\
        3134179335455713408 & 11.9  $\pm$1.2   & 26.7  $\pm$6.6$^{a}$   &     $\sim$3--10  &         - & IRS\,1    &  \\
		 3327203588170236672 & 11.1  $\pm$0.9  & -7.1  $\pm$5.8$^{a}$  & $\sim$4--15      &        - & IRS\,2     &  \\
        3326689807000188032 & 0.7   $\pm$0.8   & 8.7   $\pm$5.7$^{a}$   &  $\sim$2--10 & F5$^{4}$& IRS\,1 \\
        3326689807000188032 & 0.6   $\pm$0.7   & 8.6   $\pm$5.7$^{a}$   &  $\sim$2--10 & F5$^{4}$& IRS\,2\\
       \hline
	\end{tabular}
\end{table*}

\section{Future visitors to NGC 2264}\label{App_visitors_fut}

Tables~\ref{tab:Visitors_list_fut_RW} and \ref{tab:Visitors_list_fut_WW} provide information on sources that can be successfully traced forward in 3D to NGC 2264 in the future 5--8 Myr (these values are based on the assumption of a lifetime of 10 Myr for each subcluster).

\begin{table*}
	\caption{Future 3D-RW visitors to NGC 2264. Column 2+3: velocity in the respective NGC 2264 rest frame [rf]; Column 3: RV sources -- $^{a}$\textit{Gaia} DR2; Column 4: age from PARSEC isochrones \citep{RN225}; Column 5: from literature sources -- $^{1}$\citet{1985cbvm.book.....V}, $^{2}$\citet{1959ApJS....4...23M}, $^{3}$\citet{1993yCat.3135....0C}, $^{4}$\citet{2014A&A...570A..82V}, $^{5}$\citet{2019MNRAS.490.3158C}.}
	\label{tab:Visitors_list_fut_RW}
	\begin{tabular}{lccccc} 
		\hline
		\textit{Gaia} DR2 source-id & 2D-velocity rf & Radial velocity rf & Iso. age &  Spectral type \\
		 & (km\,s$^{-1}$) & (km\,s$^{-1}$) & (Myr)   & \\
		\hline
        Visitors at RW velocities\\		
		\hline
		S\,Mon \\
		\hline
        3129232293342932864 & 41.4  $\pm$0.7   & 30.0  $\pm$5.0$^{a}$ & $\sim$6--11 & - \\
        3352221841389893888 & 30.9  $\pm$0.8   & -18.5 $\pm$9.4$^{a}$ & $\sim$1 & - \\
        3158670342784515584 & 30.2  $\pm$0.5   & -2.3  $\pm$4.8$^{a}$ & $\sim$1--6 & - \\
        3157467988160982912 & 26.6  $\pm$0.5   & 19.9  $\pm$3.4$^{a}$ & $\sim$13 & - \\
        3327867585815573376 & 26.3  $\pm$0.5   & 54.6  $\pm$3.2$^{a}$ & $\sim$1--10 & - \\
        3131017483608422016 & 21.8  $\pm$0.8   & -41.5 $\pm$4.6$^{a}$ & $\sim$10 & - \\
        3328077695617951360 & 17.0  $\pm$0.5   & 34.6  $\pm$3.8$^{a}$ & $\sim$2--10 & - \\
        3352147521276456320 & 16.1  $\pm$0.6   & 28.1  $\pm$7.0$^{a}$ & $\sim$12--13 & - \\
        3133971085379278080 & 12.5  $\pm$0.6   & -29.6 $\pm$5.0$^{a}$ & $\sim$1--4 & - \\
        3324391273644856448 & 9.7   $\pm$1.1   & -33.3 $\pm$14.3$^{a}$ & $\sim$7--11 & -\\
 		\hline
		IRS\,1 \\
		\hline
    	3129990986425585152 & 49.7  $\pm$0.8   & -4.1  $\pm$3.4$^{a}$   & $\sim$1--5     & -           \\
        3129232293342932864 & 42.1  $\pm$0.9   & 29.9  $\pm$5.0$^{a}$   &  $\sim$6--11     &   -         \\
        3327331883137624832 & 41.9  $\pm$0.8   & 71.3  $\pm$2.9$^{a}$   &     $\sim$0.5--1 &  -           \\
        3157240462272160128 & 37.3  $\pm$0.9   & 8.5   $\pm$6.6$^{a}$   &     $\sim$10--12  &      -          \\
        3132766016335294976 & 31.4  $\pm$0.8   & 7.7   $\pm$3.6$^{a}$   &   $\sim$13    &           -    \\
        3327867585815573376 & 26.3  $\pm$0.8   & 54.9   $\pm$3.2$^{a}$   &     $\sim$1--10  &         -       \\
        3356469564045105664 & 25.1  $\pm$0.9   & -18.9 $\pm$3.1$^{a}$   &    $\sim$7--13   &          -      \\
        3131017483608422016 & 22.6  $\pm$1.1   & -41.5 $\pm$4.6$^{a}$  & $\sim$10             &        - \\
        3324293520190299904 & 22.4  $\pm$1.0   & 22.1  $\pm$7.2$^{a}$  &     $\sim$10--12         & -       \\
        3157132675772801280 & 19.4  $\pm$0.9   & 33.1  $\pm$5.9$^{a}$   &     $\sim$12--13         & -      \\
        3134392335771085568 & 18.2  $\pm$0.9   & -31.2 $\pm$3.1$^{a}$   &     $\sim$2--10         &   -\\
        3351325357155320320 & 15.8  $\pm$0.9   & 30.8  $\pm$10.9$^{a}$  &      $\sim$10--14             &-  \\
        3352147521276456320 & 15.6  $\pm$0.9   & 28.4  $\pm$7.0$^{a}$   &   $\sim$12--13                  &-  \\
        3326892250283336448 & 15.2  $\pm$0.8   & 26.7 $\pm$4.7$^{a}$   &        $\sim$4--5            &  F5/7$^{5}$\\
        3351375835908945024 & 14.6  $\pm$1.1   & 26.5  $\pm$12.3$^{a}$  &       $\sim$9--11            &  -\\
        3324391273644856448 & 11.3  $\pm$1.3   & -33.2 $\pm$14.3$^{a}$   &  $\sim$7--11               &  -\\
		\hline
		IRS\,2 \\
 		\hline
        3129990986425585152 & 48.4  $\pm$0.7   & -4.1  $\pm$3.4$^{a}$   &    $\sim$1--5          &     -    \\
        3129232293342932864 & 40.9  $\pm$0.8   & 29.9  $\pm$5.0$^{a}$   &    $\sim$6--11          &    -     \\
        3157240462272160128 & 36.2  $\pm$0.8   & 8.5   $\pm$6.6$^{a}$   &     $\sim$10--12         &    -     \\
        3132766016335294976 & 30.4  $\pm$0.7   & 7.7   $\pm$3.6$^{a}$   &    $\sim$13           &    -     \\
        3327867585815573376 & 27.5  $\pm$0.6   & 54.8  $\pm$3.2$^{a}$   &      $\sim$1--10        &    -     \\
        3134392335771085568 & 17.0  $\pm$0.8   & -31.2 $\pm$3.1$^{a}$   &    $\sim$2--10          &   -      \\
        3352147521276456320 & 16.8  $\pm$0.8   & 28.3  $\pm$7.0$^{a}$  &     $\sim$12--13         &   -     \\
        3324391273644856448 & 11.4  $\pm$1.2   & -33.3 $\pm$14.3$^{a}$  &    $\sim$7--11          &    -     \\
 		\hline
 	\end{tabular}
\end{table*}

\begin{table*}
	\caption{Future WW visitors to NGC 2264. Column 2+3: velocity in the respective NGC 2264 rest frame [rf]; Column 3: RV sources -- $^{a}$\textit{Gaia} DR2; Column 4: age from PARSEC isochrones \citep{RN225}; Column 5: from literature sources -- $^{1}$\citet{1985cbvm.book.....V}, $^{2}$\citet{1959ApJS....4...23M}, $^{3}$\citet{1993yCat.3135....0C}, $^{4}$\citet{2014A&A...570A..82V}, $^{5}$\citet{2019MNRAS.490.3158C}, $^{6}$\citet{2018A&A...620A..87M}.}
	\label{tab:Visitors_list_fut_WW}
	\begin{tabular}{lcccc} 
		\hline
		\textit{Gaia} DR2 source-id & 2D-velocity rf & Radial velocity rf & Iso. age & Spectral type \\
		 & (km\,s$^{-1}$) & (km\,s$^{-1}$) & (Myr) &   \\
		\hline 		
	    Visitors at WW velocities\\
		\hline
		S\,Mon\\
		\hline
        3330828020872121472 & 23.2  $\pm$0.7   & -3.4  $\pm$6.1$^{a}$ & $\sim$7--10 & - \\
        3324278745502788992 & 22.8  $\pm$0.7   & 6.9   $\pm$6.6$^{a}$ & $\sim$9--11 & - \\
        3355407985568310912 & 21.7  $\pm$1.5   & -6.6  $\pm$19.7$^{a}$ & $\sim$11-13 & - \\
        3130189418211411712 & 20.9  $\pm$0.8   & -9.1  $\pm$4.6$^{a}$ & $\sim$9--11 & - \\
        3327020240307406464 & 19.9  $\pm$0.5   & -7.7  $\pm$5.2$^{a}$ & $\sim$10--11 & - \\
        3133552068369881984 & 15.7  $\pm$0.5   & 17.8  $\pm$3.6$^{a}$ & $\sim$1--10 & - \\
        3327620642374686848 & 13.9  $\pm$0.6   & -23.8 $\pm$6.8$^{a}$ & $\sim$6--11 & B9$^{1}$ \\
        3325784565337792896 & 11.3  $\pm$0.5   & 9.9   $\pm$4.7$^{a}$ & $\sim$20 & - \\
        3351090405265460736 & 9.2   $\pm$0.7   & -7.5  $\pm$14.5$^{a}$ & $\sim$1--10 & - \\
        3158447073205530240 & 7.6   $\pm$0.5   & -4.2  $\pm$4.6$^{a}$ &$\sim$13 &  -\\
        3327865528528223872 & 7.0   $\pm$0.7   & 10.2  $\pm$5.3$^{a}$ & $\sim$1--5 & -\\
		\hline
		IRS\,1 \\
		\hline
	    3157987649142772480 & 25.0  $\pm$0.9   & 10.3  $\pm$3.3$^{a}$   &  $\sim$5--10     &   -   \\
        3153460646235577728 & 24.9  $\pm$1.0   & -3.0  $\pm$4.2$^{a}$   &   $\sim$5--12    &       - \\
        3130189418211411712 & 21.8  $\pm$1.0   & -9.1  $\pm$4.6$^{a}$ & $\sim$9--11 & - \\
        3327020240307406464 & 19.9  $\pm$0.8   & -7.4  $\pm$5.2$^{a}$   &    $\sim$10--11   &      -  \\
        3358899312943684992 & 19.2  $\pm$0.9   & 6.1   $\pm$3.2$^{a}$   &   $\sim$0.3--2    &        -\\
        3158958277392687744 & 19.0  $\pm$0.9   & -9.7  $\pm$5.4$^{a}$   &   $\sim$6--9    &       - \\
        3355361462483283968 & 18.4  $\pm$0.8   & 15.6  $\pm$4.2$^{a}$   &    $\sim$10-11   &      -   \\
        3353006308573860736 & 18.0  $\pm$0.8   & -7.8  $\pm$3.2$^{a}$   &  $\sim$0.3--2     &     -    \\
        3327106792487405696 & 17.2  $\pm$0.9   & 12.5  $\pm$6.9$^{a}$   &   $\sim$5--12    &    -    \\
        3133552068369881984 & 16.5  $\pm$0.8   & 17.8  $\pm$3.6$^{a}$ & $\sim$1--10 & - \\
        3133854739004375040 & 14.6  $\pm$0.8   & 2.1   $\pm$10.4$^{a}$   &   $\sim$5--6    &       A5/7$^{5}$\\
        3327963415126752896 & 12.9  $\pm$0.9   & 17.7   $\pm$3.5$^{a}$   &  $\sim$9--11     &    -    \\
        3330780157758632448 & 12.0  $\pm$1.4   & 1.0   $\pm$12.1$^{a}$  &    $\sim$11-13   &   -      \\
        3134016607734184832 & 11.5 $\pm$1.1  & -1.9  $\pm$18.0$^{a}$  &   $\sim$11-13    &    -     \\
        3356244778336418176 & 9.1 $\pm$2.1  & -10.9  $\pm$19.6$^{a}$  &   $\sim$5--10    &     -    \\
        3131647714228692224 & 8.7 $\pm$1.6  & -25.0  $\pm$14.1$^{a}$  &   $\sim$10--11    &        A2$^{1}$ \\
        3129542694918227072 & 8.1 $\pm$1.4  & 18.9  $\pm$12.2$^{a}$  &  $\sim$10--12     &     -   \\
        3153947145772918016 & 8.1   $\pm$1.1   & -5.7  $\pm$7.1$^{a}$  &  $\sim$14     &   -    \\
        3132900332851633024 & 7.6   $\pm$0.9  & -2.0  $\pm$3.3$^{a}$  &   $\sim$10--11    &     -  \\
        3331424505931639680 & 7.1   $\pm$1.1   & -11.2 $\pm$7.0$^{a}$   & $\sim$5--8      &  -     \\
        3351516195439957632 & 7.1   $\pm$0.9   & 2.6 $\pm$4.3$^{a}$   &   $\sim$4--10    &    -   \\
        3133726581481375872 & 5.5   $\pm$0.8  & 11.4  $\pm$3.1$^{a}$   &  $\sim$2--5     &   -  \\
        3330679453658587904 & 5.5   $\pm$1.0  & 2.6 $\pm$7.4$^{a}$   & $\sim$10--12      &     -    \\
        3326687878559500288 & 2.0   $\pm$0.9   & 9.8   $\pm$13.4$^{a}$  & $\sim$0.5--13      &   K0$^{6}$  \\
		\hline
		IRS\,2 \\
		\hline
		    3324278745502788992 & 24.3  $\pm$0.8   & 7.0   $\pm$6.6$^{a}$   & $\sim$9--11      &   -     \\
            3157987649142772480 & 24.2  $\pm$0.8   & 10.3  $\pm$3.3$^{a}$   &  $\sim$5--10     &  -    \\
            3153460646235577728 & 23.7  $\pm$0.9   & -3.0  $\pm$4.2$^{a}$   & $\sim$5--12      & -      \\
            3327020240307406464 & 21.2  $\pm$0.7   & -7.5  $\pm$5.2$^{a}$   & $\sim$10--11      &       - \\
            3358899312943684992 & 20.2  $\pm$0.8   & 6.0   $\pm$3.2$^{a}$   &  $\sim$0.5--2     &      -   \\
            3355361462483283968 & 19.4  $\pm$0.7   & 15.5  $\pm$4.2$^{a}$   &  $\sim$10-11     &     -   \\
            3133854739004375040 & 13.8 $\pm$0.8   & 2.1   $\pm$10.4$^{a}$   & $\sim$5--6      &    -     \\
            3325784565337792896 & 13.5  $\pm$0.7   & 10.0  $\pm$4.7$^{a}$   & $\sim$20      &   -    \\
		\hline
	\end{tabular}
\end{table*}		
		
\begin{table*}
	\begin{tabular}{lcccc} 
	\textbf{Table D2} - continued\\
		\hline
		\textit{Gaia} DR2 source-id & 2D-velocity rf & Radial velocity rf & Iso. age & Spectral type \\
		 & (km\,s$^{-1}$) & (km\,s$^{-1}$) & (Myr) & \\
		\hline
            3330780157758632448 & 13.1  $\pm$1.3   & 1.0   $\pm$12.1$^{a}$  &  $\sim$11-13     &  -      \\
            3130581905206339200 & 10.6  $\pm$1.1   & 17.7  $\pm$8.7$^{a}$   &  $\sim$10     &      A0$^{5}$  \\
            3134016607734184832 & 10.3  $\pm$1.0   & -1.9  $\pm$18.0$^{a}$  &  $\sim$11-13     & -        \\
            3356244778336418176 & 10.2  $\pm$2.0  & -11.0  $\pm$19.6$^{a}$  &  $\sim$5--10     &    -     \\
            3331411036914202368 & 8.6  $\pm$1.1  & -1.6  $\pm$10.7$^{a}$  &   $\sim$9--11    &       -\\
            3331424505931639680 & 8.5   $\pm$0.9   & -11.2 $\pm$7.5$^{a}$   & $\sim$5--8      &    -    \\
            3326046042943212544 & 8.0   $\pm$0.9   & 0.5 $\pm$10.4$^{a}$   &  $\sim$6--11     &    -    \\
            3351516195439957632 & 7.0   $\pm$0.7   & 2.5 $\pm$4.3$^{a}$   & $\sim$4--10      &    -   \\
            3132900332851633024 & 6.3   $\pm$0.8  & -2.1 $\pm$3.3$^{a}$  &  $\sim$10--11     &    -     \\
            3158447073205530240 & 5.2   $\pm$0.7   & -4.1  $\pm$4.6$^{a}$ &$\sim$13  & - \\
            3326687878559500288 & 2.7   $\pm$0.7   & 9.7   $\pm$13.4$^{a}$  &  $\sim$0.5--13     &  -    \\ 
		\hline
	\end{tabular}
\end{table*}

\section{Red giants visiting NGC 2264}\label{red_giants}

Table~\ref{tab:Red_giant_visitors} provides information on sources that can be successfully traced to NGC 2264 in 2D and/or 3D, however, are located on or near the red giant branch and are therefore at the end of their stellar evolution. We exclude these stars from the main CAMDs in our analysis, but state these here explicitly as some of these stars show large errors that could potentially turn them into viable younger candidates.



\begin{table*}
	\caption{2D trace-backs to NGC 2264 that are located on or near the red giant branch. Column 2+3: velocity in the respective NGC 2264 rest frame [rf]; Column 3: RV sources -- $^{a}$\textit{Gaia} DR2.}
	\label{tab:Red_giant_visitors}
	\begin{tabular}{lccccc} 
		\hline
		\textit{Gaia} DR2 source-id & 2D-velocity rf & Radial velocity rf    \\
		 & (km\,s$^{-1}$) & (km\,s$^{-1}$) &  \\
		\hline 		
		S\,Mon\\
		\hline
		3128637079593202048 & 34.0  $\pm$0.6   & 179.9 $\pm$2.9$^{a}$   \\
		3133974899309720960  & 28.0  $\pm$0.5    & 63.5   $\pm$3.5$^{a}$         \\
        3352423666192288896 & 18.2  $\pm$0.6   & -37.7 $\pm$4.8$^{a}$    \\
        3326929874196650496 & 11.3  $\pm$0.5   & 3.9   $\pm$4.2$^{a}$    \\
		\hline
		IRS\,1 \\
		\hline
		3324925468797373184 & 73.3  $\pm$0.8   & 36.7  $\pm$2.9$^{a}$       \\
		3331183854621309696  & 41.2  $\pm$0.9    & 26.7   $\pm$2.9$^{a}$          \\
	    3128637079593202048  & 33.5  $\pm$0.8    & 180.2  $\pm$2.9$^{a}$      \\
		3131775150199250176  & 29.8  $\pm$0.8    & 26.9   $\pm$2.9$^{a}$        \\   
		3134350554328161152  & 17.0    $\pm$1.1    & -52.4  $\pm$2.9$^{a}$           \\
		3133698303416819712 & 16.5  $\pm$1.0   & -5.6  $\pm$2.9$^{a}$    \\
		3351991940380416000 & 14.9  $\pm$1.2   & 4.2   $\pm$2.9$^{a}$     \\
		3326646788607684480 & 10.8  $\pm$0.9   & -8.6  $\pm$2.9$^{a}$        \\
		\hline
		IRS\,2 \\
		\hline
		3324925468797373184 & 73.0  $\pm$0.7   & 36.7  $\pm$2.9$^{a}$      \\
		3128637079593202048  & 35.1  $\pm$0.8    & 180.1  $\pm$2.9$^{a}$       \\
		3132441802143684864  & 49.0    $\pm$0.7    & -              \\
		3133974899309720960  & 28.5  $\pm$0.8    & 63.7   $\pm$3.5$^{a}$    \\
		3134350554328161152  & 17.8  $\pm$0.9    & -52.5  $\pm$2.9$^{a}$       \\
        3133698303416819712 & 17.1  $\pm$0.9   & -5.6  $\pm$2.9$^{a}$     \\
        3351991940380416000 & 14.3  $\pm$1.0   & 4.2   $\pm$2.9$^{a}$      \\
        3326646788607684480 & 11.3  $\pm$0.8   & -8.7  $\pm$2.9$^{a}$       \\
		\hline
	\end{tabular}
\end{table*}


\bsp	
\label{lastpage}
\end{document}